\documentclass[aps,prd,twocolumn,nofootinbib,superscriptaddress,showpacs,amsmath,amssymb]{revtex4}
\usepackage{graphicx,epsf}
\usepackage[]{latexsym}
\newcommand{\noi}{\noindent}

\newcommand{\be}{\begin{eqnarray}}
\newcommand{\ee}{\end{eqnarray}}
\newcommand{\bea}{\begin{eqnarray}}
\newcommand{\eea}{\end{eqnarray}}

\def\comment#1{}

\newcommand{\om}{\tilde{\omega}}

\usepackage{color}

\definecolor{darkred}{rgb}{.8,0,0}

\definecolor{darkblue}{rgb}{0,0,.7}

\definecolor{darkgreen}{rgb}{0,.7,0}

\begin{document}


\title{Gravitational Collapse in Scale-Dependent Gravity}

\author{Ramin Hassannejad}
\email{r.hassannejad@ut.ac.ir}
\affiliation{Department of Physics, University of Tehran, P.O. Box 14395-547, Tehran, Iran.}
\affiliation{Department of Physics, Shahid Beheshti University, G.C., Evin, Tehran 19839, Iran}

\author{ Gaetano Lambiase}
\email{lambiase@sa.infn.it}
\affiliation{Dipartimento di Fisica “E.R. Caianiello”, Universita’ di Salerno, I-84084 Fisciano (Sa), Italy.\\
and INFN - Gruppo Collegato di Salerno, I-84084 Fisciano (Sa), Italy.}

\author{Fabio Scardigli\footnote{corresponding author}}
 \email{fabio@phys.ntu.edu.tw}
\affiliation{Dipartimento di Matematica, Politecnico di Milano, Piazza Leonardo da Vinci 32, 20133 Milano, Italy}
\affiliation{Department of Applied Mathematics, University of Waterloo, Ontario N2L 3G1, Canada}

\author{Fatimah Shojai}
 \email{fshojai@ut.ac.ir}
\affiliation{Department of Physics, University of Tehran, P.O. Box 14395-547, Tehran, Iran.}


%


\begin{abstract}
In this paper we study an Oppenheimer-Snyder (OS)-like gravitational collapse in the general framework of 
scale-dependent gravity. We explore the collapse in spherically symmetric solutions suggested both by asymptotically safe gravity (characterized by a positive $\om$-parameter) and by scale-dependent gravity (negative $\om$-parameter), when a singularity at a finite positive radial coordinate is developed. The inner geometry of the collapsing star is described, as usual, by a spatially flat Friedmann-Lemaitre-Robertson-Walker (FLRW) metric, and matter is uniformly distributed without any assumptions about its equation of state.
The outer asymptotically-safe/scale-dependent black hole metric is smoothly matched to the inner geometry, and this yields the equation of motion of the star surface, the energy density, pressure, and equation of state of the collapsing matter. We study in detail the proper-time evolution of the event and apparent horizons. Finally, the constraints of the energy conditions on the equation of state, and its properties, are considered and discussed.
\end{abstract}
\maketitle

\section{\label{sec:level1}Introduction}

One of the most remarkable predictions of general relativity is the existence of black holes (BHs). In fact, since the formulation of general relativity by Albert Einstein, many people have tried to gain a deeper understanding of this theory through the study of BHs, which play a leading role in it. BHs have some classical and quantum features. The existence of the event horizon, and of trapped surfaces that are characterized by the negative values of the null expansion parameters, are classical properties. Besides, the singularity inside the event horizon\cite{Penrose:1964wq,Hawking:1970zqf,Hawking:1976ra}, Hawking radiation and information paradox \cite{Hawking:1975vcx,Almheiri:2020cfm}, and thermodynamics of BHs \cite{Bardeen:1973gs,Hawking:1976de,Bekenstein:1973ur,Bekenstein:1974ax} are quantum features of BHs that have attracted much of attention. Moreover, according to the weak cosmic censorship conjecture \cite{Penrose:1969pc}, if the stellar matter satisfies the null energy condition, the BH singularity remains hidden behind the event horizon, which specifies the causal structure of spacetime. Thus, the singularity is inaccessible to the asymptotic observer. In general relativity, BHs describe the final state of the classical gravitational collapse of massive stars. However, in certain models, the quantum effects can generate a repulsive pressure that counteracts the gravitational attraction, so that the singularity is avoided at the end of the collapse \cite{Frolov:1981mz}. Apart from the above theoretical achievements, some experimental results confirm the correctness of the general theoretical framework, such as the first images of the shadow of a BH provided by the Event Horizon Telescope \cite{EventHorizonTelescope:2019dse,EventHorizonTelescope:2019ths}, and the first observation of the merger of a binary BH that is confirmed by the LIGO and Virgo collaborations \cite{LIGOScientific:2016aoc}.

It is believed that only a complete theory of quantum gravity would be able to resolve the curvature singularities in BH spacetimes as well as in cosmological spacetimes, and to clarify other problems in gravitational physics. To build such a theory, many approaches have been explored, which can be found in the literature \cite{Jacobson:1995ab,Connes:1996gi,Jacobson:2015hqa,Reuter:1996cp,Rovelli:1997yv,Ashtekar:2004vs,Horava:2009uw,Verlinde:2010hp}. A further important problem is the perturbative non-renormalizability of gravity, where GR and extended gravity theories share the same drawbacks
\cite{tHooft:1974toh,Bern:2018jmv,Goroff:1985sz,Feynman:1963ax,DeWitt:1967yk,DeWitt:1967ub,Deser:1974cz}. In fact, many believe that gravity as we know it, should not be quantized at all, since Einstein's gravity is an effective theory
\cite{Donoghue:1995cz} that results from the quantization of some as yet unknown fundamental theory.
Thus, Einstein's gravity can be seen as an effective theory that is valid near a certain non-zero momentum scale $k$
\cite{PhysRevD.57.971}. In theories of gravity, the scale dependence is expected to modify the horizon, the quasinormal modes spectra of classical BH backgrounds, and the thermodynamics of BHs \cite{Koch:2016uso,Rincon:2017goj,Contreras:2017eza,Rincon:2018sgd,Contreras:2018dhs,Rincon:2018lyd,Rincon:2018dsq,Rincon:2019cix,Contreras:2018gpl}. The evolution of photon trajectories \cite{Fathi:2019jid}, the Sagnac effect \cite{Rincon:2019zxk}, transverse wormhole solutions \cite{Contreras:2018swc}, and some cosmological solutions \cite{Canales:2018tbn} have also been studied. Apart from scale-dependent gravity, there is a generalized version of the renormalizability condition proposed by Steven Weinberg based on a non-trivial fixed point of the underlying renormalization group (RG) flow for gravity, which he called asymptotic safety (AS) \cite{Weinberg:1976xy,Weinberg:1980gg,Weinberg:2009ca,Weinberg:2009bg}. In this picture, the metric fields remain the fundamental degree of freedom, and the low energy regime of classical GR is connected to the high energy regime through the RG \cite{Litim:2003vp,Morris:1993qb,Wetterich:1992yh,Niedermaier:2006wt,Reuter:2001ag}. This setup avoids the virulent divergences encountered in perturbation theory and leads to well-defined physical observables, such as S-matrix elements, provided that gravity has a non-trivial high-energy fixed point under the renormalization group \cite{Falls:2010he,Litim:2006dx,Niedermaier:2006ns,Litim:2007pbw,Litim:2003vp,Litim:2011cp}.

In this paper, we study the classical gravitational collapse of a spherically symmetric solution in the framework of both scale dependent gravity (SDG) \cite{Scardigli:2022jtt}, and asymptotically safe gravity (ASG) \cite{PhysRevD.62.043008}.  We consider a spherically symmetric star, such as in the standard Oppenheimer-Snyder (OS) model \cite{Oppenheimer:1939ue}. In particular, we have ignored the gradients of the density and pressure components. We also assume that there are no interactions between the background stress energy (the source of Newton's constant) and the stellar matter other than gravitational. Besides to spherical symmetry, we also assume that at the stellar surface, particles follow the radial geodesics, and further that the inner metric is smoothly connected to the outer BH metric (see Ref.\cite{Shojai_2022}). The velocity of the particles is assumed to be zero at spatial infinity. This simplifies our calculations and means that the inner Friedmann-Lemaitre-Robertson-Walker (FLRW) geometry of the collapsing star is spatially flat.

Considering a collapsing ball of perfect fluid without any special assumption on its equation of state, we compute here the stellar density and pressure as some functions of the proper time of collapse. We also see that the pure radial
pressure at the surface of the star is zero, showing the consistency of our assumption on the motion of particles along radial geodesics. Then, for the spherically symmetric solution of the SD/AS gravity framework, we have obtained the stellar equation of state, which results from the smooth joining  of the inner and outer spacetimes. We determine the position of the stellar surface, of the event horizon, and of the apparent horizon at a given proper time. 

We define the $\omega$-parameter in Appendix \ref{Appendix A} (see also references therein), and we put $\om=\omega/(2M)^2$, where $M$ is the geometric mass of the star (see also Eqs.\eqref{G}-\eqref{dwdlmskp}). According to the possible values of the $\om$-parameter, we study the following three cases.

In the case of ASG collapse ($\om > 0$) the metric is singularity free, the radial coordinate of the surface of the star can span all the positive real numbers ($\tilde{R}\geq 0$), and it takes an infinite proper time to the surface of the star to reach the center $\tilde{R}=0$, where there is no singularity. In this case the dynamics of the gravitational collapse is similar to that described for the Hayward metric or for other non singular black holes (see e.g. \cite{Shojai_2022,Bardeen,Hayward:2005gi,Flachi:2012nv,Frolov:2016pav,Cadoni:2022chn,Malafarina:2022oka,Bonanno:2023rzk}). 

For the Schwarzschild BH ($\om = 0$), the radial coordinate of the surface of the star goes to zero in a finite proper time and at $\tilde{R}=0$ there is a singularity. All the known physics is broken in the central singularity, which results to be pointlike, with a zero volume. 

In the case of SDG collapse ($\om < 0$) we will see that, as in the Schwarzschild case, it takes a finite proper time for the collapse to complete. However, in the SDG spherically symmetric solution, the singularity is at a \textit{finite positive} radial coordinate, therefore the 'naive' euclidean volume of this ``extended'' singularity has a finite positive value. In the final stage of the collapse, the surface of the star hits this central spherical singularity, and this process takes a finite proper time. The nature of this extended central singularity will be certainly a topic for future investigations (see e.g. Refs.\cite{Casadio:2021cbv,Casadio:2023ymt}).

The outline of this paper is as follows.
In Sec.\ref{mssk}, we recall the SDG/ASG action and the 4-dimensional spherically symmetric metric. Then in Sec.\ref{dmkdk}, we develop the OS-like gravitational collapse to the well-known BH solution of 4-dimensional SDG/ASG. Section \ref{ddjdjjd} deals with the identification of the inner and outer horizons of the SDG/ASG metric. Section \ref{EEH} describes the evolution of the apparent and event horizons as they develop during the collapse process. The equation of state of the stellar matter is derived and discussed in Sec.\ref{njjvft}, and the energy conditions are discussed in detail in Sec.\ref{EC}. In Sec.\ref{pfi} we briefly sketch some possible future investigations. Finally, we summarize our results in Sec.\ref{nddj}.

Throughout this paper, the signature of the metric tensor is assumed to be $(-, +, +, +)$. Unless explicitly specified, we use geometrized units, i.e., $G_{0} = c =\hbar= 1$. A dimensionless variable is denoted by a tilde and is obtained by normalizing it with respect to the Schwarzschild radius, i.e. $\tilde{\tau} = \tau/2M$, $\tilde{r} = r/2M$, and 
$\tilde{\omega}=\omega/(2M)^2$.

\section{Asymptotically safe or scale dependent gravity}
\label{mssk}

We consider the effective average action $\Gamma_{k}[g_{\mu\nu}]$, where $k$ is a scale parameter with the dimension of a mass (or equivalently, of an energy/momentum) as follows \cite{BONANNO2017254,PhysRevD.62.043008}

\begin{align}
\label{s;[sdf}
\Gamma_{k}=\frac{1}{16\pi G(k)}\int d^4x\sqrt{-g}(R-2\Lambda(k))+S_{k}\,.
\end{align}

In the above expression,  the cosmological constant $\Lambda(k)$ and the Newton constant $G(k)$ are scale-dependent coupling constants. Furthermore, $R$ and $S_{k}$  are the Ricci scalar and the action for the matter sector, respectively. In the following, since our considerations are local, we shall ignore the cosmological constant, $\Lambda(k)\simeq 0$.

 The effective average action \eqref{s;[sdf}, $\Gamma_{k}$, describes the physical system at scale $k$, and provides a method to derive the effective field equations in terms of the effective metric \cite{BONANNO2017254},
\begin{align}
\label{dldncsp}
\frac{\delta\Gamma_{k}}{\delta g_{\mu\nu}}[\langle g \rangle_{k}]=0\,,
\end{align}
where the quantities $\langle.\rangle$ can be interpreted as averaged over (Euclidean) spacetime volumes. Substituting equation \eqref{s;[sdf} into \eqref{dldncsp}, we get the effective field equations as follows \cite{BONANNO2017254},
\begin{align}
\label{axskss}
R_{\mu\nu}[\langle g \rangle_{k}]-\frac{1}{2}R\langle g_{\mu\nu}\rangle_{k}=8\pi G(k)\langle T_{\mu\nu}\rangle_{k}
\end{align}
in which $R_{\mu\nu}$ and $T_{\mu\nu}$ are the Ricci and energy-momentum tensors, respectively. Further explanations and details are collected in Appendix \ref{Appendix A}. 

The interior geometry of our collapsing star is described in comoving coordinates by a spatially flat Friedmann-Lemaitre-Robertson-Walker (FLRW) metric 
\begin{align}
\label{mwdnqpqwd}
ds^2=-d\tau^2+a(t)^2(dr^2 + r^2 d \Omega^2)\,,
\end{align}
while the matter content of the star is depicted as a perfect fluid characterized by an energy-momentum tensor that, again in comoving coordinates, reads $T_{\mu\nu}=(\rho +p)u_{\mu}u_{\nu}+pg_{\mu\nu}$, so that 
$T^{\mu}_{\ \nu}=g^{\mu\sigma}T_{\sigma\nu}=diag[-\rho,p,p,p]$, since $g_{\mu\nu}u^{\mu}u^{\nu}=-1$. Substituting the above FLRW metric into the field equations \eqref{axskss}, we obtain the improved Friedmann and continuity equations as follows \cite{BONANNO2017254}
\begin{align}
\label{dcwpdw[e}
&H^2=\left(\frac{\dot{a}}{a}\right)^2=\frac{8\pi G(t)}{3}\rho\\
&\dot\rho+3\left(\frac{\dot{a}}{a}\right)(\rho+p)=-\frac{\rho\dot{G}}{G(t)}
\label{dmcwldk}
\end{align}
where the cutoff identification $k\rightarrow k(t)$ is applied. The continuity equation has been obtained as usual from the Bianchi identity $\nabla_{\mu}\big[8\pi G(t)T^{\mu\nu}\big]=0$ satisfied by the effective Einstein's equations (where now 
$G=G(t)$).

As mentioned before, the main idea in ASG is to solve the beta function for the gravitational coupling in order to compute the Newton constant as a function of some energy scale $k$. In Appendix \ref{Appendix A} we explain how to find 
the scale dependent gravitational constant knowing the beta function. The running Newton constant is displayed in equation \eqref{dwodnwdweo}. So, from now on, as customary in the ASG literature, we replace the Newton constant $G_{0}$ everywhere with the running gravitational constant $G(k)$. Further, to study the BH solutions of ASG, we need also to explicit the radial dependence of the scale parameter $k(r)$, and this is described by equation \eqref{dwdwod} in Appendix \ref{Appendix B}.
Finally, the running gravitational constant in asymptotically safe gravity can be written as \cite{PhysRevD.62.043008,PhysRevD.105.124054}

\begin{align}
\label{G}
G(r)=\frac{G_0 r^3}{r^3 + \omega G_0\hbar(r + \gamma G_0 M)}
\end{align}
or, with the geometrized units adopted, and in dimensionless variables, 
\begin{align}
\label{dkdmwm}
G(\tilde{r})=\frac{\tilde{r}^3}{\tilde{r}^3+\tilde{\omega}(\tilde{r}+\frac{\gamma}{2} )}
\end{align}
where $\tilde{r} = r/2M$, $\tilde{\omega}=\omega/(2M)^2$, $M$ is the geometric mass of the star, $\omega$ and $\gamma$ are the free parameters of ASG. In this paper we always assume $\gamma > 0$, in agreement with the indications of Ref.\cite{PhysRevD.62.043008}. 
We explicitly point out once again that we can properly talk of ``asymptotically safe'' gravity only when $\om > 0$ (see Eq.\eqref{dwodnwdweo}). Only in that case we have in fact $G(k) \to 0$ when $k \to \infty$ (i.e. when $r\to 0$). If we admit also values $\om < 0$ (as it is in the present article), then we have to speak more generally of ``scale dependent'' gravity. In such cases in fact we have $G(k) \to \infty$ when 
$k \sim k_{Planck}$ (see again Eq.\eqref{dwodnwdweo}). 
 
Considering the running gravitational constant \eqref{dkdmwm}, the spherically symmetric solution of the AS/SD gravity is given by
\begin{align}
\label{dwdlmskp}
&d\tilde s^2=-f(\tilde{r})d\tilde t^2+\frac{d\tilde{r}^2}{f(\tilde{r})}+\tilde{r}^2d\Omega^2\notag\\
&f(\tilde r)=1-\frac{\tilde{r}^2}{\tilde{r}^3+\tilde{\omega}(\tilde{r}+\frac{\gamma}{2})}\,,
\end{align}
where $\tilde s=s/2M$. 
In non-rescaled variables we would have $ds^2=-f(r)dt^2+f(r)^{-1}dr^2+r^2d\Omega^2$ with $f(r)=1-2MG(r)/r$. 
It is clear that in the low energy scale limit $(\tilde{r}\to\infty, \text{or}~k\to 0)$ equation \eqref{dwdlmskp} tends to the standard Schwarzschild metric \cite{PhysRevD.105.124054,Scardigli:2022jtt},
\begin{align}
\label{dwmlwp[d}
f(\tilde{r})=1-\frac{1}{\tilde{r}}
\end{align} 
which it is independent of $\om$ and $\gamma$, as expected.
At high energy scales $(\tilde{r}\to 0  \hspace{.1cm}\text{or}~k\to \infty)$, under the assumption $\tilde{\gamma}\ne 0$, we have the limit lapse function

\begin{align}
\label{dnqwekdd}
f(\tilde{r})=1-\frac{2\tilde{r}^2}{\gamma\tilde{\omega}}\,,
\end{align}
which yields a deSitter metric for $\tilde{\omega}\gamma>0$, while for $\tilde{\omega}\gamma<0$ the above corresponds to an anti-de Sitter metric \cite{PhysRevD.105.124054,Scardigli:2022jtt}. As said, hereafter we assume $\gamma > 0$, see Ref.\cite{PhysRevD.62.043008}. 

The lapse function \eqref{dwdlmskp} describes a regular space-time when $\om >0$. The radial coordinate $\tilde{r}$ can span all the non negative real values, the denominator of \eqref{dwdlmskp} never vanishes, and in particular there is no singularity at $\tilde{r}=0$. On the contrary, a straightforward graphical analysis shows that for any $\tilde{\omega}<0$ there is always a singularity at a certain radial coordinate $\tilde{r}_0 > 0$. 

\begin{figure}[h]
\centering
\includegraphics[scale=.46]{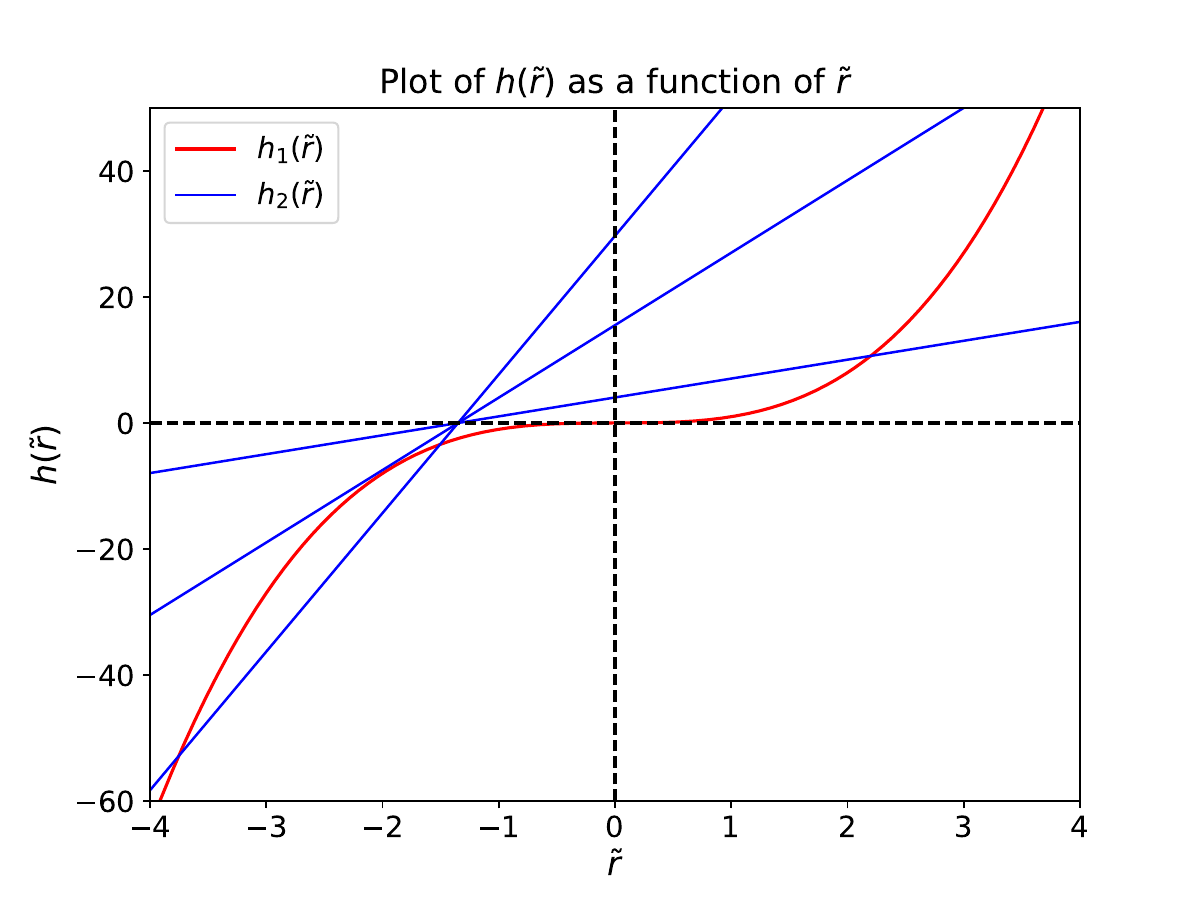}
\caption{Analysis of the denominator of \eqref{dwdlmskp} for $\gamma >0$. The plots of $h_1(\tilde{r})=\tilde{r}^3$ and $h_2(\tilde{r})=-\om(\tilde{r}+\gamma/2)$ show that for any $\om<0$ there is always one single positive zero $\tilde{r}_0>0$ of the denominator of \eqref{dwdlmskp}, namely a singularity of the metric.}
\label{fig1}
\end{figure}

In fact (see Fig.1, and also \cite{Scardigli:2022jtt}) the denominator of \eqref{dwdlmskp} has always one real positive root for any $\tilde{\omega}<0$. This singularity $\tilde{r}_0$ is actually a curvature singularity (namely, there is no way to remove it), as a computation of the Kretschmann scalar shows 
\begin{align}
\label{djjlwqqw}
&R_{\mu\nu\alpha\beta}R^{\mu\nu\alpha\beta}=4 \bigg[ \left(\tilde{r}^3+\tilde{\omega} \tilde{r} +\frac{\tilde{\omega}\gamma}{2} \right)^4\notag\\
&+\left(\tilde{r}^3+\tilde{\omega} \tilde{r} +\frac{\tilde{\omega}\gamma}{2}\right)^2 \left( \tilde{r}^3-\tilde{\omega}  (\gamma+ \tilde{r})\right)^2+\notag\\
&\left(\frac{\gamma ^2 \tilde{\omega} ^2}{4}-\tilde{r}^3 \tilde{\omega}  (\frac{7\gamma}{2} +3\tilde{r})+\tilde{r}^6\right)^2 \bigg] \bigg/\left(\tilde{r}^3+\tilde{\omega} \tilde{r} +\frac{\tilde{\omega}\gamma}{2}\right)^6\,.
\end{align}
Clearly, if $\tilde{r}_0$ solves $\tilde{r}^3+\tilde{\omega} \tilde{r} +\frac{\tilde{\omega} \gamma}{2} =0$, then the Kretschmann scalar is singular for $\tilde{r} \to \tilde{r}_0$.

Summarizing:\\
- For any $\tilde{\omega}>0$ the denominator of the metric \eqref{dwdlmskp} has always only one single negative zero, therefore the metric is singularity-free in the physical region $\tilde r >0$.\\
- For any $\tilde{\omega}<0$  the first root $\tilde{r}_{0}$ of the denominator is always positive, while, as suggested by diagrams in Fig.1, the second root $\tilde{r}_{1}$ and the third root $\tilde{r}_{2}$ are always negative, or complex conjugate. Therefore for any \textit{negative} value of $\tilde\omega$, we have only one singularity for this BH metric at 
$\tilde{r}_{0}>0$.

\section{OS-like collapse to a scale dependent black hole}
\label{dmkdk}

We study here the OS-like gravitational collapse of a star whose external metric is given by the spherically symmetric solution \eqref{dwdlmskp} of the AS/SD gravity. The line element can be more compactly written as

\begin{align}
\label{I}
&d\tilde s^2 = - \Big(1-\frac{1}{\tilde{r}\alpha(\tilde{r})}\Big)d\tilde t^2 + \Big(1-\frac{1}{\tilde{r}\alpha(\tilde{r})}\Big)^{-1} d\tilde{r}^2 + \tilde{r}^2 d \Omega^2\nonumber \\
&{\rm with} \quad \alpha(\tilde{r})=1+\frac{\tilde{\omega}(\tilde{r}+\frac{\gamma}{2})}{\tilde{r}^3}\,.
\end{align}
Like in the standard OS scenario \cite{Oppenheimer:1939ue}, we assume that the stellar matter
is described by a homogeneous and isotropic perfect fluid. However, here we don't fix a priori the equation of state of our fluid to be $p=0$ (i.e. dust), as OS did. Depending on the initial condition
of star, the interior geometry of the star is governed by a spatially closed or flat FLRW metric, 
which should be smoothly matched, at the surface of the star, to the outer BH spacetime \eqref{I}. 

To describe the collapse, we adopt here the framework 
discussed in Ref.\cite{blau2011lecture}, (in particular Sections 29.4-7). Also important are the Refs.\cite{Israel:1966rt,Martel:2000rn,Adler:2005vn,Kanai:2010ae}. 

Hence, we consider a spatially flat FLRW metric for the interior
geometry of the star and choose the time coordinate of the exterior (interior) metric to be the
proper time of freely falling (comoving) particles (see also \cite{Shojai_2022}).

According to the above choice, we model the outer Schwarzschild-like metric \eqref{I} in terms of Painlev\'e-Gullstrand (PG) coordinates \cite{Painlev,Gullstrand,Martel:2000rn} as
\begin{align}
\label{BB}
ds^2 = - d\tau^2 + \left(dr+ \sqrt{\frac{2M}{r\alpha(r)}} d\tau\right)^2 + r^2 d\Omega^2  \,,
\end{align}
where for sake of clarity we restore for a while the black hole mass $M$, with $\alpha(r)=1+\omega(r+\gamma M)/r^3$.
These are coordinates adapted to a radially infalling observer who starts at rest from infinity, with $dr = -\sqrt{2M/r \alpha(r)}d\tau$ (so that here $\tau$ is  the proper time of the infalling observer).

On the other hand, we model the interior geometry via a spatially flat FLRW metric
\be
\label{FRW}
ds^2 = - d \tau^2_c + a^2(\tau_c)(d r^2_c + r^2_c d\Omega^2)
\ee
where now $\tau_c$ is the the proper time of comoving observers, $a(\tau_c)$ is the scale factor, and $r_c$ is the usual comoving radial coordinate of the FLRW metric. By introducing the interior radial coordinate defined as
\be
r_i(\tau_c) = a(\tau_c)r_c 
\ee 
and noticing that $a(\tau_c)dr_c = dr_i-\dot{a}(\tau_c)r_c d\tau_c$,
we can recast the metric \eqref{FRW} in PG-like coordinates as
\be
\label{EE}
ds^2 = - d\tau^2_c + (dr_i - r_i H(\tau_c) d\tau_c)^2 + r_i^2d\Omega^2
\ee
where $H(\tau_c) = \dot{r_i}(\tau_c)/r_i(\tau_c) = \dot{a}(\tau_c)/a(\tau_c)$ is the Hubble parameter. In this form the metric is adapted to comoving observers 
(fixed $r_c$), which obey the Hubble relation $dr_i/d\tau_c = r_i H(\tau_c)$.

Comparison of the two metrics, the exterior Schwarzschild-like metric in Painlev\'e-Gullstrand
form \eqref{BB}, and the interior FLRW metric in Painlev\'e-Gullstrand-like form \eqref{EE}, makes it
manifest that, in order to have the continuity of the metric across the surface of the star, we should identify not only the exterior PG proper time $\tau$ of free falling observers with the interior proper time $\tau_c$ of comoving observers $(\tau \equiv \tau_c)$, but also the PG-Schwarzschild radial coordinate $r$ with the cosmological radial coordinate 
$r_i = a(\tau_c)r_c$, ($r\equiv r_i$).
Moreover, if we call $R(\tau)$ the common value of the two radial coordinates $r(\tau)$ and $r_i(\tau)$ at the star surface, namely $R(\tau):=r(\tau)|_{\rm star \ surface}=r_i(\tau)|_{\rm star \ surface}$, then demanding the continuity of the metric on the surface implies
\be
\dot{R}(\tau) = -\sqrt{\frac{2M}{R(\tau)\alpha(R)}}\,,
\label{FunEq0}
\ee
since on the star surface we can write
\be
\label{Hubble0}
H(\tau) = \frac{\dot{a}(\tau)}{a(\tau)} = \frac{\dot{a}(\tau)R_c}{a(\tau)R_c} = \frac{\dot{R}(\tau)}{R(\tau)}\,,
\ee
where $R_c$ is the comoving radial coordinate of the star surface.
To get a smooth transition between the inner and the outer geometries, we further demand the continuity of the first derivatives of the metric across the surface of the star. This is equivalent to require that the interior and exterior extrinsic curvatures be equal at the surface, namely the continuity of the extrinsic curvature $K_{\alpha\beta}$ across the surface of the star. These are the well known \textit{Israel junction conditions} \cite{Israel:1966rt}. 
Following now \cite{Israel:1966rt,blau2011lecture}
a straightforward calculation gives the components of the extrinsic curvature on the star surface as
\begin{align}
\label{EWHH}
^{(in)}K^{\tau}_{\tau}&=0 \quad \quad \quad \ \ ^{(in)}K^{\theta}_{\theta}  = \ \ ^{(in)}K^{\phi}_{\phi}=1/R  \notag\\
^{(out)}K^{\tau}_{\tau}&=\dot{\xi}/R \quad \quad ^{(out)}K^{\theta}_{\theta}  = \ ^{(out)}K^{\phi}_{\phi}=\xi/R
\end{align}
where 
\be
\xi=\sqrt{\dot{ R}^2+1-2M/(R\,\alpha(R))}
\ee 
with 
\be
\alpha(R)=1+\omega(R+\gamma M)/R^3\,.
\ee 
Demanding a smooth joining of the $(in)$ and $(out)$ geometries in Eq.\eqref{EWHH} implies $\xi=1$, and this yields once again the following differential equation for the star surface coordinate $R(\tau)$
\begin{align}
\label{FunEq}
\dot{R}(\tau)^2 = \frac{2M}{R(\tau)\,\alpha(R)}\,.
\end{align}
We see that Eq.\eqref{FunEq} is the same as Eq.\eqref{FunEq0}, so the condition \eqref{FunEq} guarantees at once the continuity of the metric and of its first derivatives. We shall see that, by demanding a smooth transition of the metric across the boundary of the star, namely imposing the Israel junction conditions, we have the absence of distributionally localized energy-momentum at the surface of the star, i.e. no energy layer and no additional pressure appears at the surface of the star, and the particles can freely follow radial geodesics.

In terms of the dimensionless variables $\tilde{R}=R/2M$, $\tilde{\tau}=\tau/2M$, $\om=\omega/(2M)^2$ we have
\be
\label{FunEqTilde}
\dot{\tilde{R}}(\tilde{\tau})^2 = \frac{1}{\tilde{R}\,\alpha(\tilde{R})} = 
\frac{\tilde{R}^2}{\tilde{R}^3+\om(\tilde{R}+\frac{\gamma}{2})}\,,
\ee
where $\alpha(\tilde{R})=1+\om(\tilde{R}+\frac{\gamma}{2})/\tilde{R}^3$ as in \eqref{I}.

It is important and physically enlightening to note that Eq.\eqref{FunEq} can be obtained also through the Newtonian limit of metric \eqref{I}. 
In fact, consider particles of (small) mass $m$ in radial free fall from rest at infinity, each one moving along a radial timelike geodesic of the spherically symmetric space-time \eqref{I}. For $r$ large, the Newtonian limit of \eqref{I} reads 
(see e.g. \cite{Weinberg:1972kfs})
\be
g_{00} = - (1 + 2\Phi_{Newton})
\ee
with
\be
\Phi_{Newton} = - \frac{M}{r \alpha(r)}
\ee 
where $\alpha(r)=1+\omega(r+\gamma M)/r^3$. The total energy of a free falling particle of mass $m$ is 
\be
K+U \ = \ \frac{1}{2}m \dot{r}^2 \ + \ m\Phi(r) \ = \ E_0 = 0
\ee
where $E_0=0$ since the particle starts from rest at infinity. Then immediately we get 
\be
\dot{r}^2 = \frac{2 M}{r \alpha(r)}
\label{T0}
\ee
which coincides with Eq.\eqref{FunEq}. This coincidence shows that, at the surface of the star, each freely falling particle moves as if it starts from rest at infinity and runs along a radial timelike geodesic. This confirms once again, that, thanks to the Israel junction conditions, no surface energy-momentum tensor appears on the star surface (it would be precisely equivalent to a discontinuity of the extrinsic curvature) and there is a smooth transition of the metric across it. 
In dimensionless variables Eq.\eqref{T0} reads
\begin{align}
\label{T}
\dot{\tilde{r}}^2=\frac{1}{\tilde{r}\alpha(\tilde{r})}
\end{align}

%




Considering Eq.\eqref{FunEqTilde}, the differential equation properly describing the surface of a collapsing star reads 
\be
\label{CC}
\dot{\tilde R}(\tilde \tau) =-\frac{\tilde R}{\sqrt{{\tilde R^3+\tilde\omega(\tilde R+\frac{\gamma}{2})}}}
\ee
where the minus sign is chosen just to describe a contraction.
Therefore, in the following we shall study various cases of Eq.\eqref{CC}, according to the values of $\om$. 

\subsection{Case $\om \ = \ 0$.}

\begin{figure}[h]
\centering
\includegraphics[scale=0.5]{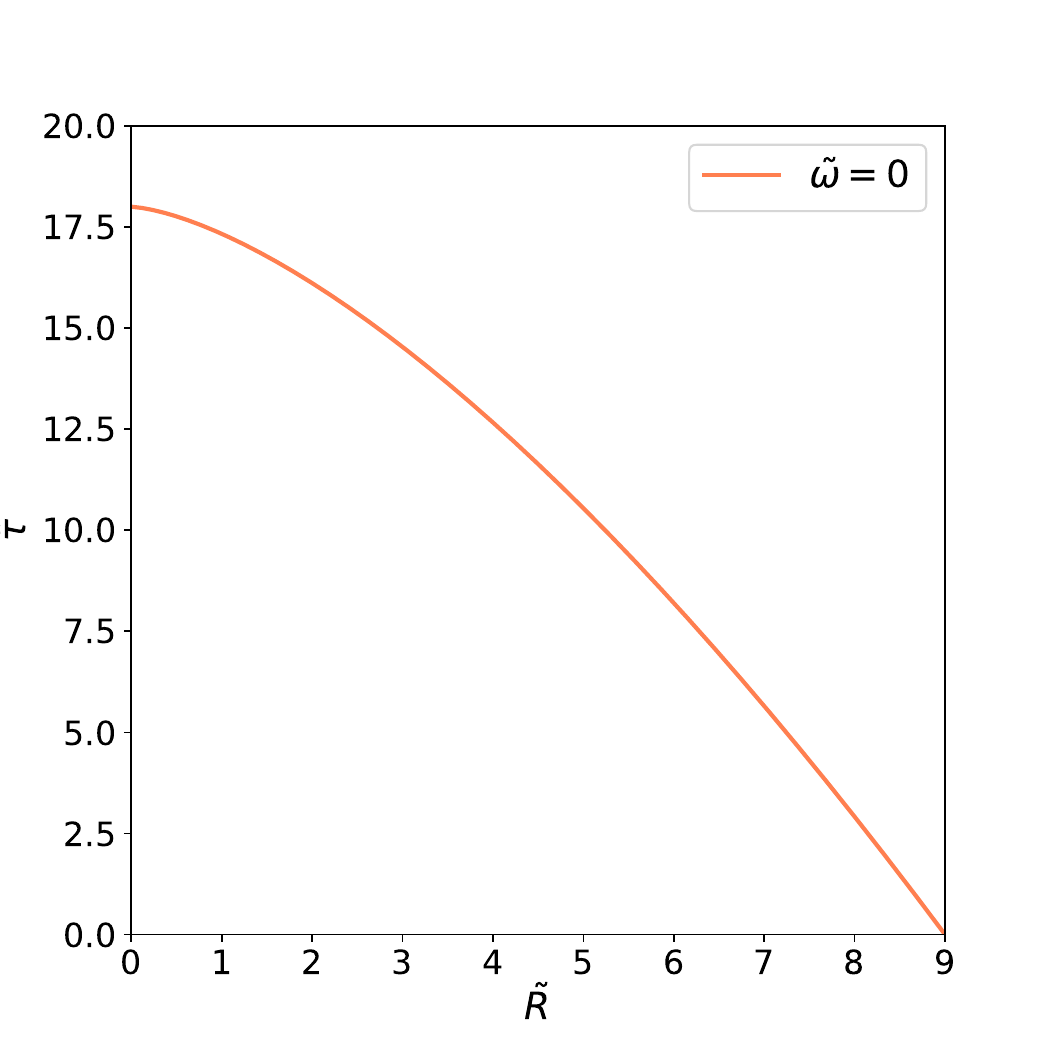}
\caption{Gravitational collapse in a Schwarzschild metric ($\om=0$). Proper time runs upward. 
The star's surface hits the central singularity at $\tilde{R}=0$ in a finite lapse of proper time $\tilde \tau$.}
\label{fig2}
\end{figure}

When $\om=0$ the metric \eqref{I} is just the usual Schwarzschild metric. The collapse is therefore the one described for the first time by OS in \cite{Oppenheimer:1939ue}. In fact, the equation of motion for the star surface reads
\be
\label{SCHW}
\dot{\tilde{R}}(\tilde{\tau}) =-\frac{1}{\sqrt{\tilde{R}}}\,,
\ee  
which, once integrated with the obvious initial condition $\tilde R = \tilde R_0$ at $\tilde \tau =0$, yields
\be
\label{SchColl}
\tilde{\tau} = \frac{2}{3}\left(\tilde{R}_0^{3/2} -  \tilde{R}^{3/2}\right)
\ee
or
\be
\label{SchColl2}
\tilde{R} = \left(\tilde{R}_0^{3/2} -  \frac{3}{2}\tilde{\tau}\right)^{2/3}\,.
\ee  
As it is well known (see Fig.\ref{fig2}), in a finite proper time $\tilde \tau_0 = \frac{2}{3}\tilde{R}_0^{3/2}$ the surface $\tilde{R}(\tilde\tau)$ hits the central singularity at $\tilde{R}=0$.

\subsection{Case $\om \ > \ 0$.}\label{CollOmPos}

In this case the metric \eqref{I} is singularity free. We are in the area of what is properly called \textit{asymptotic safety}. Also Eq.\eqref{CC} is singularity free, namely $\tilde R$ can span all the non-negative real numbers, i.e. $\tilde R \geq 0$. Eq.\eqref{CC} can be integrated explicitly here also. However the exact solutions are given in terms of elliptic integrals (see Appendices \ref{Appendix S}, \ref{Appendix C}), and therefore are not particularly appealing or clarifying. It is both easier and more instructive to look at the asymptotic behavior of the solutions. \\   

I) For $\tilde{R}$ large and $\tilde\tau$ small, Eq.\eqref{CC} reads
\be
\dot{\tilde R}(\tilde \tau) \simeq -\frac{1}{\sqrt{\tilde{R}}}
\left(1 - \frac{\om}{2\tilde{R}^2} - \frac{\gamma\om}{4\tilde{R}^3}\right) \,;
\label{approx}
\ee 
neglecting the $1/\tilde{R}^2$, $1/\tilde{R}^3$ terms, we have $\dot{\tilde{R}} \simeq -1/\sqrt{\tilde{R}}$, which yields again the Schwarzschild collapse solution \eqref{SchColl}, namely
$\tilde{\tau} = (2/3)\left(\tilde{R}_0^{3/2} -  \tilde{R}^{3/2}\right)$.\\

\begin{figure}[h]
\centering
\includegraphics[scale=0.5]{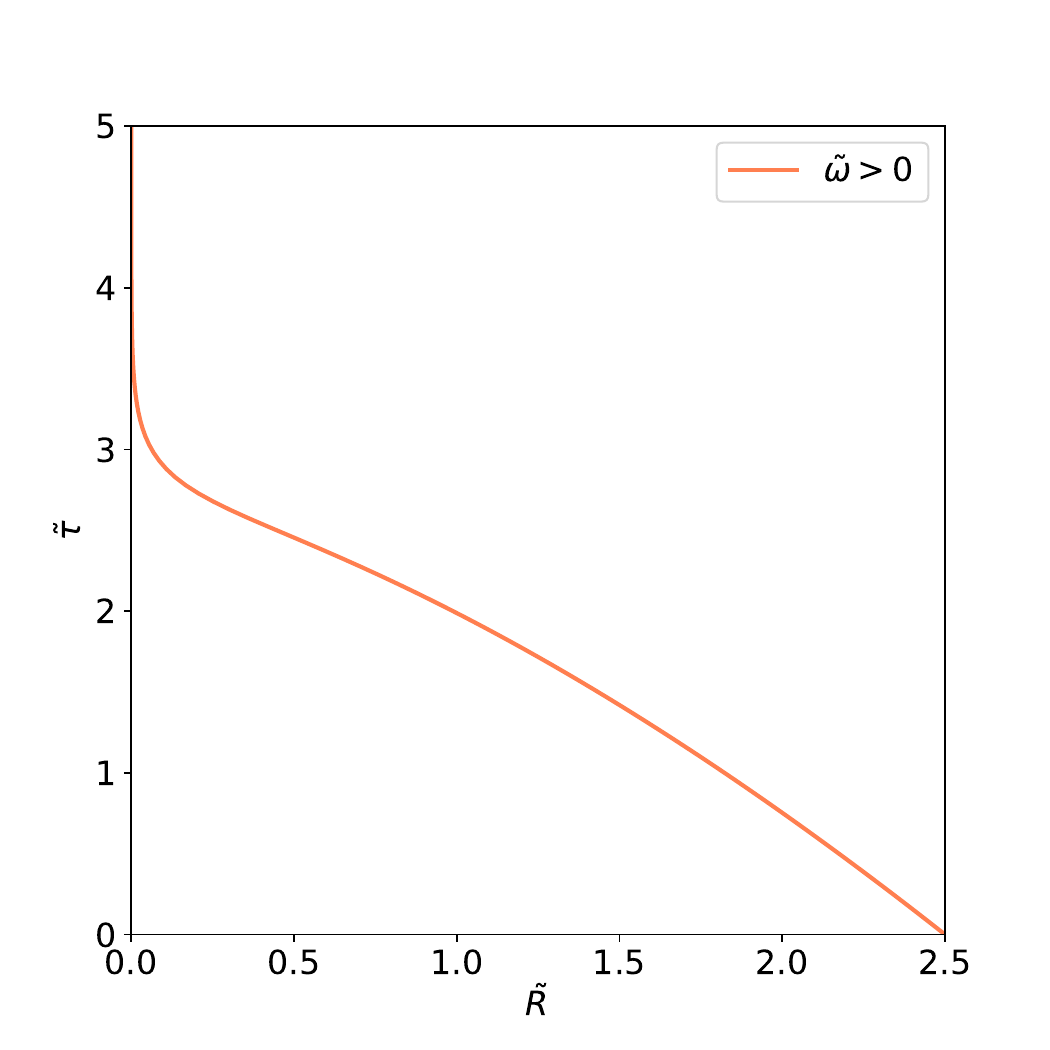}
\caption{Gravitational collapse in the metric \eqref{I} with $\om>0$. 
The star's surface hits the center $\tilde{R}=0$ (where there is no singularity) in an infinite lapse of proper time $\tilde \tau$.}
\label{fig3}
\end{figure}

II) For $\tilde{R}$ small and $\tilde\tau$ large, we have
\be
\label{appr}
\dot{\tilde R}(\tilde \tau) \simeq - \frac{\tilde R}{A}
\ee
where $A=(\om\gamma/2)^{1/2}$. The solution presents here a very interesting physical feature. An integration with the initial condition $\tilde{R}(\tilde{\tau}=0)=\tilde{R}_0$ yields
\be
\tilde{\tau} = A \log \frac{\tilde{R}_0}{\tilde{R}}
\ee
or
\be
\tilde{R} = \tilde{R}_0 \, e^{-\tilde{\tau}/A}\,.
\label{exp}
\ee
Of course, such solution cannot be held as completely reliable, given that the asymptotic equation \eqref{appr} holds when 
$\tilde{R}$ is small, while instead we chose the initial condition $\tilde{R}(\tilde{\tau}=0)=\tilde{R}_0$, where in principle 
$\tilde{R}_0$ could be large. However the above solution may give a sharp idea of the asymptotic behavior of the ``true'' solution. And in fact a numerical integration of Eq.\eqref{CC} yields the diagram in Fig.\ref{fig3}, so confirming the above analytical insights.

Clearly, it emerges from Eq.\eqref{exp} that the surface of the star takes an infinite proper time to reach the center $\tilde R =0$, where there is no singularity at all! This could be regarded as a quite general property of singularity-free metrics. In fact, further examples of gravitational collapses to singularity-free metrics, like Hayward metric, or Bardeen metric, can be found e.g. in Ref.\cite{Shojai_2022}, and they all behave in a similar way.

\subsection{Case $\om \ < \ 0$.}\label{omneg} 

The metric \eqref{dwdlmskp}, or \eqref{I}, with $\om  <  0$, presents a particular interest, since it seems to be a good candidate to describe some specific features of the so called Planck stars (see Ref.\cite{Scardigli:2022jtt}). Of course, as already previously explained, we are here in the realm of \textit{scale-dependent gravity}. We showed in Sec.\ref{mssk} Fig.\ref{fig1} that the metric \eqref{dwdlmskp}-\eqref{I}, as well as the equation of motion \eqref{CC}, have always (for any $\om<0$) one singularity at a positive radial coordinate $\tilde r_0 > 0$. In particular, for $\om<0$ we can write
\be
D(\tilde R) \ := \ \tilde R^3+\om\left(\tilde R+\frac{\gamma}{2}\right) \ = \ (\tilde R - \tilde r_0)(\tilde R^2 + b\tilde R + c)\notag
\ee 
where $\tilde r_0, b, c$ are functions of $\om,\gamma$, and the trinomial $\tilde R^2 + b\tilde R + c$ can have only negative zeros, if any. Since in Eq.\eqref{CC} we take the square root, necessarily we require 
$D(\tilde R)\geq 0$. Moreover it is physically meaningful to consider $\tilde R \geq 0$ only. Therefore, all this implies $\tilde R \geq \tilde r_0$. 

Here also the explicit integration of \eqref{CC} is quite involved, so it is much better to study the asymptotic behavior of the solutions. \\

I) For $\tilde{R}$ large ($\tilde{R} \gg \tilde r_0$) and $\tilde\tau$ small we have, as in Eq.\eqref{approx},
\be
\dot{\tilde R}(\tilde \tau) \simeq - \frac{1}{\sqrt{\tilde R}}
\ee
and the solution resembles again the Schwarzschild solution \eqref{SchColl}.\\

\begin{figure}[h]
\centering
\includegraphics[scale=0.5]{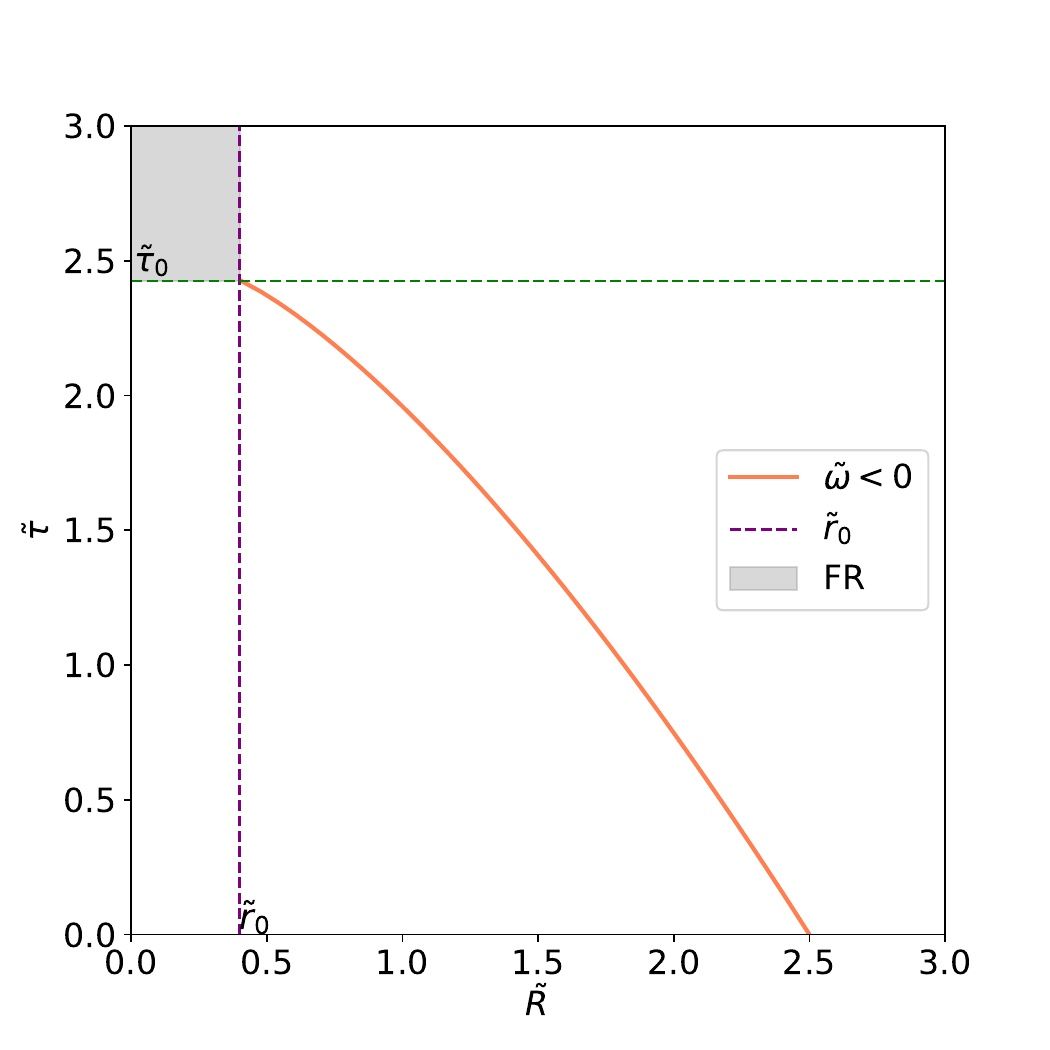}
\caption{Gravitational collapse in the metric \eqref{I} with $\om<0$. 
The star's surface hits the singularity at $\tilde{R}=\tilde r_0>0$ in a finite lapse of proper time $\tilde \tau$. The grey area is the so-called Forbidden Region (see Sec.\ref{Sb5E} for more details).}
\label{fig4}
\end{figure}

II) When instead $\tilde R$ is close to the limit value $\tilde r_0$ ($\tilde R \to \tilde r_0$), and $\tilde\tau$ is large, we have
\be
\dot{\tilde R}(\tilde \tau) &=& - \frac{\tilde R}{\sqrt{(\tilde R - \tilde r_0)(\tilde R^2 + b\tilde R + c)}} \notag\\
&\simeq& -\frac{\tilde r_0}{B\sqrt{\tilde R - \tilde r_0}}
\ee
with $B=(\tilde r_0^2+b\tilde r_0+c)^{1/2}$. An integration with the usual initial condition 
$\tilde{R}(\tilde{\tau}=0)=\tilde{R}_0$ yields
\be
\label{pt}
\tilde\tau = \frac{2}{3}\frac{B}{\tilde r_0}\left[(\tilde R_0 - \tilde r_0)^{3/2} - (\tilde R - \tilde r_0)^{3/2} \right]
\ee
or
\be
\tilde R = \tilde r_0 + \left[(\tilde R_0 - \tilde r_0)^{3/2} - \frac{3}{2}\frac{\tilde r_0}{B}\tilde\tau\right]^{2/3}
\ee
The above solution clearly indicates that in a finite proper time the surface of the star $\tilde R (\tilde\tau)$ hits the singularity at $\tilde R = \tilde r_0 >0$. Of course, the exact proper time lapse computed through the \eqref{pt} is not completely reliable, since \eqref{pt} is an approximate solution valid for $\tilde R \simeq \tilde r_0$. However a numerical integration of \eqref{CC} confirms the circumstance suggested by the analytic methods (see Fig.\ref{fig4}).

Furthermore, for $\tilde{\omega}<0$, in contrast with the regular BHs \cite{Shojai_2022}, the collapsing velocity goes to infinity at the singularity $\tilde R=\tilde r_{0}$. This is similar to the conventional OS collapse, where the collapsing star reaches the singularity of the Schwarzschild BH with a divergent speed. On the contrary, for $\om>0$ the collapsing velocity goes to zero, and this is similar to the behavior of the collapse of regular BHs in general relativity \cite{Shojai_2022}, and of BHs in 4-dimensional Einstein-Gauss-Bonnet gravity \cite{Shojai_2023}.

For sake of completeness, an exact explicit integration of Eq.\eqref{CC} is given in Appendix \ref{Appendix C}.

\section{Horizons}\label{ddjdjjd}

In this section, we study the location of the inner and outer horizons developed by the metric 
\eqref{dwdlmskp}. The horizons are defined by the zeros of the lapse function \eqref{dwdlmskp}
\be
f(\tilde r)=1-\frac{\tilde{r}^2}{\tilde{r}^3+\tilde{\omega}(\tilde{r}+\frac{\gamma}{2})}
\label{M}
\ee
namely by the equation 
\begin{equation}
\label{HH}
\tilde r^3-\tilde r^2+\om\tilde r+\frac{\om\gamma}{2}=0\,.
\end{equation}
We can extract the relevant and useful information about the horizons of \eqref{dwdlmskp} by keeping in mind the behavior of the denominator $D(\tilde r)=\tilde{r}^3+\tilde{\omega}(\tilde{r}+\frac{\gamma}{2})$, already displayed in Fig.\ref{fig1}, and through the analysis of the function $f(\tilde r)$. \\

$\bullet$ For any $\om$ we have
\be
f(\tilde r) &\simeq& 1 - \frac{2\tilde r^2}{\om \gamma} \quad {\rm when} \quad \tilde r \to 0 \notag \\
f(\tilde r) &\simeq& 1 - \frac{1}{\tilde r} \quad {\rm when} \quad \tilde r \to \pm\infty \notag \\
f'(\tilde r) &=& \frac{\tilde r}{D(\tilde r)^2}\left[\tilde r^3 - \om(\tilde r + \gamma)\right]
\label{f'} 
\ee

$\bullet$ For $\om >0$, from Fig.\ref{fig1} we infer that there is one single negative real number $\beta<0$ such that
$D(\tilde r=\beta)=0$. Further, $D(\tilde r)>0$ when $\tilde r>\beta$; $D(\tilde r)<0$ when $\tilde r<\beta$. Therefore the two limits hold: $f(\tilde r) \to \mp\infty$ for $\tilde r \to \beta^{\pm}$. \\

$\bullet$ For $\om >0$, a straightforward graphical argument applied to $f'(\tilde r)$, Eq.\eqref{f'}, similar to that of Fig.\ref{fig1}, suggests that there is only one single positive real number $\tilde r_m>0$ such that $f'(\tilde r_m)=0$. Besides we have $f'(\tilde r)<0$ for $0<\tilde r <\tilde r_m$, and $f'(\tilde r)>0$ for $\tilde r >\tilde r_m$. Therefore $\tilde r  = \tilde r_m$ is a minimum for $f(\tilde r)$ in the relevant physical region $\tilde r >0$. \\

\begin{figure}[h]
\centering
\includegraphics[scale=0.45]{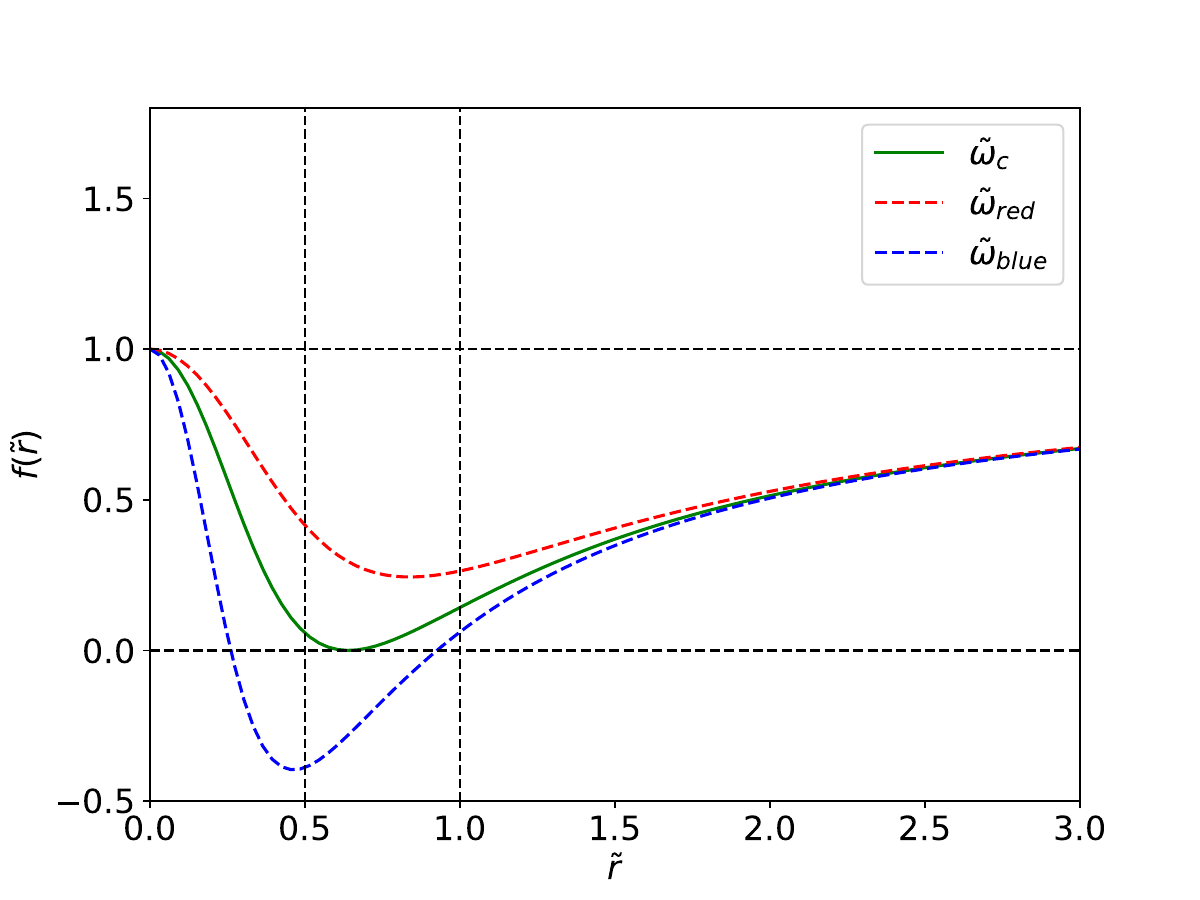}
\caption{Lapse function $f(\tilde r)$ for $\om>0$. Horizons are the zeros of $f(\tilde r)$, depicted for increasing values of $\om$, $0<\om_{\rm blue} < \om_{\rm green}\equiv\om_c < \om_{\rm red}$, i.e. for decreasing values of the mass $M$.}
\label{fig5}
\end{figure}

The above information suffices to plot a general graph for the lapse function $f(\tilde r)$, when $\om>0$, in the physical region $\tilde r >0$, see Fig.\ref{fig5}.      

In Fig.\ref{fig5} we see once again that the lapse function \eqref{M} is singularity-free for $\om>0$ (and $\tilde r \geq 0$). However the lapse can develop two distinct horizons (zeros) (name them $\tilde{r}_-$, $\tilde{r}_+$), or two coincident, or no horizons at all, according to the mass $M$ of the central body (namely according to the parameter $\om = \omega/(2M)^2$). An analysis of Eq.\eqref{HH} shows that when $\om \to 0^+$ then $\tilde{r}_+\to 1^-$ (on this, see Sec.4A of Ref.\cite{Scardigli:2022jtt}), i.e. $\tilde{r}_+$ goes to the normalized Schwarzschild radius.
We don't need to solve a $3^{rd}$ degree equation to compute the critical value $\om_c\equiv \om_{\rm green}$ (see Fig.\ref{fig5}) for which the two horizons coincide, and the related coordinate $\tilde r =\tilde r_c$ of the minimum of $f(\tilde r)$. In fact, looking at the horizon equation \eqref{HH}, we see that the pair 
$(\tilde r_c, \om_c)$ should solve the system
\be \left\{ \begin{array}{ll}
y(\tilde r)  &= \ \tilde r^3-\tilde r^2+\tilde\omega\left(\tilde r+\frac{\gamma}{2}\right) \ = \ 0  \\  \\
y'(\tilde r)  &= \ 3\tilde r^2-2\tilde r+\om \ = \ 0
\end{array}
\right. 
\ee
which is easily reducible to a $2^{nd}$ degree equation in $\tilde r$. Of course, in general we get the two pairs of solutions 
$[\tilde r_c(\gamma), \om_c(\gamma)]$ as functions of $\gamma$. For the specific numerical value $\gamma=9/2$ (suggested in Ref.\cite{PhysRevD.62.043008}) we get
\be
(\tilde r_{c}, \ \om_{c}) &\simeq& (0.6401, \ 0.0510) \notag \\
(\tilde r_{c2}, \ \om_{c2}) &\simeq& (-3.5151, -44.0979) \notag
\ee
Since we are in the case $\om>0$, we are here interested in the first solution $(\tilde r_{c}, \ \om_{c})$. The second solution will be relevant for the following case.\\

$\bullet$ For $\om <0$, again from Fig.\ref{fig1}, we infer that $D(\tilde r)$ has always one single positive real zero 
$\tilde r_0>0$ such that $D(\tilde r_0)=0$. Further, there could be two, or one, or no one, negative zeros for $D(\tilde r)$, according to different values\footnote{Here also it is easy to compute the threshold value of $\om$ for which $D(\tilde r)$ has two coincident negative zeros. In fact, in such case, the pair $(\tilde r, \om)$ should satify both the equations $D(\tilde r)=0$ and $D'(\tilde r)=0$. The solution is the pair $(\tilde r_t = -3\gamma/4, \ \om_t=-27\gamma^2/16)$. For $\om_t<\om<0$, $D(\tilde r)$ has one single positive zero; for $\om=\om_t$, one positive zero, and two coincident negative zeros; for $\om<\om_t<0$ one positive zero, and two distinct negative zeros. Of course, from the physical point of view, we are interested only in the positive zero, $\tilde r_0>0$, which is a singularity of the metric \eqref{M}.} of $\om<0$. Moreover, again from Fig.\ref{fig1}, it is evident that $D(\tilde r)<0$ when $0<\tilde r<\tilde r_0$; and $D(\tilde r)>0$ when $\tilde r > \tilde r_0$. Therefore the two limits hold: $f(\tilde r) \to \mp\infty$ for $\tilde r \to \tilde r_0^{\pm}$.\\

$\bullet$ For $\om <0$, the inspection of \eqref{f'} reveals that $f'(\tilde r)>0$ for any $\tilde r>0$.\\

\begin{figure}[h]
\centering
\includegraphics[scale=0.48]{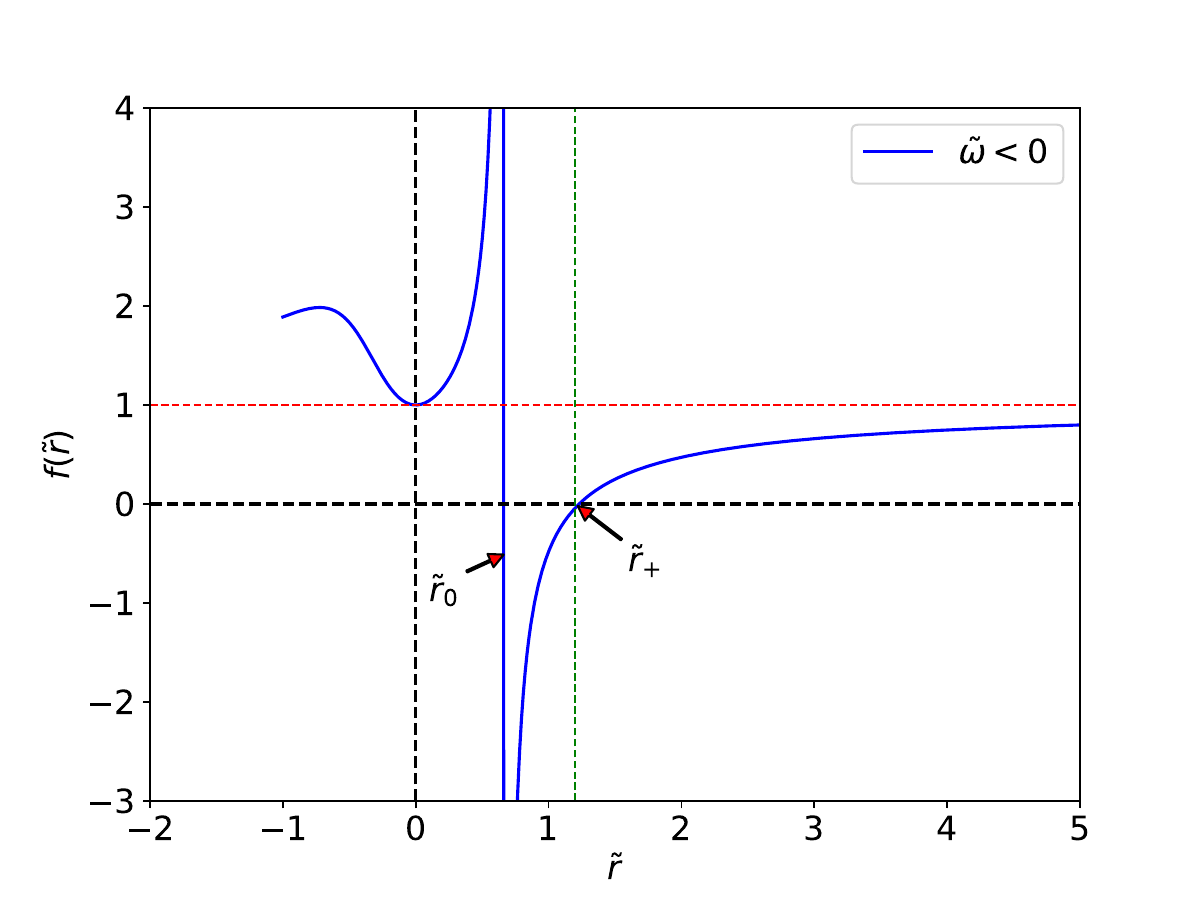}
\caption{The lapse function $f(\tilde r)$ for $\om<0$ in the physical region $\tilde r >0$: there is a singularity at $\tilde r = \tilde r_0$, and a horizon at $\tilde r=\tilde r_+$, where $f(\tilde r_+)=0$.}
\label{fig6}
\end{figure}

The above information is sufficient to plot a general graph (Fig.\ref{fig6}) for the lapse function $f(\tilde r)$, when $\om<0$, in the physical region $\tilde r >0$.   

So, in the physical region $\tilde r >0$, for any $\om<0$, the lapse function has always one singularity at $\tilde r = \tilde r_0$, and one horizon at $\tilde r=\tilde r_+ >\tilde r_0$, where $f(\tilde r_+)=0$. 
\footnote{Noticing that 
$\om_{c2}(\gamma)<\om_t(\gamma)<0$ for any $\gamma>0$, for sake of completeness we can say that in the unphysical region $\tilde r <0$, the lapse has no singularities when $\om_t<\om<0$ and two singularities when $\om\leq\om_t$; moreover, again in the region $\tilde r <0$, the lapse has no real zeros for $\om_{c2}<\om<0$, and two zeros for $\om\leq\om_{c2}<0$.} 

Finally, in the ``classical'' limit $\om\to 0^-$, again from Eq.\eqref{HH} we immediately see that $\tilde r_+ \to 1^+$, namely the horizon tends to the normalized Schwarzschild radius (see Ref.\cite{Scardigli:2022jtt}, Sec.4A).
In the Appendix \ref{Appendix D} the reader can find an explicit expression of the horizon $\tilde r_+$ as a root of Eq.\eqref{HH}.

%

\section{Evolution of the apparent and event horizons}\label{EEH}

In Sec.\ref{dmkdk} we showed that the space-time geometry of a collapsing star can be described, in an Oppenheimer-Snyder -like construction, by an exterior geometry given by a Schwarzschild-like metric, and by an interior geometry which is a solution of the Friedmann equations for a collapsing sphere of perfect fluid. We showed that Painlev\'e-Gullstrand(-like) coordinates allow us to write the metric in a unified way as
\be 
\label{PGI}
ds^2 &=& - d\tau^2 + (dr - r H(\tau) d\tau)^2 + r^2d\Omega^2 \\
ds^2 &=& - d\tau^2 + \left(dr+ \sqrt{2M/r\alpha(r)} d\tau\right)^2 + r^2 d\Omega^2 \qquad \quad 
\label{PGE}   
\ee
respectively for $r<R(\tau)$ and for $r>R(\tau)$, being $R(\tau)$ the coordinate of the surface of the star, and $H(\tau)$ the Hubble parameter.
Since the exterior geometry (namely for $r>R(\tau)$) is just a deformed Schwarzschild geometry, in the exterior metric the event horizon is described by the mathematical surface $r = r_+$ (defined in the previous section), which however usually lies well inside the star. From being a virtual object, the event horizon comes into existence (i.e. appears in the exterior region) at the instant $\tau = \tau_f$ when the collapsing star crosses its (outer) horizon , i.e. at the time $\tau_f$ such that $R(\tau_f) = r_+$, or, in dimensionless coordinates, 
\be
\tilde{R}(\tilde{\tau}_{f})=\tilde{r}_{+} \,.
\ee
Once appeared in the exterior region (namely when the collapsing star has crossed its ``deformed'' Schwarzschild radius, or outer horizon, $r_+$) the event horizon remains ``frozen'' at its radial coordinate $r=r_+(M)$ (unless further mass falls into the black hole). Therefore, it is of the utmost interest to study the formation and evolution of the horizons in the interior of the star, during the collapse process.

\subsection{Apparent horizon}\label{AppH}

\begin{figure}[h]
\centering
\includegraphics[scale=0.2]{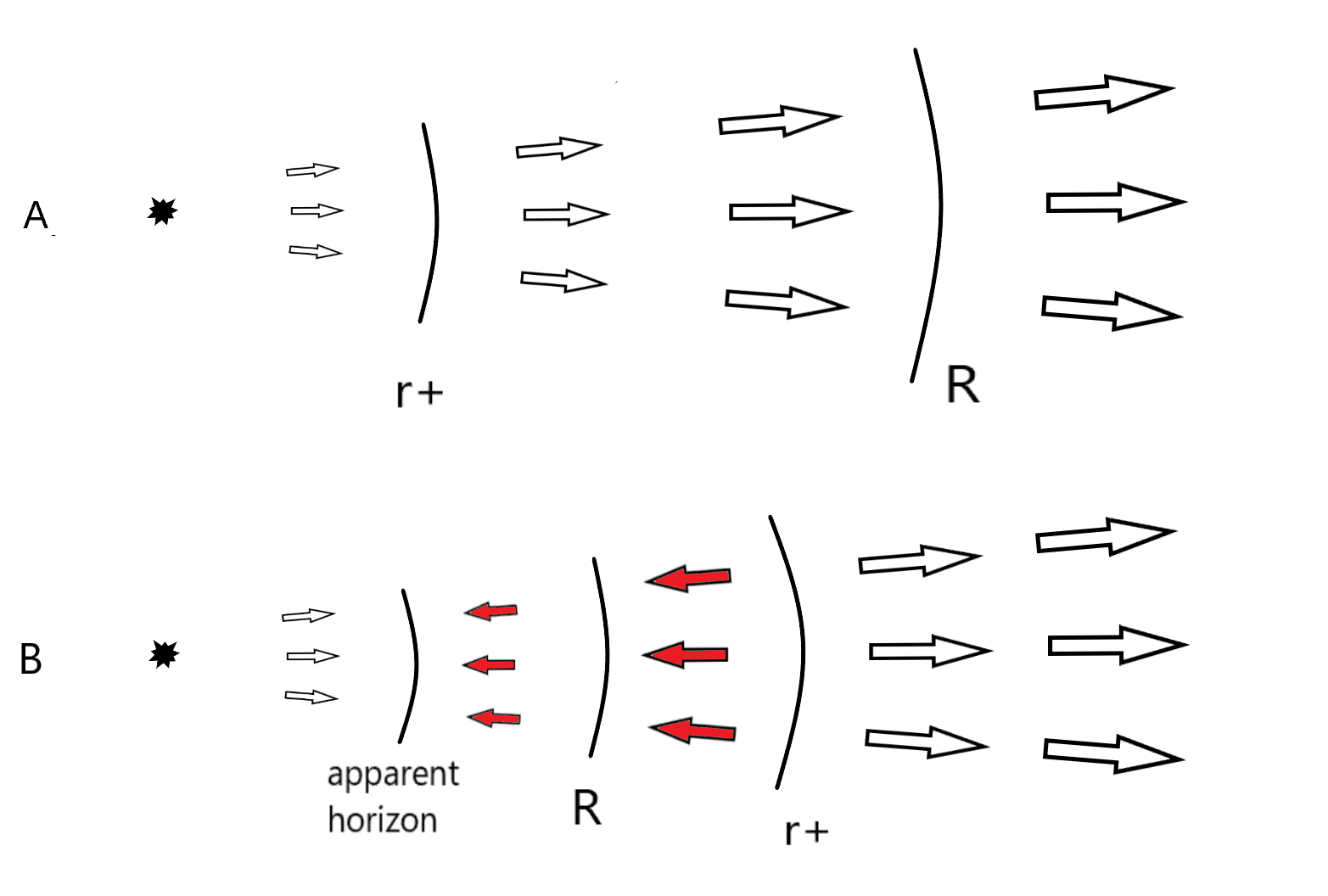}
\caption{Birth and behavior of the apparent horizon for $\om=0$. Outward null geodesics (i.e. light rays) travel from left to right (diagrams A and B). The apparent horizon ($ah$) comes to exist only when the surface of the star falls below the outer horizon, $\tilde R<\tilde r_+$ (diagram B). Then, in the region $\tilde r_{ah}<\tilde r_+$ even outward light rays are forced to travel inward. Any sphere with $\tilde r_{ah}<\tilde r<\tilde r_+$ is a trapped surface.}
\label{fig8}
\end{figure}

To explore the definition and the evolution of the apparent and event horizons inside the star, we have to look at the behavior of the internal radial null geodesics, namely the internal radial light rays. For the metric \eqref{PGI}, they must obey the equation
\be
0=-d\tau^2+(dr-rH(\tau)d\tau)^2 \notag
\ee
namely
\be
\label{IN}
dr &=& (-1+rH(\tau))d\tau \quad \quad {\rm or} \\  
dr &=& (+1+rH(\tau))d\tau
\label{OUT}
\ee
Reminding that $H(\tau)<0$, it is clear that Eq.\eqref{IN} can describe only ingoing null geodesics, since
\be 
\dot{r}=(-1+rH(\tau))<0 \,.
\ee
Instead, Eq.\eqref{OUT} will still describe ingoing null light rays when $(+1+rH(\tau))<0$, i.e. for light rays in the region $r>-1/H$, but on the contrary light rays in the region $r<-1/H$ can actually travel in the outgoing direction, since in this case 
\be
\dot{r}=(+1+rH(\tau))>0 \,. 
\ee
So inside the star there is a surface, called \textit{apparent horizon}, ($ah$), defined by
\be
\label{AH}
r_{ah} = -\frac{1}{H(\tau)}\,,
\ee 
whose properties are illustrated in Fig.\ref{fig8}, in particular for the case $\om=0$ (the Schwarzschild case).

By conducting the analysis in dimensionless variables for ease of calculation ($\tilde{R}=R/2M$, $\tilde{r}=r/2M$, $\tilde{\tau}=\tau/2M$, $\om=\omega/4M^2$, $\tilde{H}=2M H$, $\tilde{r}_+=1$ i.e. the normalized Schwarzschild radius), 
from Eqs.\eqref{FunEq0},\eqref{Hubble0},\eqref{PGI},\eqref{PGE}, we see that null outward geodesics (i.e. outward light rays) must obey the equations
\be
\label{LRIN}
\frac{d\tilde{r}}{d\tilde{\tau}} &=& 1 - \frac{\tilde r}{\tilde R}\sqrt{\frac{1}{\tilde R \alpha(\tilde R)}} \ \quad \quad \tilde r < \tilde R  \\
\frac{d\tilde{r}}{d\tilde{\tau}} &=& 1 - \sqrt{\frac{1}{\tilde r \alpha(\tilde r)}}  \quad \quad \quad \quad \tilde r > \tilde R 
\label{LROUT}
\ee
where, as usual $\alpha(\tilde{X})=1+\om(\tilde{X}+\frac{\gamma}{2})/\tilde{X}^3$. Then we have the following properties:
\begin{enumerate}
\item The horizons are located at $\tilde r$ such that $\tilde r \alpha(\tilde r)=1$, equivalent to Eq.\eqref{HH};
\item In particular the outer horizon obeys $\tilde r_+ \alpha(\tilde r_+)=1$;
\item If $\om>0$, then $\alpha(\tilde r)>1$ for any $\tilde r$, then $\tilde r_+ < 1$;
\item If $\om<0$, then $\alpha(\tilde r)<1$ for any $\tilde r$, then $\tilde r_+ > 1$;
\item $\tilde r \alpha(\tilde r)>1$ $\Leftrightarrow$ 
$\tilde r^3-\tilde r^2+\om(\tilde r+\frac{\gamma}{2})>0$ $\Leftrightarrow$ $\tilde r > \tilde r_+$ or $0<\tilde r < \tilde r_-$.
\end{enumerate}
We also notice that 
\be
&&\frac{\tilde r}{\tilde R}\sqrt{\frac{1}{\tilde R \alpha(\tilde R)}} \ < \ 1 \quad \quad \quad \Longleftrightarrow  \notag\\
&&\tilde r \ < \ \tilde R\sqrt{\tilde R \alpha(\tilde R)} = -\frac{\tilde R}{\dot{\tilde R}} = -\frac{1}{\tilde H} = \tilde{r}_{ah}\,. \notag
\ee
Armed with the above properties, and with Eqs.\eqref{LRIN}, \eqref{LROUT}, we are able to infer the features displayed in Fig.\ref{fig8} for the case $\om=0$. Specifically, when $\tilde R(\tilde \tau)> \tilde r_+$ there is no trace of an apparent horizon (case A, Fig.\ref{fig8}); it begins to exist only when $\tilde R(\tilde \tau)\leq \tilde r_+$, namely at the moment $\tilde \tau=\tilde \tau_f$ when the surface of the star crosses the outer horizon $R(\tilde \tau_f) = \tilde r_+$ (case B, Fig.\ref{fig8}). When $\tilde R(\tau)< \tilde r_+$ radial light rays in the region 
$0<\tilde r<\tilde r_{ah}$ can actually travel outward (of course just until $\tilde r_{ah}$), while light rays in the region  $\tilde r_{ah}<\tilde r<\tilde r_+$ are forced to travel only inward. All the spherical surfaces with center in $\tilde r=0$ and radius $\tilde r_{ah}<\tilde r<\tilde r_+$ are called \textit{trapped surfaces}. Further, $\tilde{r}_{ah}$ lies always inside the star. In fact (see property n.5 above)
\be
{\rm if} \quad \tilde R < \tilde r_+ \ \Rightarrow \ \tilde R \alpha(\tilde R)<1 \ \Rightarrow \
 \tilde r_{ah} = \tilde R \sqrt{\tilde R \alpha(\tilde R)} < \tilde R\,.  \notag
\ee
The cases $\om>0$ and $\om<0$ will be discussed in detail in the next sections, however we can anticipate here that some of the basic features of the above analysis remain valid also in those cases. 

\subsection{Event horizon}\label{EvH}

Speaking about ``evolution'' of the event horizon ($eh$) can perhaps sound a bit misleading 
since the outer horizon has actually a constant radial coordinate $r=r_+$, fixed by the zeros of $f(r)$, and it does not evolve in time (if the total mass $M$ is constant).
What we actually are talking about, and we can compute, is the trajectory of the last light ray able to leave the star, namely able to leave the surface $R(\tau)$. Clearly, when $R(\tau)<r_+$, no light ray can leave $R(\tau)$, also because $R(\tau)$ itself is then a trapped surface. So the last ray can leave the star at the instant in which the surface of the star passes through the outer horizon $r_+$, namely at time $\tau_f$ such that $R(\tau_f)=r_+$ (see Ref.\cite{Adler:2005vn}). In other words, the last light ray is described by the specific outgoing null geodesic arriving at the surface of star just at the time $\tau_f$ when $R(\tau_f)=r_+$. 

Since the outgoing null geodesics obey the equation $\dot{r}=1+rH(\tau)$, then the trajectory of the last light ray will satisfy the Cauchy problem
\be
\label{CP}
&&\frac{dr_{eh}}{d\tau} =  1 + r_{eh}H(\tau)   \\
&&R(\tau_f) = r_{eh}(\tau_f) = r_+             \notag 
\ee
The Hubble parameter can be computed at any radial coordinate inside the star, so we can also compute $H(\tau)$ on the star surface $R(\tau)=a(\tau)R_c$
\be
\label{Hubble}
H(\tau)=\frac{\dot{a}R_c}{aR_c}=\frac{\dot{R}(\tau)}{R(\tau)}\,.
\ee
Then, rewriting
\be
\frac{dr_{eh}}{d\tau}=\frac{dr_{eh}}{dR}\frac{dR}{d\tau}=\dot{R}\,\frac{dr_{eh}}{dR}\,,
\ee
we can recast the Cauchy problem \eqref{CP} as
\be
\label{EH}
&&\frac{dr_{eh}}{dR} \ - \ \frac{r_{eh}}{R} \ = \ \frac{1}{\dot{R}}     \\
&&r_{eh}(R=r_+) \ = \ r_+                                         \notag
\ee
If we know the explicit form of $R(\tau)$, then the Cauchy problem \eqref{CP} will yield the trajectory of the last ray in the form $r_{eh}(\tau)$. If instead we don't know explicitly $R(\tau)$ (or we don't want to use it), then the Cauchy problem \eqref{EH} will yield the last null geodesic able to leave the star in the form $r_{eh}(R)$.

In general $1/\dot{R}$ is a function of $R$, i.e. $1/\dot{R}=F(R)$. Therefore Eqs.\eqref{CP}, \eqref{EH} are both first order non-homogeneous linear differential equations, in the unknown function $r_{eh}(R(\tau))$, of the kind
\be
\label{LE1st}
y' + p(x)y = q(x)
\ee
whose general integral is
\be
\label{Int}
y(x)=\left(e^{-\int p(x)dx} \right)\left(\int q(x) \, e^{\int p(x)dx}\, dx + c\right)\,,
\ee
where the integration constant $c$ is fixed through the initial condition $y(x_0)=y_0$.
Therefore,  the solution of \eqref{EH} is:
\be
\label{EH1}
r_{eh}(R)=R\left(1+\int_{r_+}^{R} d\bar{R}\frac{F(\bar{R})}{\bar{R}}\right)
\ee

Let us now call $\tau_i$ the time such that $r_{eh}(\tau_i)=0$. We know that $r_{eh}(\tau_f)=r_+$. Now, let's choose an arbitrary time $\tau_x$ such that $\tau_i<\tau_x<\tau_f$. Then any light ray (or null geodesic) passing through the point $r(\tau_x)$ such that $r(\tau_x)<r_{eh}(\tau_x)$ is bounded to be confined into the sphere $S(0,r_+)$ (i.e. into the black hole). On the contrary, any light ray (or null geodesic) passing through the point $r(\tau_x)$ such that $r(\tau_x)>r_{eh}(\tau_x)$ will eventually escape from the star. So, at any given time $\tau_x$ the null surface $r=r_{eh}(\tau_x)$ behaves like an event horizon, and actually it may be called an ``evolving'' event horizon, to stress the time evolution of the event horizon inside the star (see e.g. Refs.\cite{Shojai_2022,blau2011lecture,Adler:2005vn,Nielsen2005,Shojai_2023}). So, in this specific sense the event horizon is a null hypersurface that at a certain time $\tau_i$ starts growing from $r=0$ in the interior of the star, it grows until it reaches the outer horizon $r_+$, and eventually ``eats'' the entire star (see e.g. Fig.\ref{fig9}). 

Having established the equations for the apparent horizon, Eq.\eqref{AH}, and for the event horizon, 
Eq.\eqref{CP} or \eqref{EH}, we proceed in the following to apply such equations to the various kinds of collapse, according to the different values of the parameter $\om$. For ease of calculation, we will return to the dimensionless variables $\tilde{R}=R/2M$, $\tilde{r}=r/2M$, $\tilde{\tau}=\tau/2M$, $\om=\omega/(2M)^2$, $\tilde{H}(\tilde{\tau})=2M H(\tau)$.
%
%

\subsection{\textbf{Case $\om \ = \ 0$.}}

This is the case of the standard gravitational collapse in a Schwarzschild spacetime. Remembering Eqs.\eqref{SCHW}, \eqref{AH} and \eqref{Hubble}, we can write for the apparent horizon
\be
\label{dnkebdedf}
\tilde{r}_{ah} = \tilde{R}(\tilde{\tau})^{3/2}\,.
\ee 
Since in this case we have the explicit solution \eqref{SchColl2} of the equation of motion \eqref{SCHW} of the star surface, we can write the proper-time evolution of the apparent horizon as
\be
\tilde{r}_{ah} = \tilde{R}_0^{3/2} -  \frac{3}{2}\tilde{\tau}\,.
\label{ahtau}
\ee
We defined $\tau_f$ as the instant in which the star surface crosses the outer horizon, $R(\tau_f) = r_+$. Since in the Schwarzschild case $r_+=2M$, then in dimensionless variables we have $\tilde{R}(\tilde{\tau}_f) = \tilde{r}_+ = 1$.
Therefore from Eq.\eqref{SchColl}
\be
\tilde{\tau}_f = \frac{2}{3}\left(\tilde{R}_0^{3/2}-1\right)
\ee
and
\be
\tilde{r}_{ah}(\tilde{\tau}_f) = 1\,,
\ee
as expected, and anticipated in Subsec.\ref{AppH}. \\
\\
For the ``evolving'' event horizon, we can rewrite in dimensionless variables the Cauchy problem \eqref{CP} as
\be
\label{CP1}
&&\frac{d\tilde{r}_{eh}}{d\tilde{\tau}} =  1 + \tilde{r}_{eh}\frac{\dot{\tilde{R}}}{\tilde{R}}   \\
&&\tilde{r}_{eh}(\tilde{\tau}_f) = \tilde{r}_+ =1 = \tilde{R}(\tilde\tau_f)            \notag 
\ee  
where we used $\tilde{H}(\tilde{\tau})=\dot{\tilde{R}}/\tilde{R}$; or also the Cauchy problem \eqref{EH}
\be
\label{EH1}
&&\frac{d\tilde{r}_{eh}}{d\tilde{R}} \ - \ \frac{\tilde{r}_{eh}}{\tilde{R}} \ = \ \frac{1}{\dot{\tilde{R}}}     \\
&&\tilde{r}_{eh}(\tilde{R}=1) \ = \ \tilde{r}_+ = 1 \,. \notag
\ee
From Eq.\eqref{SCHW} we have $1/\dot{\tilde{R}}=-\sqrt{\tilde{R}}$, and hence, using \eqref{Int}, the solution of the last Cauchy problem \eqref{EH1} reads
\be
\label{dmdkjdj}
\tilde{r}_{eh} = -2\tilde{R}^{3/2} + 3\tilde{R} \,,
\ee
or also, in terms of the explicit form \eqref{SchColl2} of $\tilde{R}(\tilde\tau)$, we can write
\be
\label{EHfunction}
\tilde{r}_{eh} = -2\left(\tilde{R}_0^{3/2} -  \frac{3}{2}\tilde{\tau}\right) + 
3\left(\tilde{R}_0^{3/2} -  \frac{3}{2}\tilde{\tau}\right)^{2/3} \,.
\ee 
Of course, the previous solution can also be obtained by directly integrating the Cauchy problem \eqref{CP1}.
From Eq.\eqref{SchColl} we already computed the time $\tilde\tau_0$ at which the star surface hits the central singularity, 
$\tilde{R}(\tilde\tau_0)=0$, namely $\tilde \tau_0 = \frac{2}{3}\tilde{R}_0^{3/2}$. From Eq.\eqref{EHfunction} we can easily compute also the time $\tilde\tau_i$ at which the trajectory of the last light ray (i.e. null geodesic) crosses the center of the star, namely $\tilde{r}_{eh}(\tilde\tau_i)=0$. It results
\be
\tilde{\tau}_i = \frac{2}{3}\left(\tilde{R}_0^{3/2}-\frac{27}{8}\right)\,.
\ee

\begin{figure}[h]
\centering
\includegraphics[scale=0.45]{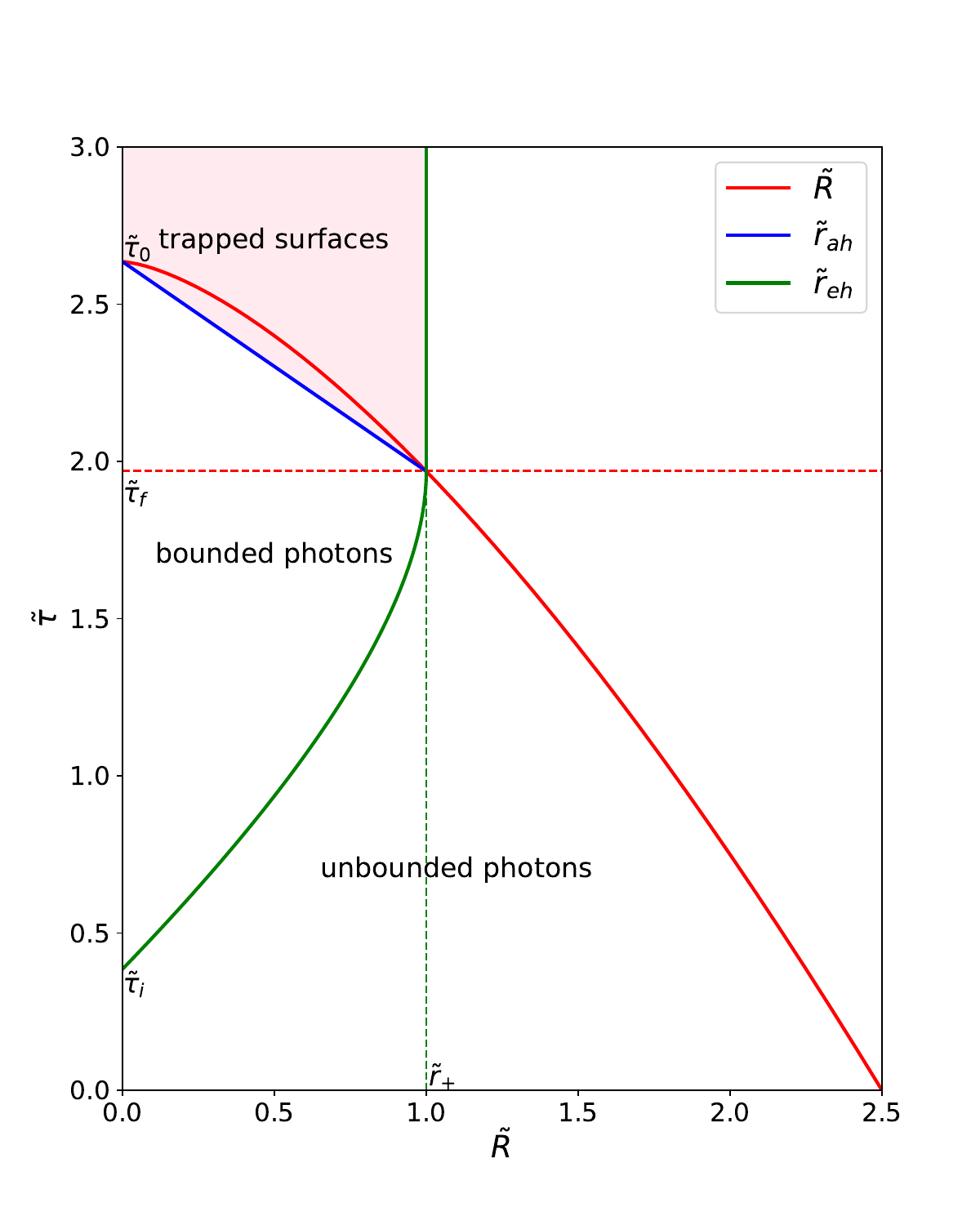}
\caption{Gravitational collapse in the Schwarzschild metric ($\om=0$). Evolution of the star surface (red), of the event horizon (green), and of the apparent horizon (blue). At any time $\tilde\tau>\tilde\tau_f$, the spherical surfaces with $\tilde r$ in $\tilde r_{ah}(\tilde\tau)<\tilde r<1=\tilde r_+$ are trapped surfaces (pink area).}
\label{fig9}
\end{figure}

Fig.\ref{fig9} summarizes, respectively, the evolution of the star surface, the trajectory of the last light ray able to leave the star (i.e. the evolution of the event horizon), and the evolution of the apparent horizon.
%
%
%
%
\subsection{\textbf{Case $\om \ > \ 0$.}}

For $\om>0$ we have to consider the critical value $\om_c$ for which the lapse function \eqref{M} can develop either two horizons $r_-$, $r_+$ (possibly coincident), when $0<\om\leq\om_c$, or no horizon at all, when $\om>\om_c$ (see Fig.\ref{fig5}).\\ 

\begin{figure}[h]
\centering
\includegraphics[scale=0.45]{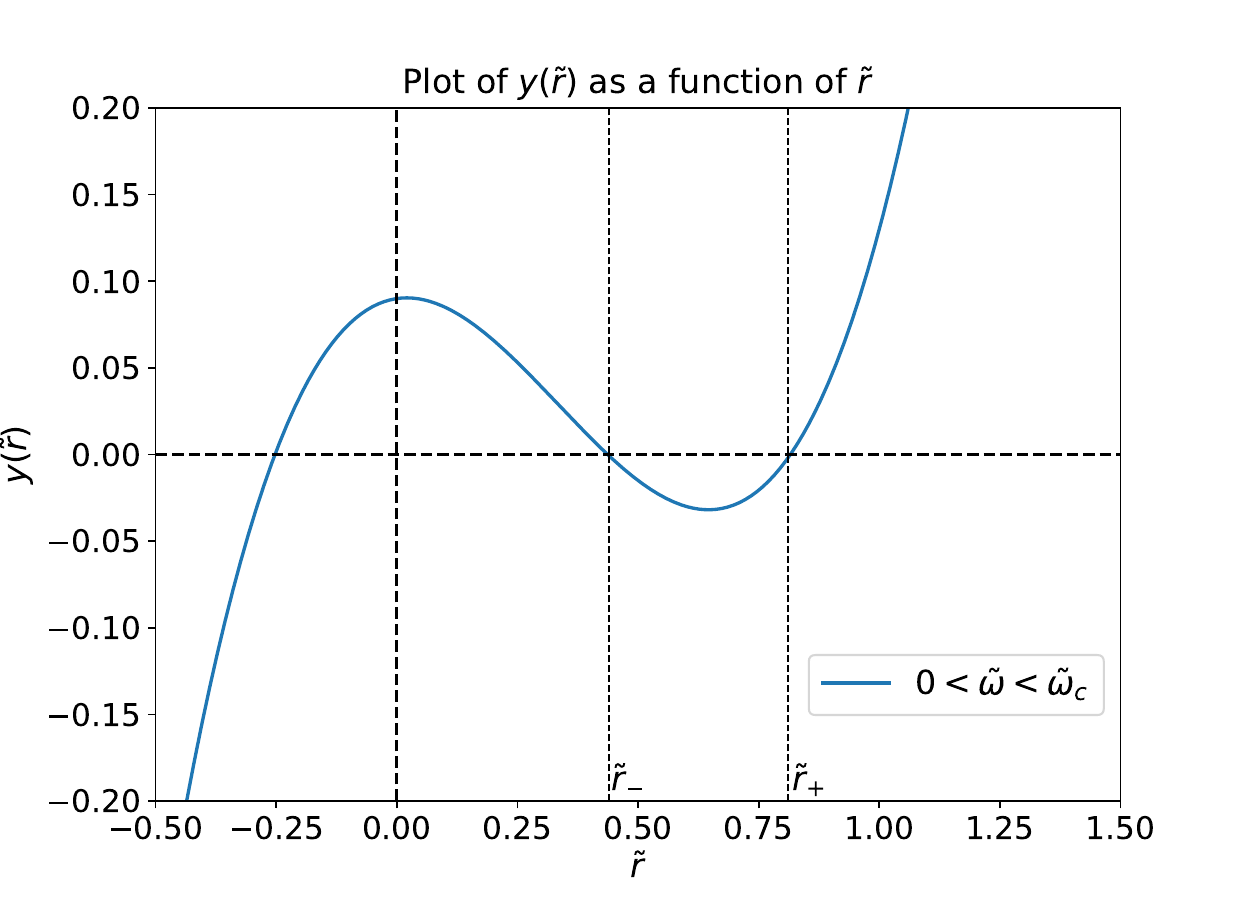}
\caption{Plot of $y(\tilde r)$ for $0<\om<\om_c$, Eq.\eqref{Y}, with the positive zeros $r_-$ and $r_+$.}
\label{fig10}
\end{figure}

\begin{figure}[h]
\centering
\includegraphics[scale=0.2]{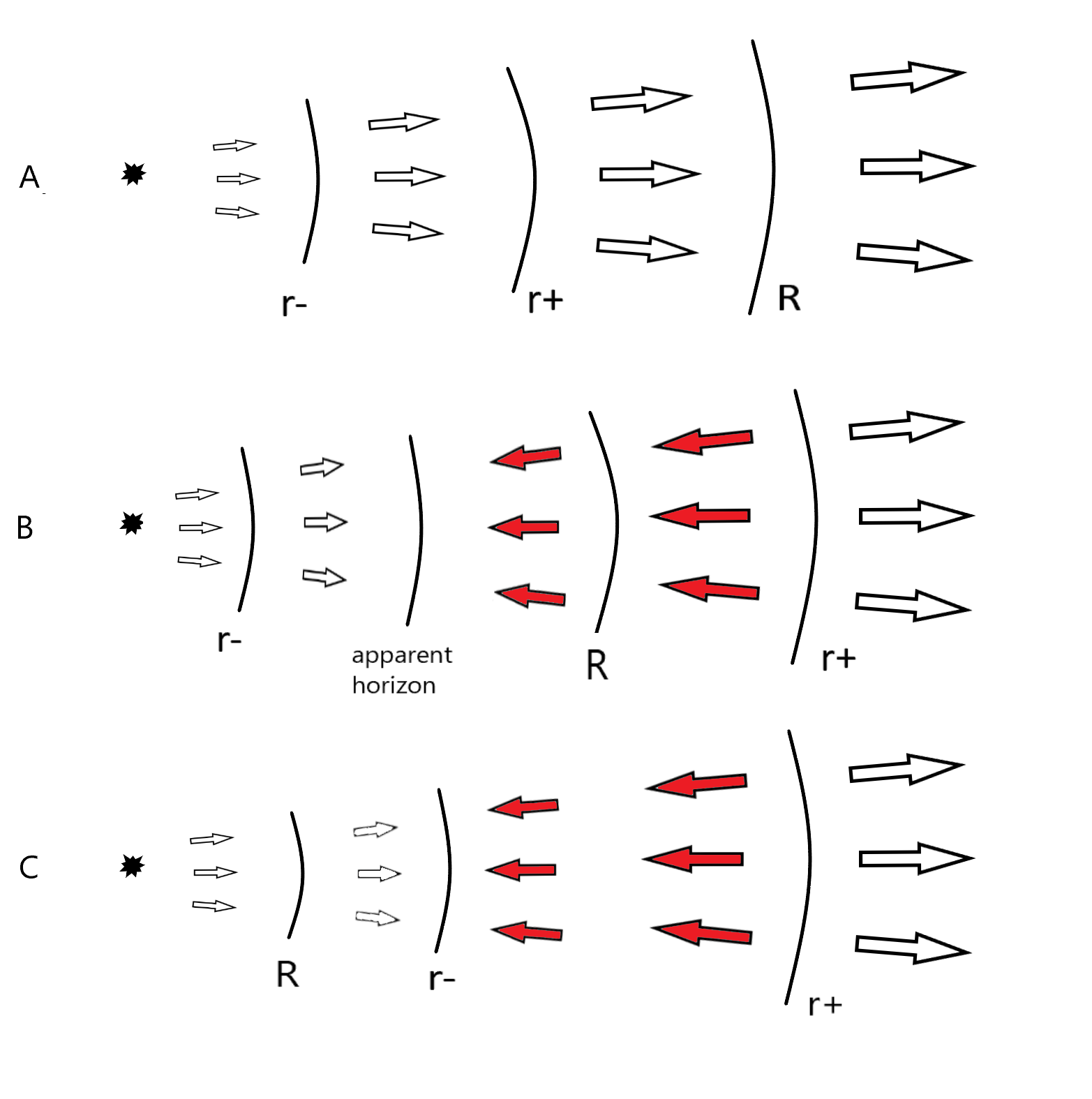}
\caption{Birth and behavior of the apparent horizon for $0<\om<\om_c$. Outward null geodesics (i.e. light rays) travel from left to right. The apparent horizon comes to exist only when the surface of the star falls below the outer horizon, $\tilde R<\tilde r_+$. Then, in the region $\tilde r_{ah}<\tilde r_+$ even outward light rays are forced to travel inward. Any sphere with $\tilde r_{ah}<\tilde r<\tilde r_+$ is a trapped surface. Once the star surface crosses the inner horizon $\tilde r_-$, the apparent horizon ceases to exist, and any sphere with $\tilde r_-<\tilde r<\tilde r_+$ is a trapped surface.}
\label{fig11}
\end{figure}

$\bullet$ Case $0<\om<\om_c$. We are in the situation of Fig.\ref{fig5}, blue (dashed, lower) line. Remembering the lapse, Eq.\eqref{M}, we note that in our analysis will be useful also the identities
\be
&&f(\tilde r) = 1-\frac{1}{\tilde r \alpha(\tilde r)}\,; \notag \\
&&\tilde r \alpha(\tilde r) - 1 = \frac{\tilde r^3-\tilde r^2+\om(\tilde r+\gamma/2)}{\tilde r^2} =: \frac{y(\tilde r)}{\tilde r^2}
\label{Y}
\ee 
as well as the plot of $y(\tilde r)$ in Fig.\ref{fig10}.
%
%
\\
With the help of Eqs.\eqref{LRIN}, \eqref{LROUT} we can now proceed as in Sec.\ref{AppH} for the analysis of the apparent horizon. Again, we can summarize our analysis with the diagram in Fig.\ref{fig11}.
When $\tilde R(\tilde\tau)>\tilde r_+$,  or $0<\tilde R(\tilde\tau)<\tilde r_-$,  there is no trace of an apparent horizon, or to be more precise, it lies outside the star surface, and therefore it is unphysical (cases A and C, Fig.\ref{fig11}); it begins to exist only at the moment $\tilde \tau_f$ when the surface of the star crosses the outer horizon 
$R(\tilde \tau_f) = \tilde r_+$ (case B, Fig.\ref{fig11}). When $\tilde R(\tilde\tau)<\tilde r_+$ radial light rays in the region $0<\tilde r<\tilde r_{ah}$ can actually travel outward (of course just until $\tilde r_{ah}$), while light rays in the region  $\tilde r_{ah}<\tilde r<\tilde r_+$ are forced to travel only inward. All the spherical surfaces with center in 
$\tilde r=0$ and radius $\tilde r_{ah}<\tilde r<\tilde r_+$ are \textit{trapped surfaces}. Further, $\tilde{r}_{ah}$ lies always inside the star. In fact
\be
&&\tilde{r}_{ah}=-\frac{\tilde R}{\dot{\tilde R}} = \sqrt{\tilde R^3+\om(\tilde R+\gamma/2)} < \tilde R \quad \Leftrightarrow \notag \\ 
&&\tilde R^3-\tilde R^2+\om(\tilde R+\gamma/2) <0 
\quad \Leftrightarrow \quad \tilde r_- < \tilde R < \tilde r_+ \,.\notag
\ee
Moreover, we can prove that $\tilde r_-<\tilde r_{ah}$. In fact
\be
&&\tilde r_-<\tilde r_{ah} = \sqrt{\tilde R^3+\om(\tilde R+\gamma/2)} \quad \Leftrightarrow \quad \notag \\
&&\tilde r_-^2 < \tilde R^3+\om(\tilde R+\gamma/2) \quad \Leftrightarrow \quad \notag \\
&&\tilde r_-^3+\om(\tilde r_-+\gamma/2) < \tilde R^3+\om(\tilde R+\gamma/2) \quad \Leftrightarrow \quad \notag \\
&&\tilde r_- < \tilde R\,, \notag
\ee
where the last holds since $f(x)=x^3+\om(x+\gamma/2)$ is an increasing function of $x$ for $\om>0$.
Together with the previous result we can state that in general
\be
\tilde r_- \ < \ \tilde r_{ah} \ < \ \tilde R \ < \ \tilde r_+ \,.
\ee
%
%
%
A further important and general result is that the star surface crosses the outer horizon exactly at the same instant $\tilde \tau_f$ in which the apparent horizon comes to existence. Namely
\be
\tilde R(\tilde \tau_f) =\tilde r_+ \quad \Leftrightarrow \quad \tilde r_{ah}=\tilde r_+\,.
\ee
To prove the above statement is sufficient to remember $\tilde r_{ah}(\tilde \tau) = \sqrt{\tilde R(\tilde \tau)^3+\om(\tilde R(\tilde \tau)+\gamma/2)}$, and to compute such equality at $\tilde \tau =\tilde \tau_f$. Then
\be
\tilde r_{ah}(\tilde \tau_f) &=& \sqrt{\tilde R(\tilde \tau_f)^3+\om(\tilde R(\tilde \tau_f)+\gamma/2)} \notag \\
&=& \sqrt{\tilde r_+^3+\om(\tilde r_+ +\gamma/2)} = \sqrt{\tilde r_+^2} = \tilde r_+        \notag
\ee
Analogously, we define $\tilde \tau_0$ as the instant in which $\tilde R$ crosses the inner horizon. Then we can prove that 
\be
\tilde R(\tilde \tau_0) =\tilde r_- \quad \Leftrightarrow \quad \tilde r_{ah}=\tilde r_-\,.
\ee
In a few words, the above equations are telling us that the evolution of the apparent horizon $r_{ah}$ starts at $r_+$ and ends at $r_-$.

\begin{figure}[h]
\centering
\includegraphics[scale=0.48]{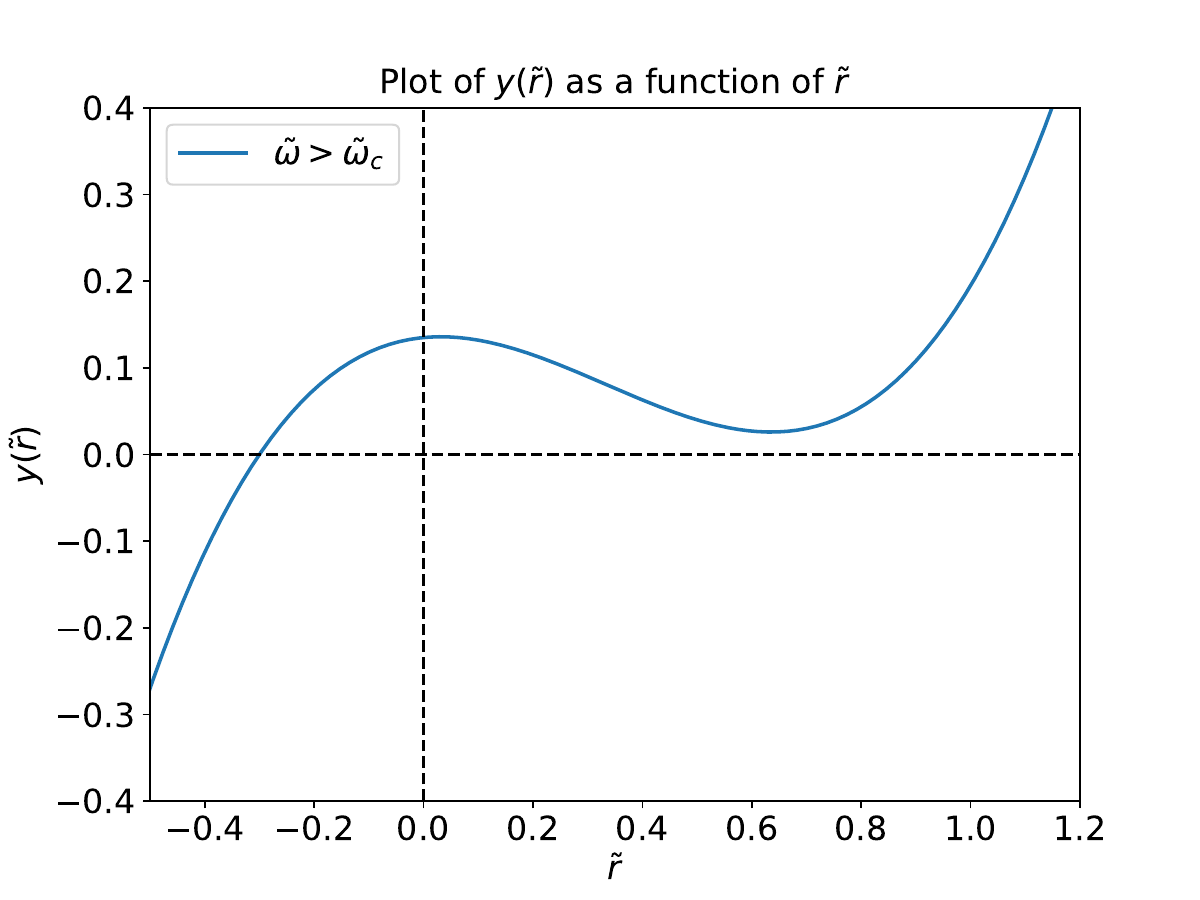}
\caption{Plot of $y(\tilde r)$, Eq.\eqref{Y}, for $\om>\om_c$: no positive zeros, no horizons at all.}
\label{fig12}
\end{figure}

\begin{figure}[h]
\centering
\includegraphics[scale=0.25]{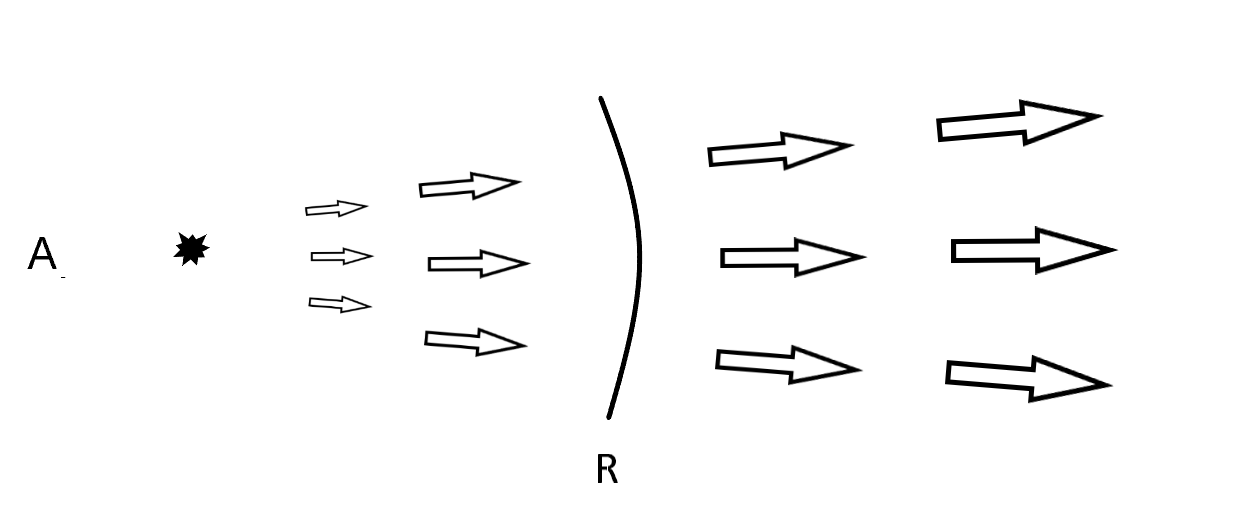}
\caption{No apparent horizon and no trapped surfaces appear when $\om>\om_c$.}
\label{fig13}
\end{figure}

$\bullet$ Case $\om>\om_c$. With reference to Fig.\ref{fig5}, we are considering the red (dashed, upper) line. There are no horizons at all. The plot of the $y(\tilde r)$ in this case is in Fig.\ref{fig12}. We can prove that now the apparent horizon lies always outside the star, namely it is unphysical. In fact, with reference to Fig.\ref{fig12}, we have 
\be
&&y(\tilde R) = \tilde R^3-\tilde R^2+\om(\tilde R+\gamma/2) >0  \quad \Leftrightarrow \quad \notag \\
&&\sqrt{\tilde R^3+\om(\tilde R+\gamma/2)} > \tilde R \quad \Leftrightarrow \quad \tilde{r}_{ah} > R 
\ee 
%
%
Coherently, the usual analysis conducted with the help of Eqs. \eqref{LRIN}, \eqref{LROUT} shows the complete absence of trapped surfaces (see Fig.\ref{fig13}).
%
%

We can now focus on the evolution of the event horizon. As we know, the trajectory of the last ray able to leave the star is a solution of the Cauchy problem \eqref{CP}. Such a photon will take an infinite amount of time to reach the far observer, since it comes exactly from the surface of the event horizon. Remembering Eq.\eqref{Hubble}, we can rewrite it as 
\be
&&\frac{d\tilde r_{eh}}{d\tilde \tau} =  1 + \tilde r_{eh}\frac{\dot{\tilde R}(\tilde \tau)}{\tilde R(\tilde \tau)}   \\
&&\tilde R(\tilde \tau_f) = \tilde r_+; \quad  \tilde r_{eh}(\tilde \tau_f) = \tilde r_+      \,.       \notag 
\ee
Actually, since $\tilde R(\tilde \tau)$ should obey to Eq.\eqref{CC} of the collapsing star, the above can be more properly re-written as a system of first order non linear coupled differential equations 
\begin{subequations}
\begin{align}
\label{73a}
\dot{\tilde R} & = -\frac{\tilde R}{\sqrt{{\tilde R^3+\om(\tilde R+\frac{\gamma}{2})}}}\,; \quad \quad \tilde R(\tilde \tau_f) = \tilde r_+       \\
\dot{\tilde r}_{eh} & =  1 - \frac{\tilde r_{eh}}{\sqrt{\tilde R^3+\om(\tilde R+\frac{\gamma}{2})}}\,; \quad  \tilde r_{eh}(\tilde \tau_f) = \tilde r_+ 
\label{73b}
\end{align} 
\end{subequations}
where $\dot{\tilde X}:=d\tilde X/d\tilde \tau$. It should be kept in mind that Eq.\eqref{73b} is just 
Eq.\eqref{LRIN}, namely it is valid inside the star (here we are interested in the evolution of the inner null geodesics only). 
The asymptotic behaviors of Eq.\eqref{73a} for $\tilde R (\tilde \tau)$, as well as its explicit integration, have already been studied, for $\om>0$, respectively, in Sec.\ref{CollOmPos} and in Appendix \ref{Appendix C}. Once we know the function $\tilde R (\tilde \tau)$, analytically or numerically, then the evolution of the apparent horizon is given by
\be
\label{AHom}
\tilde r_{ah}(\tilde \tau) = \sqrt{\tilde R(\tilde \tau)^3+\om(\tilde R(\tilde \tau)+\gamma/2)}\,,
\ee
valid in general, as we saw, when $\tilde r_-<\tilde R<\tilde r_+$.   
An analytical expression for the inverse function $\tilde \tau (\tilde r_{ah})$ of \eqref{AHom}, coming from an explicit integration of \eqref{73a}, can be found in Appendix \ref{Appendix E}. \\

The evolution of the event horizon inside the star, $\tilde{r}_{eh}(\tilde \tau)$, is controlled by Eq.\eqref{73b}. As we have seen in Sec.\ref{EvH}, and in Eq.\eqref{EH1}, with the help of the (trivial) relation $\dot{\tilde{r}}=(d\tilde{r}/d\tilde{R})\dot{\tilde{R}}$ and of Eq.\eqref{73a}, such Eq.\eqref{73b} can be re-written as
\begin{equation}
\label{MM}
\frac{d\tilde{r}_{eh}}{d\tilde{R}}-\frac{\tilde{r}_{eh}}{\tilde{R}}+\frac{\sqrt{{\tilde R^3+\tilde\omega(\tilde R+\frac{\gamma}{2})}}}{\tilde{R}}=0
\end{equation}
which gives $\tilde{r}_{eh}$ as a function of $\tilde{R}$. This is a first-order linear inhomogeneous differential equation of the kind \eqref{LE1st}. In Appendix \ref{Appendix F} the reader can find its explicit integration in terms of elliptic functions, under the initial condition $\tilde{r}_{eh}(\tilde r_{+})=\tilde r_{+}$.  
%
\begin{figure}[h]
\centering
\includegraphics[scale=0.45]{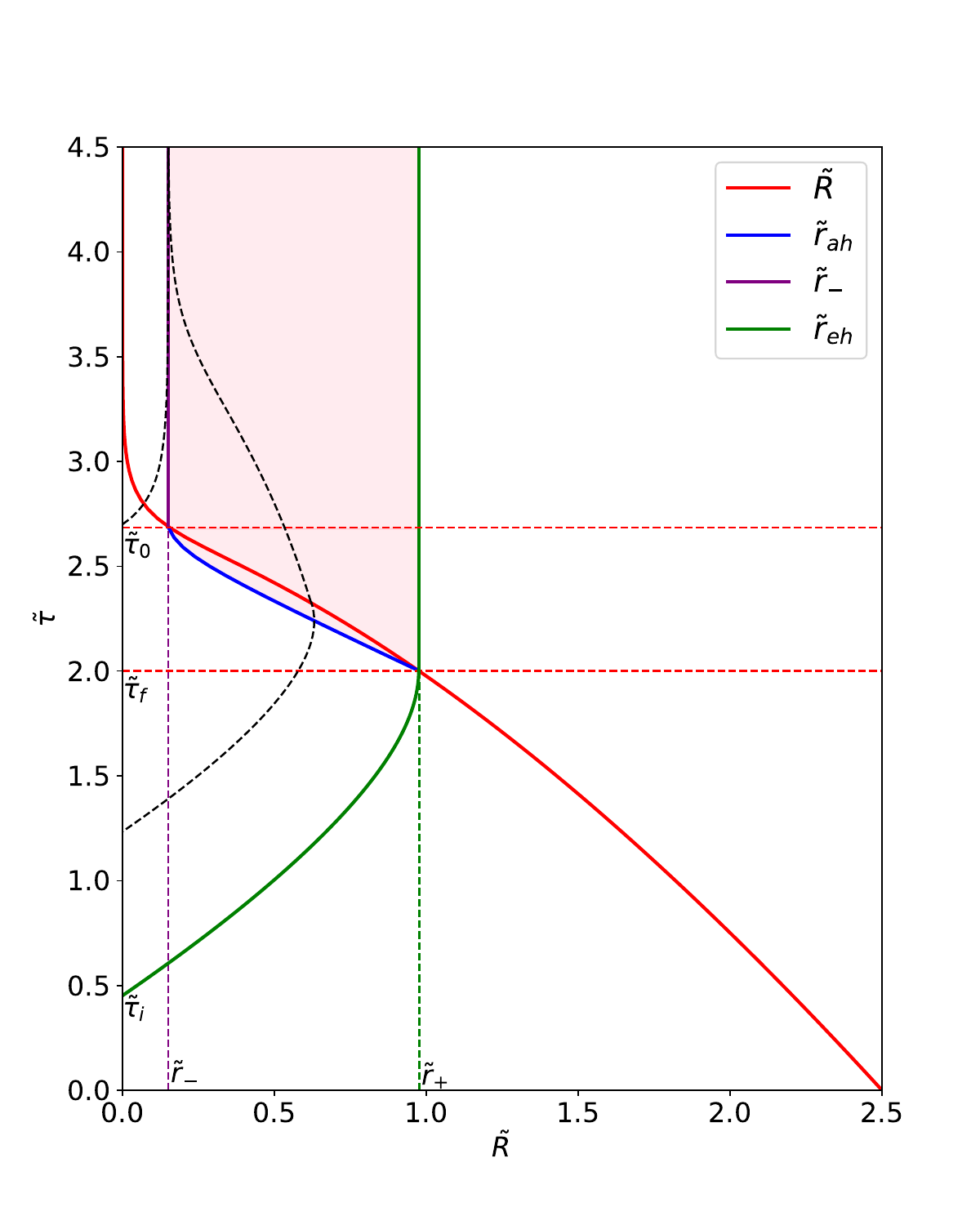}
\caption{For $0<\om<\om_c$, here we represent the evolution of the star surface (red), the event/outer horizon $\tilde r_+$ (green), the inner horizon $\tilde r_-$ (purple), the apparent horizon (blue). Notice that $\tilde r_+<1$. At any time $\tilde\tau_f<\tilde\tau<\tilde\tau_0$ the surfaces $\tilde r_{ah}(\tilde\tau)<\tilde r<\tilde r_+$ are trapped. Then, for $\tilde\tau>\tilde\tau_0$, all the surfaces $\tilde r_-<\tilde r<\tilde r_+$ are trapped (pink area). Notice that the surfaces $0<\tilde r<\tilde r_-$ are \textit{not trapped}. Trajectories of light rays inside $\tilde r_+$ and inside $\tilde r_-$ are depicted in thin dashed black lines.}
\label{fig14}
\end{figure}
%

It is however much more insightful (see Fig.\ref{fig14}) to proceed to a numerical integration of the system of differential equations \eqref{73a}\eqref{73b}, in particular when the parameter $\om$ is in the range $0<\om<\om_c$ (so that the horizons $r_-$, $r_+$ are real, and therefore also the apparent horizon $\tilde r_{ah}$ is such). Several observations are now in order. For example, when 
$\tilde \tau \to \tilde \tau_f$, then $\tilde R(\tilde \tau) \to \tilde r_+$ and $\tilde r_{eh}(\tilde \tau) \to \tilde r_+$, so according to \eqref{73b} we have 
\be
\frac{d\tilde r_{eh}}{d\tilde \tau} \ \simeq \ 1 - \frac{\tilde r_+}{\tilde r_+} \ \to \ 0\,,
\ee
therefore the inner trajectory of the last ray attaches smoothly to the outer horizon $\tilde r_+$. 
Moreover, for sake of completeness, we can also have a look at the null trajectories once they left the star surface. Then they have to obey Eq.\eqref{LROUT}, namely
\be
\dot{\tilde r}  =  1 - \frac{\tilde r}{\sqrt{\tilde r^3+\om(\tilde r+\frac{\gamma}{2})}}\,, \quad \tilde r > \tilde R\,.
\ee 
Without entering in too many details, a qualitative analysis of the above differential equation confirms that any light ray leaving the star surface at a time $\tilde \tau<\tilde \tau_f$ can escape to infinity; light rays leaving the star surface at times $\tilde \tau_f<\tilde \tau<\tilde \tau_0$ are bounded to be confined in the spherical shell $S(\tilde r_-,\tilde r_+)$, and precisely their trajectories ``glide'' smoothly on the inner horizon null surface $\tilde r_-$, namely $\tilde r(\tilde \tau) \to \tilde r_-$ when $\tilde \tau \to +\infty$; null trajectories leaving the star surface in the last phase of the collapse, when $\tilde \tau>\tilde \tau_0$, can actually move outward, but only until approaching smoothly, from inside, the inner horizon surface $\tilde r_-$. Such an analysis confirms what already presented in Fig.\ref{fig11}. In particular the surfaces inside the inner horizon, $0<\tilde r<\tilde r_-$, are \textit{not} trapped surfaces. Pictorially, we can say that our black hole, for this range of the parameter $0<\om<\om_c$, consists of a luminous core surrounded by a thick spherical black shell.\\

%
\subsection{\textbf{Case $\om \ < \ 0$.}}
\label{Sb5E}
%

For any $\om \ < \ 0$, we proved in Sec.\ref{ddjdjjd} that the lapse function $f(\tilde r)$ has always, in the physical region $\tilde r >0$, one singularity at $\tilde r = \tilde r_0>0$ and one horizon at $\tilde r =\tilde r_+ >1$, with $\tilde r_0<\tilde r_+$. In the unphysical region $\tilde r<0$ things can get more complicted, but that region is indeed non-physical, and therefore we ignore it.  

\begin{figure}[h]
\centering
\includegraphics[scale=0.45]{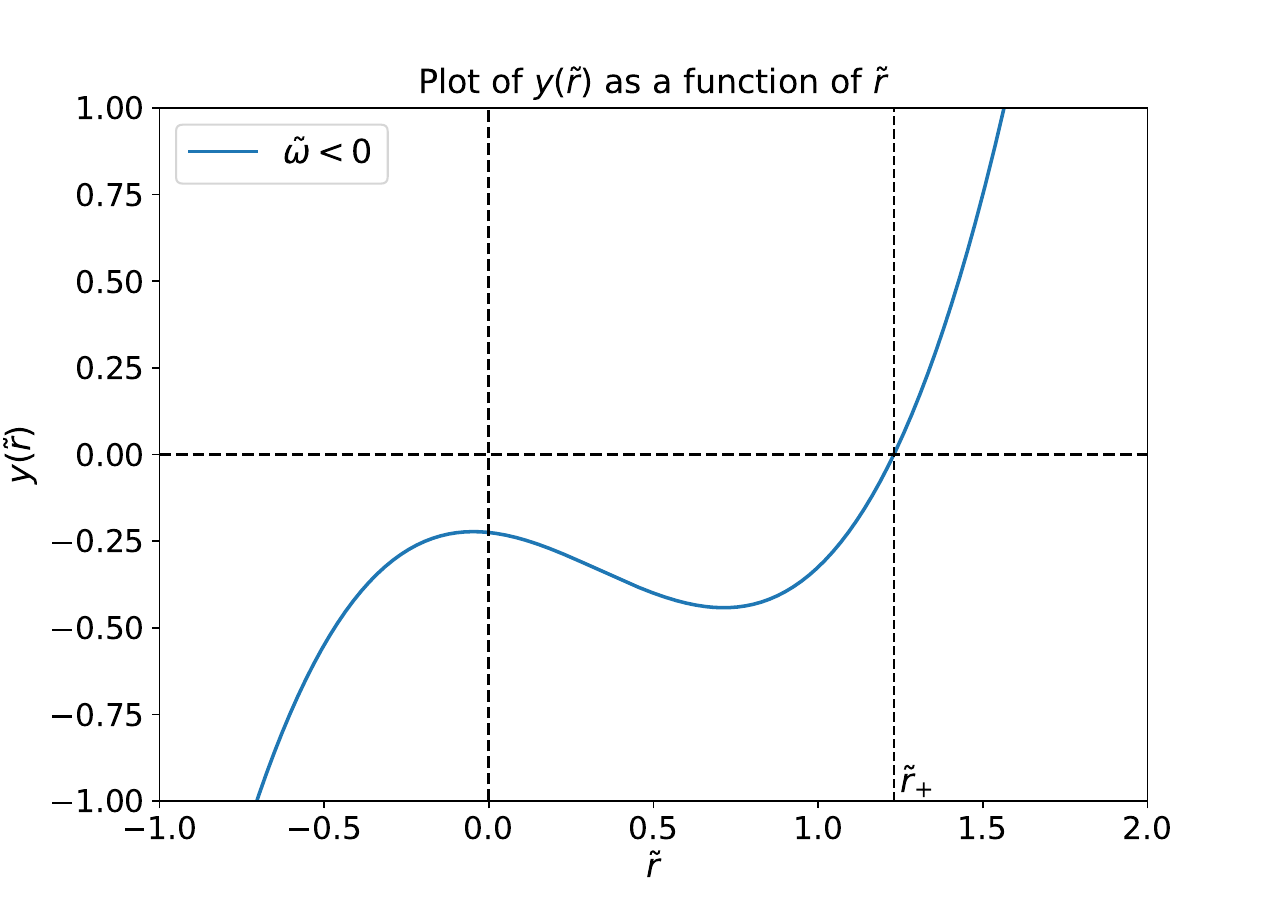}
\caption{Plot of $y(\tilde r)$ for any $\om<0$: always one positive zero, i.e. horizon, $\tilde r_+>1$.}
\label{fig15}
\end{figure}

As usual (see Sec.\ref{AppH}), to understand the behavior of the apparent horizon during the various phases of the collapse we make use of Eqs.\eqref{LRIN}, \eqref{LROUT}, as well of the ``horizon'' function $y(\tilde r)$, Eq.\eqref{Y}. A useful plot of $y(\tilde r)$ for $\om<0$ is in Fig.\ref{fig15}. 
%
%

\begin{figure}[h]
\centering
\includegraphics[scale=0.2]{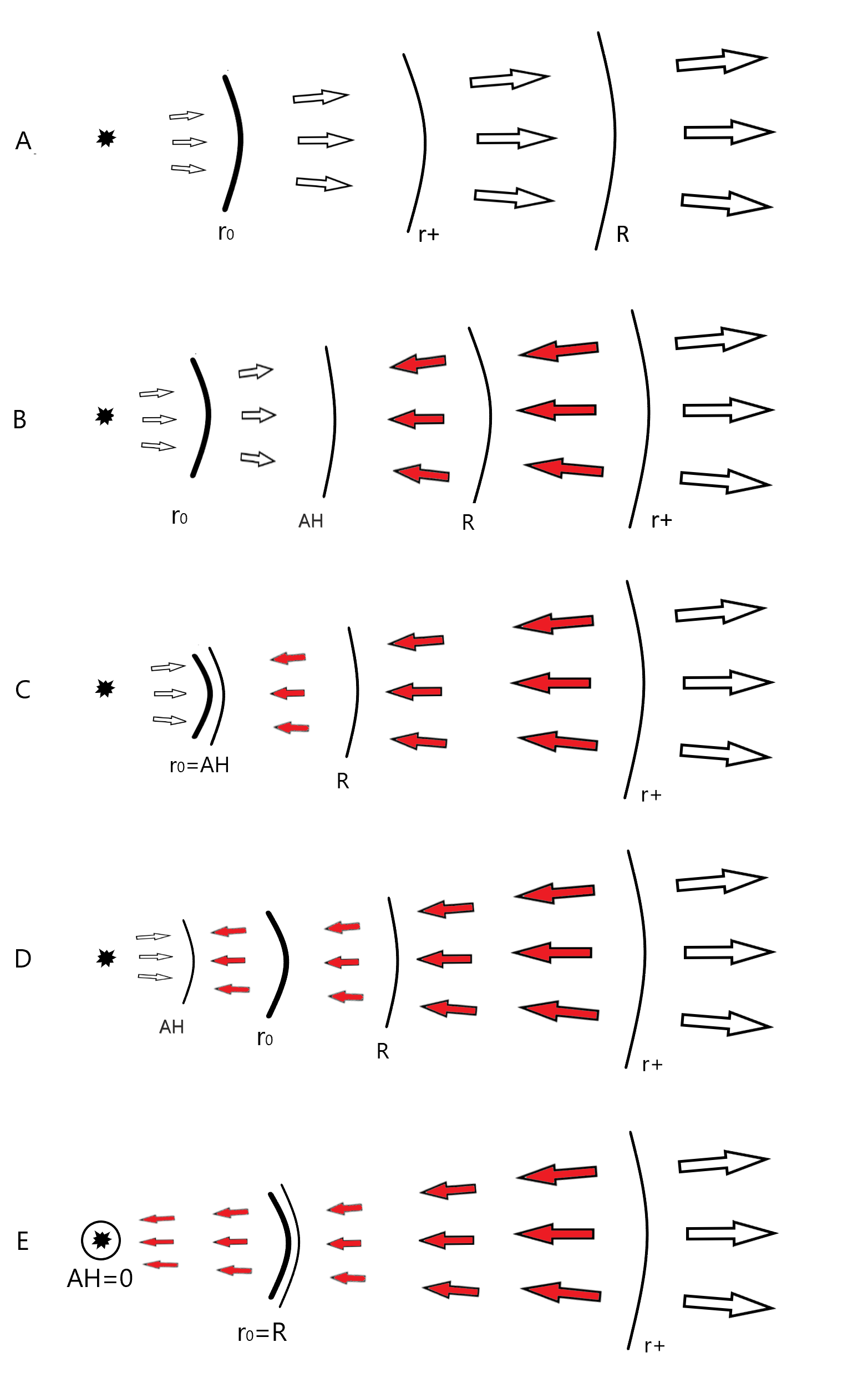}
\caption{Various steps of the collapse for $\om<0$. Light rays (i.e. null geodesics) move outward from left to right. Trapped surfaces regions are full of red arrows. Apparent horizon crosses $\tilde r =\tilde r_0$ singular surface in C. Collapse stops when stellar surface $\tilde R$ hits upon $\tilde r_0$, in E.}
\label{fig16}
\end{figure}

For example, when $\tilde R(\tilde \tau)>\tilde r_+$, namely when the surface of the star has not yet crossed the outer horizon, the analysis of the various regions proceeds as follow:\\
$\bullet$ in the region $\tilde r>\tilde R$ we apply Eq.\eqref{LROUT}. Since $\tilde r > \tilde r_+$, then 
$\tilde r \alpha(\tilde r)-1 = y(\tilde r)>0$. Therefore $\tilde r \alpha(\tilde r)>1$. Hence from Eq.\eqref{LROUT} we infer 
$d\tilde r/ d\tilde \tau >0$, so light rays can actually travel outward;\\
$\bullet$ in the region $\tilde r_+<\tilde r<\tilde R$ we should apply Eq.\eqref{LRIN}. Since $\tilde R>\tilde r_+$ then $y(\tilde R)=\tilde R \alpha(\tilde R)-1 >0$. Moreover $\tilde r/\tilde R<1$. Therefore from Eq.\eqref{LRIN} we infer $d\tilde r/ d\tilde \tau >0$, so here also light rays can travel outward;\\
$\bullet$ analogous steps are followed for the regions $\tilde r_0<\tilde r<\tilde r_+$ and $0<\tilde r<\tilde r_0$.\\
In so doing we arrive at the behavior schematized in Fig.\ref{fig16}, point A.\\
Similar considerations apply to all the other possible positions of the star surface $\tilde R$ relative to $\tilde r_+$, 
$\tilde r_0$. The results of such analysis are summarized in Fig.\ref{fig16}. 
%
%
We see that, once again, the apparent horizon comes to existence only when $\tilde R (\tilde \tau)$ crosses the outer horizon 
$\tilde r_+$, namely at $\tilde\tau=\tilde\tau_f$ such that $\tilde R(\tilde\tau_f)=\tilde r_+$. For $\tilde\tau<\tilde\tau_f$ there is no trace of an apparent horizon (case A). Instead, when $\tilde\tau>\tilde\tau_f$ we have the actual presence of an apparent horizon, whose radial coordinate is always smaller than the radial coordinate of the star surface (cases B, C, D, E). Precisely $\tilde r_{ah}(\tilde\tau)<R(\tilde\tau)$ for any $\tilde\tau$ such that $\tilde\tau_f < \tilde\tau < \tilde\tau_0$, where $\tilde\tau_0$ is the instant in which the star surface $\tilde R$ hits the singularity $\tilde r_0$, 
$\tilde R(\tilde\tau_0)=\tilde r_0$. Here also we can say that, when $\tilde R(\tau)< \tilde r_+$, radial light rays in the region $0<\tilde r<\tilde r_{ah}$ can actually travel outward (of course just until $\tilde r_{ah}$), while light rays in the region  $\tilde r_{ah}<\tilde r<\tilde r_+$ are forced to travel only inward. All the spherical surfaces with center in 
$\tilde r=0$ and radius $\tilde r_{ah}<\tilde r<\tilde r_+$ are \textit{trapped surfaces} (pictorially indicated with red arrows). On the contrary, the surfaces with radius $0<\tilde r<\tilde r_{ah}$ are \textit{not} trapped. 

It is important to point out that the spherical singular surface at $\tilde r=\tilde r_0$ has not at all the character of an ``impenetrable'' physical barrier or of a singularity actually present at any moment. Light rays can cross it (in both senses), and it remains a \textit{virtual} singular surface, but \textit{only until when} the collapsing stellar surface $\tilde R$ hits upon it. After that instant, $\tau>\tau_0$, the whole region $S(0, \tilde r_0)$ becomes actually singular (and, obviously, trapped).

A similar behavior can be observed in the more familiar Schwarzschild metric, where the central singularity remains virtual until the star surface collapses upon it. The central singular point of the Schwarzschild metric does not represent an actual physical singularity present at any moment, just because the physical metric describing the inner of a star is \textit{not} a Schwarzschild metric. In principle, photons coming from the center of a star can keep exiting the center (and traveling outward until the apparent horizon), even during the collapse, at least until when the collapsing surface of the star itself does hit upon the center.  

The above general considerations (based on Eqs. \eqref{LRIN}, \eqref{LROUT}) are confirmed by further specific calculations. For example, $\tilde r_{ah}$ lies always inside the star. In fact
\be
&&\tilde r_{ah} < \tilde R \quad \Leftrightarrow \quad \notag \\  
&&\tilde r_{ah}=-\frac{1}{\tilde H}=-\frac{\tilde R}{\dot{\tilde R}} = \sqrt{\tilde R^3+\om(\tilde R+\gamma/2)} < \tilde R 
\notag \\  
&&\Leftrightarrow \quad \tilde R^3-\tilde R^2+\om(\tilde R+\gamma/2) <0 
\quad \Leftrightarrow \quad \tilde R < \tilde r_+ \,,\notag 
\ee 
(where the last inference is supported by Fig.\ref{fig15}).
Therefore \begin{center} $\tilde r_{ah} < \tilde R < \tilde r_+$. \end{center}
Moreover, the natural request of having to do with real solutions only, forces $\tilde R(\tilde \tau)$ to be always larger than the singularity radial coordinate $\tilde r_0$, namely $\tilde R(\tilde \tau) \geq \tilde r_0$. In fact
\be
\tilde R^3+\om(\tilde R+\frac{\gamma}{2}) \geq 0 \quad \Leftrightarrow \quad \tilde R(\tilde \tau) \geq \tilde r_0
\ee 
at least for the physically acceptable values $\tilde R>0$ (this is clear since Fig.\ref{fig1}). We point out that the collapse stops when the star surface $\tilde R$ arrives at $\tilde r_0$. $\tilde R$ cannot absolutely become smaller that $\tilde r_0$, otherwise $\tilde r_{ah}=[\tilde R^3+\om(\tilde R+\gamma/2)]^{1/2}$ would become imaginary, together (even worse) with 
$\dot{\tilde R}(\tilde \tau)$, see Eq.\eqref{CC}. Further we note that $\tilde r_{ah} \to 0$ when 
$\tilde R(\tilde\tau) \to \tilde r_0$, for $\tilde\tau \to \tilde\tau_0$ (cases C, D, E of Fig.\ref{fig16}). Light rays, and apparent horizon, can actually travel along the region $0<\tilde r <\tilde r_0$, at least until when $\tilde R$ has reached the hard singularity $\tilde r_0$. Finally, when $\tilde R \to \tilde r_0^+$ then 
\be
\tilde R\alpha(\tilde R)=\frac{\tilde R^3+\om(\tilde R+\gamma/2)}{\tilde R^2} \to 0^+
\ee
and therefore 
\be
\frac{d\tilde{r}}{d\tilde{\tau}} = 1 - \frac{\tilde r}{\tilde R}\sqrt{\frac{1}{\tilde R \alpha(\tilde R)}} \to -\infty\,,
\ee
namely the singularity region $0<\tilde r <\tilde r_0$ becomes inaccessible once $\tilde R \equiv r_0$.\\

\begin{figure}[h]
\centering
\includegraphics[scale=0.6]{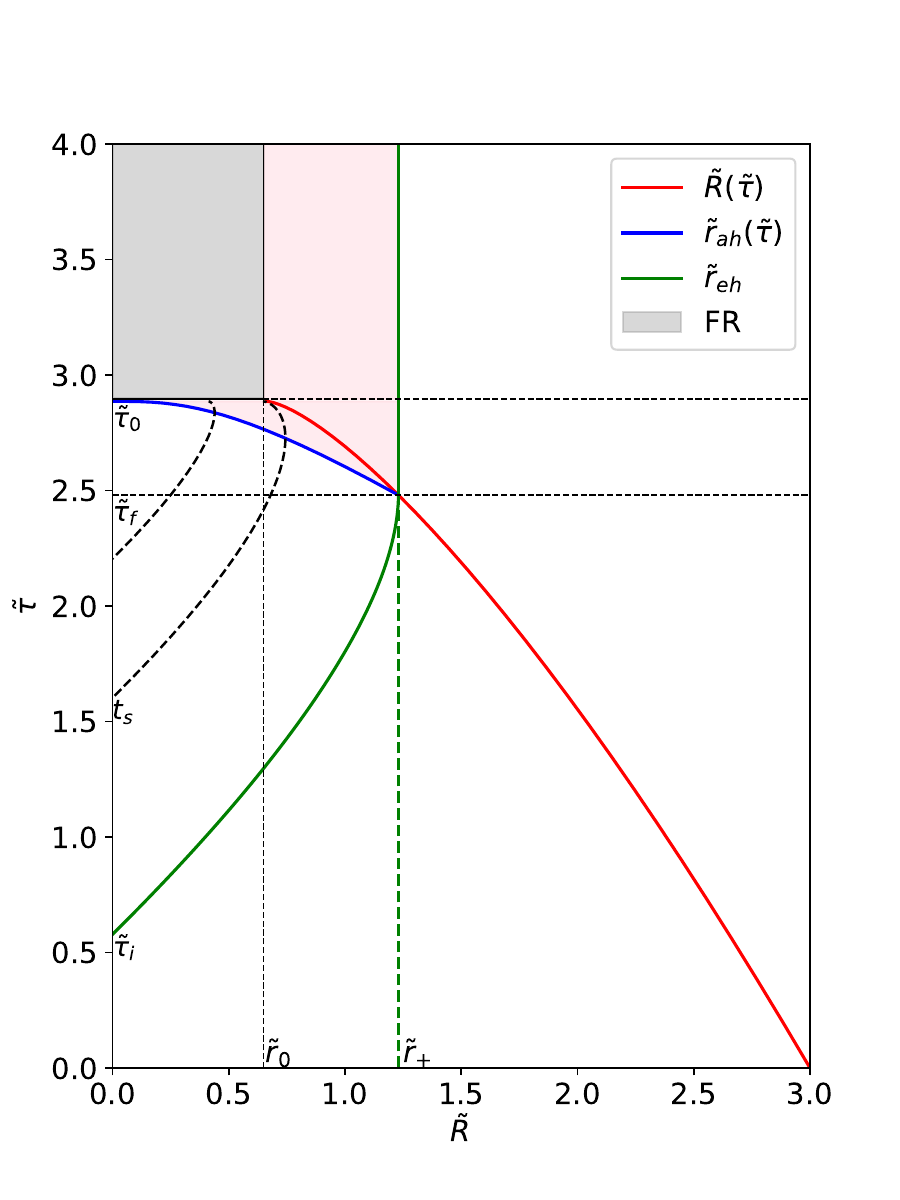}
\caption{Gravitational collapse for $\om<0$. Evolution of the star surface (red), of the event horizon (green), of the apparent horizon (blue) during the collapse. The black dashed curved line (departing from the centre at $\tilde\tau=t_s$) represents the last light ray arriving at the star surface just before this crashes on the singularity. The grey area is the extended ``central'' singularity. The black $\tilde r_0$ dashed vertical line indicates the positive radial coordinate of the singularity. All the surfaces with $\tilde r_{ah}(\tilde\tau)<\tilde r<\tilde r_+$ for $\tilde\tau>\tilde\tau_f$ are \textit{trapped surfaces} (pink and grey areas).}
\label{fig17}
\end{figure}

We look now at the evolution of the event horizon, namely at the trajectories of the light rays inside the star. As we discussed, the specific trajectory (the geodesic) of the last ray able to leave the star is governed by the systems of differential equations \eqref{73a},\eqref{73b}, with the specific conditions $\tilde R(\tilde \tau_f) = \tilde r_+$, $\tilde r_{eh}(\tilde \tau_f) = \tilde r_+$. 
We can proceed to a numerical integration of system (\ref{73a}-\ref{73b}) in the case of $\om<0$ (an analytical solution is reported in Appendix \ref{Appendix F}). The result are displayed in Fig.\ref{fig17}. 
In particular, the integration of \eqref{73a} yields the function $\tilde R(\tau)$ (red line), which describes the star surface, and therefore the evolution of the apparent horizon $r_{ah}(\tilde\tau)$, through Eq.\eqref{AHom} (blue line).
The evolution of the event horizon (green line) inside the star, $\tilde{r}_{eh}(\tilde \tau)$, is described by Eq.\eqref{73b}.
%
%
Any light ray leaving the star surface at a time $\tilde \tau<\tilde \tau_f$ can escape to infinity; light rays leaving the star surface at times $\tilde \tau_f<\tilde \tau<\tilde \tau_0$ are bounded to collapse onto the singularity 
$\tilde r=\tilde r_0>0$. Besides the black dashed vertical straight line $\tilde r=\tilde r_0$, Fig.\ref{fig17} displays also the trajectory of the last light ray able to arrive at the surface of the star before this crashes onto the singularity, namely the black dashed curved line, let's call it $\tilde r_s(\tilde\tau)$, passing through the center of the star at $\tilde\tau=t_s$.
For any given time $\tilde\tau_x$ such that $\tilde\tau_i<\tilde\tau_x<\tilde\tau_0$, a null geodesic passing through the point $\tilde r(\tilde\tau_x)$ such that 
${\rm Max}\{0,\tilde r_s(\tilde\tau_x)\}<\tilde r(\tilde\tau_x)<{\rm Min}\{\tilde r_{eh}(\tilde\tau_x),\tilde R(\tilde\tau_x)\}$
is bounded to arrive at the surface of the star, leave the star, and finally crash back onto the singularity $\tilde r_0$. On the contrary, any light ray passing through the point $\tilde r(\tilde\tau_x)$ such that
$0<\tilde r(\tilde\tau_x)<{\rm Max}\{0,\tilde r_s(\tilde\tau_x)\}$ is not even able to arrive at the star surface, but instead directly crashes into the singularity at $\tilde\tau=\tilde\tau_0$ (see the black curve dashed line on the left side of Fig.\ref{fig17}).
For any time $\tau>\tau_f$ the surfaces $S(0,\tilde r)$ with $\tilde r_{ah}(\tilde\tau)<\tilde r<\tilde r_+$ are \textit{trapped surfaces} (pink and grey areas in Fig.\ref{fig17}).

Finally, we can comment by saying that the three types of collapses above described, and summarized by the Figures \ref{fig9}, \ref{fig14}, \ref{fig17}, share some general properties: (a)- For a stationary spacetime outside the star (as it is in our case), the apparent horizon coincides with the event horizon, outside the star (see e.g. Ref.\cite{Hawking:1973uf}); (b)- The event horizon starts its evolution inside the star always before the star surface reaches $r_+$. This also means that the event horizon always begins to form inside the star before the apparent horizon appears; (c)- The event horizon always evolves from a zero radius and reaches $r_+$, while the apparent horizon starts its evolution from a non-zero radius; (d)- The apparent horizon always appears placed inside the event horizon. According to Ref.\cite{Nielsen} (section 3.2), this last feature is a direct consequence of the null energy condition (NEC), which in fact in our system always holds, for any $\om$ (see Sec.\ref{EC}). 

%
%

\section{Equation of state of the collapsing star}\label{njjvft}

Before proceeding further with technical calculation, we would like to remind the general framework in which we move. 
To describe the interior of a collapsing star, in (improved) Friedmann models we have usually a FLRW metric (in our case flat), and three variables, represented respectively by the scale factor $a(\tau)$, the pressure $p(\tau)$ and energy density $\rho(\tau)$ of the ``matter'' content of the system (a perfect fluid, in our case). The system is then closed by three fundamental independent equations: the (00) component of the improved Einstein field equation $G_{\mu\nu}= 8\pi G(k)T_{\mu\nu}$; the energy-momentum conservation equation, coming directly from the Bianchi identity $[8\pi G(k)T^{\mu\nu}]_{;\nu}=0$; the equation of state of the ``matter'' in question, usually expressed as $p=p(\rho)$.

For example, in the classical OS collapse \cite{Oppenheimer:1939ue}, Authors assumed, besides an internal FLRW flat metric and an external Schwarzschild metric, also a perfect fluid of negligible pressure, namely a ``dust''. In other words they fixed the third equation $p=0$, and then they proceeded to find the other two variables $a(\tau)$, and $\rho(\tau)$, using the remaining two equations (see also Ref.\cite{Weinberg:1972kfs} p.342, p.472). In fact they found $\rho \sim a^{-3}$, and $\dot{a}^2 \sim 1/a$, which is the motion we studied in Sec.\ref{dmkdk}A, Eq.\eqref{SCHW}. In different approaches, as for example in Ref.\cite{Shojai_2022}, and in the present paper, the evolution of $a(\tau)=R(\tau)/R_c$ is fixed by the requirement of a smooth joining between the external and internal geometries (see Eq.\eqref{FunEq}). Then, once $a(\tau)$ is known, the remaining two equations allow us to compute the energy density $\rho(\tau)$, the pressure $p(\tau)$, and therefore the equation of state $p=p(\rho)$.

We are now in a position to compute the equation of state of the perfect fluid composing our collapsing star. 

We are going to use Eqs.\eqref{dcwpdw[e}, \eqref{dmcwldk}. For $G(k)$ the cutoff identification in this case will be (see also Appendix \ref{Appendix B}) $k \to k(\tilde r (\tilde \tau))$, therefore we deal with the $G(\tilde r)$ of Eq.\eqref{dkdmwm}. We compute Eq.\eqref{dcwpdw[e} on the surface of the star, therefore with 
$\tilde r(\tilde \tau) = \tilde R(\tilde \tau) = a(\tilde \tau) \tilde R_c$, it reads
\be
\label{dnwdnpwdf}
\left(\frac{\dot{a}}{a}\right)^2=\left(\frac{\dot{a}\tilde R_c}{a\tilde R_c}\right)^2=
\left(\frac{\dot{\tilde R}}{\tilde R}\right)^2=\frac{8\pi}{3}G(\tilde R)\tilde\rho\,.
\ee
Observing that $G(\tilde R)=1/\alpha(\tilde R)$, and using Eq.\eqref{FunEqTilde} (dictated by the junction condition of the metric across the star surface), we have
\be
\label{ax,cpk}
\tilde{\rho}(\tilde\tau)=\frac{3}{8\pi \tilde{R}(\tilde \tau)^3}
\ee
where $\tilde{\rho}(\tilde\tau)=(2M)^2\rho(\tau)$. 
To arrive at the pressure we can now insert \eqref{ax,cpk} into the continuity (or energy-momentum conservation) equation \eqref{dmcwldk}, which, computed in terms of tilded variables and at the surface of the star, reads
\be
\dot{\tilde\rho}+3\left(\frac{\dot{\tilde R}}{\tilde R}\right)(\tilde\rho+\tilde p) = 
-\frac{\tilde\rho\dot{G}(\tilde R)}{G(\tilde R)}\,.
\ee
Reminding that $\dot{G}(\tilde R)=G'(\tilde R)\dot{\tilde R}$, 
we finally obtain the stellar surface pressure as
\be
\label{edeidneeo}
\tilde{p}(\tilde\tau)=-\frac{\om(2\tilde{R}+3\gamma/2)}{8\pi \tilde{R}^3\big(\tilde{R}^3+\om(\tilde{R}+\gamma/2)\big)}
\ee
where $\tilde{p}(\tilde\tau)=(2M)^2p(\tau)$. Eliminating $\tilde R$ from the resulting expressions \eqref{ax,cpk}, \eqref{edeidneeo} leads to the equation of state of the star. The result is
\be
\label{smkpeoo}
\tilde p(\tilde \rho)=-\frac{4\pi\tilde\omega\tilde\rho^2}{3}\left\{\frac{4\left(\frac{3}{8\pi\tilde\rho}\right)^{1/3}+ \ 3\gamma}{3+4\pi\tilde\omega\tilde\rho\left[2\left(\frac{3}{8\pi\tilde\rho}\right)^{1/3}+ \ \gamma\right]}\right\}
\ee
It is interesting to examine the properties of this equation of state. First of all, we note that if $\om=0$, namely if we consider the collapse in a Schwarzschild background with a motion given by Eq.\eqref{SCHW}, we find $\tilde p=0$, coherently with the OS collapse of dust. 

Second, for $\om>0$, the expression in the curly brackets is always positive (since of course $\tilde \rho \geq 0$) and therefore the pressure is always negative, $\tilde p<0$. It is quite well known that a negative pressure, as in 
Eq.\eqref{smkpeoo}, can act as an anti-gravitational effect able to resist against the gravitational collapse 
\cite{Brustein:2018web}. To avoid the singularity at the end of the collapsing process, some exotic matter with a negative pressure is needed. This kind of matter appears for example in the Gravastar model \cite{Mazur:2001fv}, or in Schwarzschild-AdS black hole models with a minimal length \cite{Miao:2016ipk}. In general, models with a minimal length also describe non singular gravitational collapses \cite{Montani}.
Of course negative pressures may lead to violation of some energy condition, as we shall discuss in the next section.


Moreover, both for $\om>0$ and for $\om<0$, the following asymptotic behaviors hold
\be
&&{\rm for \ \tilde\rho \ large}: \quad \{...\} \sim \frac{3}{4\pi\om\tilde\rho} \ \ \Rightarrow \ \ \tilde p \sim -\tilde\rho \notag\\
&&{\rm for \ \tilde\rho \ small}: \quad \{...\} \sim \frac{4}{3}\left(\frac{3}{8\pi\tilde\rho}\right)^{\frac{1}{3}} \ \ \Rightarrow 
\ \ \tilde p \sim -K\om\,\tilde\rho^{\frac{5}{3}} \notag
\ee
where $K$ is a positive numerical constant. Again in both cases, $\om \lessgtr 0$, the pressure, although always negative when $\om>0$, goes like $\tilde p \propto \tilde\rho^\lambda$, so our star behaves as a polytrope with index, respectively, $\lambda=1$ or $\lambda=5/3$ (according to the Weinberg definition Ref.\cite{Weinberg:1972kfs} p.309). 

Here, however, we should distinguish between two different physical situations. 

When $\om>0$, $\tilde R$ can span any positive value, the pressure is always negative (see Eqs.\eqref{edeidneeo},\eqref{smkpeoo}), $\tilde R$ can go to zero, $\tilde\rho$ can diverge to $+\infty$, and the pressure can diverge to $-\infty$ as $-\tilde\rho$.
  
\begin{figure}[h]
\centering
\includegraphics[scale=0.45]{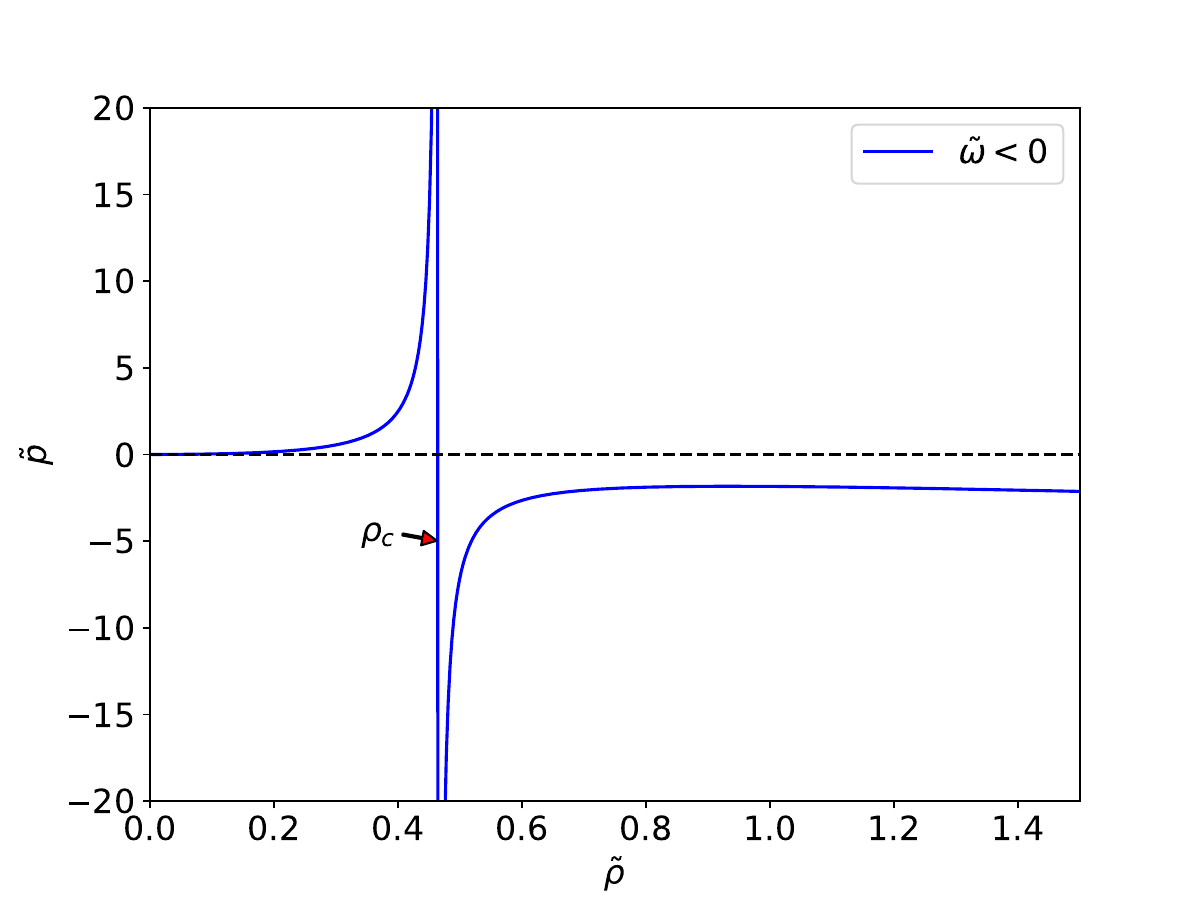}
\caption{For $\om<0$, the behavior of the star pressure $\tilde p(\tilde\rho)$, Eq.\eqref{smkpeoo} is plotted  against density. The density cannot be larger than $\tilde\rho_c$, since $\tilde R>\tilde r_0$. For $0<\tilde\rho<\tilde\rho_c$ the pressure is always positive. When $\tilde\rho \to \tilde\rho_c$, where there is a curvature singularity, the density is finite and the pressure goes to infinity. The region $\tilde\rho > \tilde\rho_c$ is unphysical.}
\label{fig18}
\centering
\includegraphics[scale=0.45]{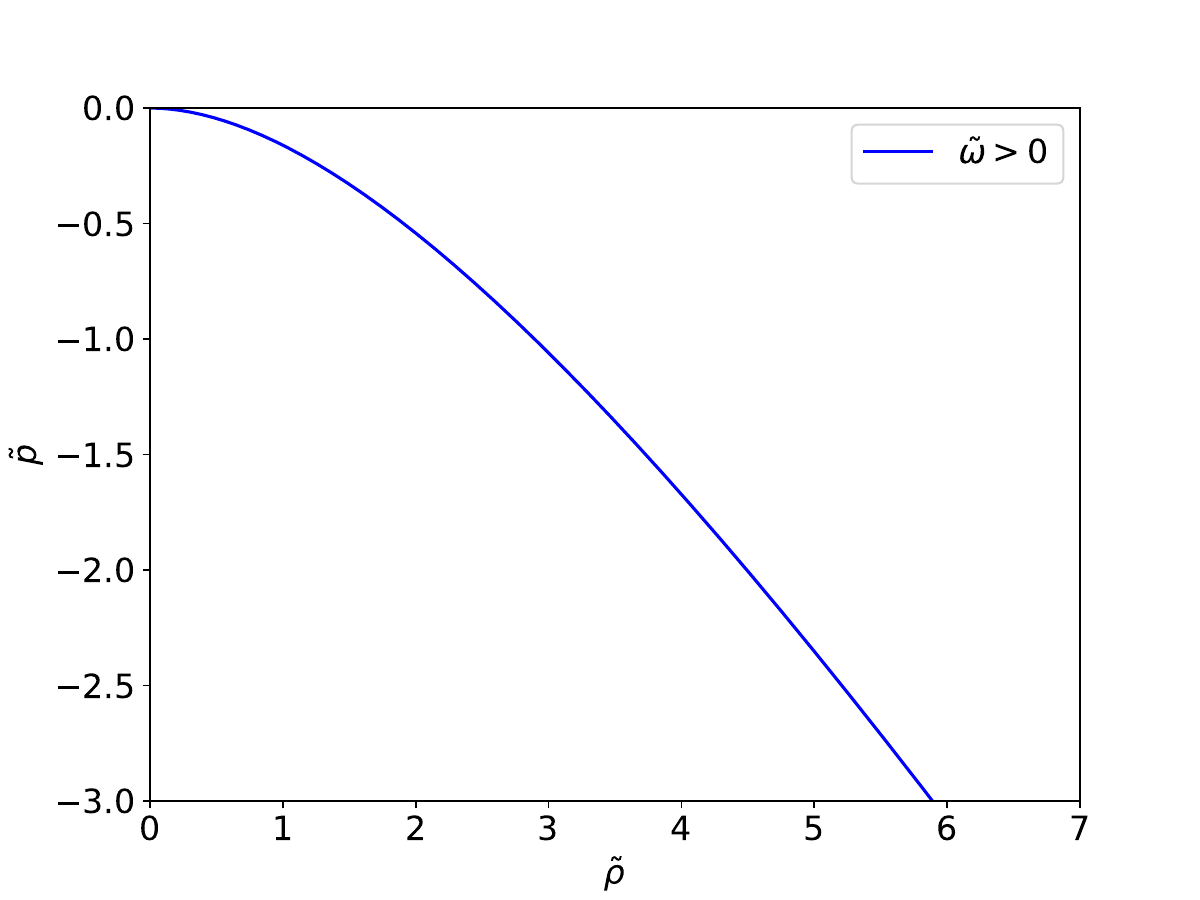}
\caption{For $\om>0$, the behavior of the star pressure $\tilde p(\tilde\rho)$, Eq.\eqref{smkpeoo}, is plotted against density. The pressure is always negative, for any $\tilde\rho>0$ and any $\om>0$.}
\label{fig19}
\end{figure}

When $\om<0$, then we know that $\tilde R \geq \tilde r_0 >0$ should hold (where $\tilde r_0$ is the positive solution of the singularity equation $\tilde r_0^3+\om(\tilde r_0+\gamma/2)=0$, see also Sec.\ref{omneg}). Therefore the minimum value for $\tilde R$ is $\tilde R \to \tilde r_0$, and the density can at most reaches the maximum \textit{finite} value 
\be
\tilde \rho_c = \frac{3}{8\pi\tilde r_0^3} \,.
\label{rhocr}
\ee
Notice that when $\tilde R > \tilde r_0$, and $\om<0$ then the pressure \eqref{edeidneeo} is positive. In terms of density, this can be restated for Eq.\eqref{smkpeoo} by saying that for $0<\tilde\rho<\tilde\rho_c$ (and $\om<0$), then 
$\tilde p(\tilde\rho)>0$. In particular, therefore, the limit $\tilde\rho \to +\infty$ cannot be taken when $\om<0$, simply because it must hold $0<\tilde\rho<\tilde\rho_c$, being $\tilde R \geq \tilde r_0$ always. Of course, when 
$\tilde R \to \tilde r_0$ or equivalently $\tilde\rho \to \tilde\rho_c$, namely when the star surface reaches the curvature singularity in $\tilde r_0$, then the pressure diverges, $\tilde p(\tilde\rho) \to +\infty$ 
\footnote{It is straightforward to show that $\tilde p(\tilde\rho) \to +\infty$ in Eq.\eqref{smkpeoo} when $\tilde\rho \to \tilde\rho_c=3/(8\pi\tilde r_0^3)$. In fact, it is sufficient to remember that $\tilde r_0^3=-\om(\tilde r_0+\gamma/2)$, by definition of singularity $\tilde r_0$.}. 
The behavior of the function $\tilde p(\tilde\rho)$ is displayed, respectively, in Fig.\ref{fig18} for $\om<0$, and in Fig.\ref{fig19} for $\om>0$. 

%

Looking at the behavior of the pressure, a physical consideration arises spontaneously. When $\om>0$ the pressure is always negative, even for a vanishingly small density of the ``matter''. Now, a form of ``matter'' able to produce negative pressures, even at small densities, should be a quite exotic one, certainly not so common in the observable universe. This could be a strong argument in favor of a \textit{negative} value of the $\om$-parameter. In fact, for such $\om<0$ the pressure keeps being positive in all the range $0<\tilde\rho<\tilde\rho_c$, and it diverges to positive infinity when the star reaches the singularity in $\tilde r_0$. \\      

We assumed that the exterior of the star $\tilde r > \tilde R$ is described by the metric \eqref{I}.
In particular, once $\tilde R \to \tilde r_0$ (for $\om<0$), the whole black hole spacetime $\tilde r > \tilde r_0$ is described by the metric \eqref{I}. Of course, also in the exterior of the star the effective Einstein field equations \eqref{axskss} hold, namely $G_{\mu\nu}=8\pi G(k)T_{\mu\nu}$. It is then possible to use in reverse the Einstein equations in order to \textit{define} and compute the energy-momentum tensor of an ``effective'' matter fluid, which virtually permeates the region outside the star, and it is able to simulate the effect of the quantum fluctuations generating the metric \eqref{I} (see in particular Refs.\cite{PhysRevD.62.043008,Barraco:2003jq,Brustein:2018web}). Since the metric \eqref{I} is static, isotropic, and diagonal, then the $G_{\mu\nu}$ computed from it will be diagonal, and so will be $T_{\mu\nu}$ too. Therefore, in analogy with the energy-momentum tensor for a perfect fluid, and according to our metric signature, we can adopt for our ``effective'' energy-momentum tensor the form
\be
T^\mu_{\ \nu} = {\rm diag}(-\tilde\rho, \ \tilde p_r, \ \tilde p_T, \ \tilde p_T)
\ee
where we allow in principle for an anisotropy between the radial pressure $p_r$ and the transverse components 
$p_T=p_\theta=p_\phi$ (tangential stress). 

The metric \eqref{I} is just a particular case of 
\be
ds^2 = -e^{\nu(r)} dt^2 + e^{\lambda(r)} dr^2 + r^2d\Omega^2\,,
\ee
where we dropped the tildes, for agility of notation. 
According to what we said above, and following for example Ref.\cite{Stephani:2004ud}, we can then write the field equations
\be
&&8\pi G(r)T^0_{\ 0} = 8\pi G(r)(-\rho) = -e^{-\lambda}\left(\frac{\lambda'}{r} - \frac{1}{r^2}\right) - \frac{1}{r^2}\notag\\
&&8\pi G(r)T^1_{\ 1} = 8\pi G(r) p_r = e^{-\lambda}\left(\frac{\nu'}{r} + \frac{1}{r^2}\right) - \frac{1}{r^2}\notag\\
&&8\pi G(r)T^2_{\ 2} = 8\pi G(r) p_T = \notag\\ 
&&=\frac{1}{2}e^{-\lambda}\left(\nu''+\frac{\nu'^2}{2}+\frac{\nu'-\lambda'}{r}-\frac{\nu'\lambda'}{2}\right)\notag
\ee
where we implemented the cutoff $k \to k(r)$, so that we deal with the $G(r)$ of Eq.\eqref{dkdmwm}. 
Note that $G(r)=1/\alpha(r)$. Remembering that 
\be
e^{\nu(r)} = \left(1 - \frac{1}{r\alpha(r)}\right) = e^{-\lambda(r)}\,,
\ee
we finally arrive at the density and anisotropic pressure of the ``effective'' matter/fluid content of our model
\be
\tilde\rho^{\rm eff} = -\tilde p_r^{\rm eff} = -\frac{\alpha'(\tilde r)}{8\pi \ \tilde r^2 \ \alpha(\tilde r)}
\ee
\be
\tilde p_T^{\rm eff} = \frac{\alpha(\tilde r)\alpha''(\tilde r) - 2(\alpha'(\tilde r))^2}{16 \pi \ \tilde r \ \alpha(\tilde r)^2}
\ee
If now we insert the expression \eqref{I} for $\alpha(\tilde r)$ in the previous two relations, then
the density and anisotropic pressure of the matter/fluid content take the form
\begin{align}
\label{ssksoeldc}
\tilde{\rho}^{\rm eff}&=-\tilde{p}^{\rm eff}_{r}=\frac{\tilde{\omega} \big(2\tilde{r}+\frac{3\gamma}{2}\big)}{8\pi\tilde{r}^3\left[ \tilde{r}^3+\om(\tilde{r}+\frac{\gamma}{2})\right]}\\
\tilde p_T^{\rm eff}&=\frac{3\om(\tilde{r}+\gamma)}{8\pi\left[\tilde{r}^3+\om(\tilde{r}+\frac{\gamma}{2})\right]^2}\notag\\
&-\frac{\tilde{\omega}^2 \left(\tilde{r}^2+\frac{3\gamma}{2}\tilde{r}+\frac{3\gamma^2}{4}\right)}{8\pi \tilde{r}^3\left[\tilde{r}^3+\om(\tilde{r}+\frac{\gamma}{2})\right]^2}
\end{align}
Once again, we notice that if we take $\om=0$, we then have $\alpha(\tilde r)=1$, and therefore $\tilde\rho^{\rm eff}=-\tilde p_r^{\rm eff}=\tilde p_T^{\rm eff}=0$, namely we go coherently back to the standard Schwarzschild vacuum solution.
We can compute the total radial pressure on the stellar surface by using equations \eqref{edeidneeo} and \eqref{ssksoeldc}, and the result is 
\be
\tilde{p} \ - \ \tilde{p}_r^{\rm eff}|_{\tilde r=\tilde R} \ = \ 0\,. 
\ee
This confirms that there is no additional pressure on the surface of the star and the particles follow the radial free falling geodesics, in agreement with equation \eqref{CC}. 

Finally, equation \eqref{ax,cpk} yields the following mass function
\be
\label{mcsksms}
M=\frac{4\pi}{3}R^3\rho=M_{s}\,.
\ee
Here $M$ is the geometric mass of the star, namely the mass parameter which appears into the metric. The previous equation is therefore telling us that the geometric mass $M$ coincides actually with the physical mass of the star, which of course is obtained by definition by $M_{s} = V_{olume}\cdot \rho$ (for $\rho$ uniform). Notice that this is not always necessarily true. For example, in the gravitational collapse to a regular BH, the geometric mass $M$ results larger than the mass $M_s$ of the star (see e.g. Ref.\cite{Shojai_2022}, eq.11; see also Ref.\cite{Hayward:2005gi}). The deep reason of the equality $M=M_s$ in the present model, is that here the gravitational ``constant'' is not constant, as in Ref.\cite{Shojai_2022}, but instead ``running'', it depends on the cutoff $k$, namely $G(\tilde r)= 1/\alpha(\tilde r)$, see Eq.\eqref{dkdmwm}.

\section{Energy conditions}\label{EC}

When considering Einstein equations, it appears clear that they completely leave open the question on 
what specific properties matter should have, in order to be considered physical. Actually, this aspect was 
already well known to Einstein himself, who, very wisely, intentionally gave only a general phenomenological representation 
of the matter in his theory, in terms of an energy momentum tensor, indeed so avoiding to assume risky hypothesis on a yet unknown (then, as well as today) complete microscopic description of matter.  

An easy way to understand the gravity of this (potential) ``arbitrariness'' hidden in $T_{\mu\nu}$ is the following: without
any restrictions on $T_{\mu\nu}$, any metric $g_{\mu\nu}$ whatsoever can be considered a solution of the Einstein equations, with an ``energy-momentum tensor'' defined by $T_{\mu\nu}=G_{\mu\nu}[g_{\mu\nu}]/(8\pi G_N)$. The energy-momentum tensor so computed will be automatically conserved by the Bianchi identity. Obviously, the problem with this approach is that generically this candidate ``energy-momentum tensor'' will not have any of the very general properties usually associated with reasonable forms of matter: positivity of energy density, causal propagation of energy/matter, attractive (rather than repulsive) gravitational force.

To limit the above paradoxical arbitrariness, scholars have essentially followed two (not mutually exclusive) paths: when looking for exact solutions in certain specific situations, they usually specify the matter content and the energy momentum tensor (either phenomenologically or microscopically) with reasonable and well-motivated physical properties, and then furthermore impose some symmetry conditions on the unknown metric. When instead they look for general properties of solutions to the Einstein equations, scholars limit the arbitrariness of $T_{\mu\nu}$ by imposing some simple general constraints on the energy-momentum tensor known as \textit{Energy Conditions}. Usually these conditions are given in the form of pointwise constraints on the contraction of an energy-momentum tensor with causal (i.e. timelike or null, or non-spacelike) vectors. 

For sake of readability, here we report the main energy conditions (see also Refs.\cite{blau2011lecture,Hawking:1973uf,Krommydas:2017ydo}), together with their implementation for the matter we consider in our model of collapse, namely the energy-momentum tensor of a perfect fluid, $T_{\mu\nu}=(\rho +p)u_{\mu}u_{\nu}+pg_{\mu\nu}$, with $g_{\mu\nu}u^{\mu}u^{\nu}=-1$. Later in this section we shall check to what extent these energy conditions are obeyed or violated by our collapsing fluid. 
\begin{itemize}
\item NEC: Null Energy Condition. It requires $T_{\mu\nu}k^\mu k^\nu\geq 0$ for any null vector $k_\mu k^\mu=0$. For the energy momentum tensor of a perfect fluid this translates into 
		\be
		\rho + p \ \geq \ 0\,.
		\ee
\item WEC: Weak Energy Condition. It requires $T_{\mu\nu}t^\mu t^\nu\geq 0$ for any timelike vector $t_\mu t^\mu<0$ (without loss of generality normalized to $t_\mu t^\mu=-1$), and by continuity, also for null vectors $k_\mu k^\mu=0$. For a perfect fluid energy momentum tensor this translates into
		\be
		\rho \ \geq \ 0 \quad\quad {\rm and} \quad\quad \rho + p \ \geq \ 0\,.
		\ee
\item DEC: Dominant Energy Condition. It requires that the energy-momentum current density $P^\mu=T^\mu_\nu t^\nu$ seen by an observer with a timelike (and future-directed) 4-velocity $t^\nu$	should not exceed the speed of light, namely $P^\mu$ should be a causal future-directed non-spacelike vector, i.e. $P^\mu t_\mu \leq 0$ and $P^\mu P_\mu \leq 0$. This constraints the energy-momentum tensor of a perfect fluid as
\be
\rho \ \geq \ |p|\,.
\ee	
\item SEC: Strong Energy Condition. It requires that the geometry (the Ricci tensor) should have a focussing effect on timelike geodesic congruences (families of freely falling particles), namely it should be $R_{\mu\nu}t^\mu t^\nu \geq 0$ for any timelike vector $t_\mu t^\mu<0$. Because of the Einstein equations, this means $(T_{\mu\nu}-\frac{1}{2}g_{\mu\nu}T)t^\mu t^\nu\geq 0$, which implies, for a perfect fluid,
\be
\rho + p \ \geq \ 0 \quad\quad {\rm and} \quad\quad \rho + 3p \ \geq \ 0\,.
\ee  
\end{itemize}
Clearly we have
\be
&&{\rm DEC} \quad\quad \Rightarrow \quad\quad {\rm WEC} \quad\quad \Rightarrow \quad\quad {\rm NEC} \notag \\
&&{\rm and} \notag \\
&&{\rm SEC} \quad\quad \Rightarrow \quad\quad {\rm NEC} \notag
\ee

In the following we examine to what extent the above energy conditions are satisfied by our collapsing perfect fluid.

\subsection{\textbf{Case $\om \ > \ 0$.}}

\noi$\bullet$ DEC requires $\tilde\rho \geq |\tilde p|$, which in terms of Eqs.\eqref{ax,cpk},\eqref{edeidneeo} reads
\be
\frac{3}{8\pi \tilde{R}^3} \ \geq \ 
\frac{1}{8\pi \tilde{R}^3}\cdot\frac{\om(2\tilde{R}+3\gamma/2)}{\tilde{R}^3+\om(\tilde{R}+\gamma/2)}
\ee
which is equivalent to
\be
3\tilde{R}^3+\om\tilde{R} \ \geq \ 0\,,
\ee
which is always true for any $\om>0$ (and obviously $\tilde R \geq 0$).
So, DEC is always obeyed when $\om>0$, an therefore WEC and NEC will be obeyed too.\\

\noi$\bullet$ SEC requires $\tilde\rho + 3\tilde p \geq 0$, which according to Eqs.\eqref{ax,cpk},\eqref{edeidneeo} translates into
\be
\label{SEC}
\tilde{R}^3 \geq \om(\tilde{R}+\gamma)\,.
\ee 
In analogy to Fig.\ref{fig1}, a straightforward graphical analysis shows that for any $\om>0$ there exists a single 
$\tilde r_2(\om)>0$, root of $\tilde{r}_2^3 = \om(\tilde{r}_2+\gamma)$, such that SEC is fulfilled for any $\tilde R > \tilde r_2$, and it is violated for any $0<\tilde R <\tilde r_2$.

\begin{figure}[h]
\centering
\includegraphics[scale=0.45]{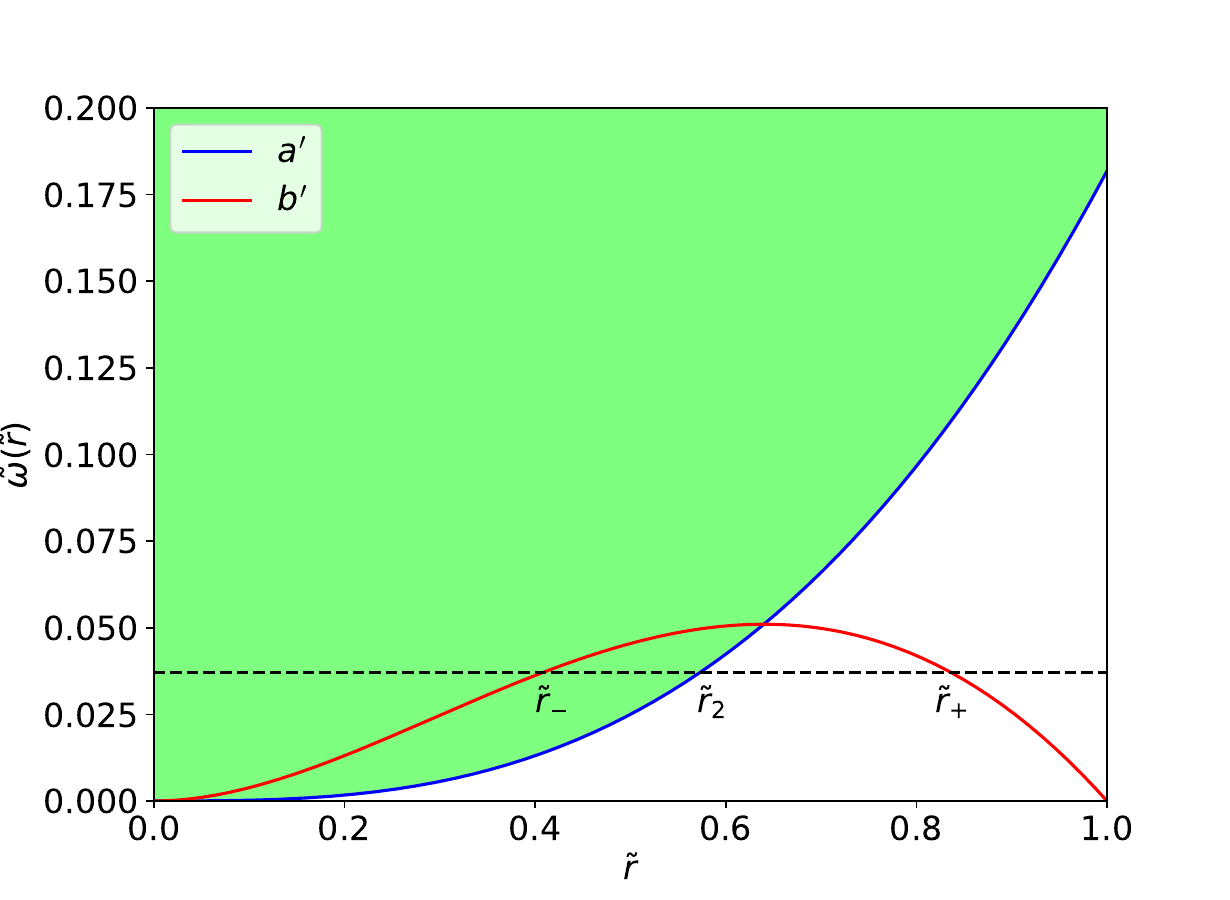}
\caption{For any given $\om>0$, SEC is violated for any $0<\tilde R<\tilde r_2(\om)$, namely in the green region, whose boundary is line $a'\equiv \om(\tilde r_2)$. For any given $0<\om<\om_c$, we have two horizons $\tilde r_-$, $\tilde r_+$, and we always have $\tilde r_-(\om)<\tilde r_2(\om)<\tilde r_+(\om)$, so the region where SEC is violated is protected by the horizon $\tilde r_+$.}
\label{fig20}
\end{figure}

It is interesting to investigate where this $\tilde r_2(\om)$ falls in respect to the two horizons $\tilde r_-$, $\tilde r_+$, which are present when $0<\om<\om_c$. For a given $\om$, we know that $\tilde r_2$ solves the following equation ($a$), while $\tilde r_-\,, \tilde r_+$ solve equation ($b$), where
\be
&&(a) \quad\quad \tilde r^3 = \om(\tilde r+\gamma) \notag \\
&&(b) \quad\quad \tilde r^3-\tilde r^2+\om(\tilde r+\gamma/2)=0\,.
\ee 
We don't need to solve complicated $3^{rd}$-degree equations, it is sufficient to invert for $\om$ the two equations ($a$) and ($b$), and plot the two functions
\be
&&(a') \quad\quad \om=\frac{\tilde r^3}{\tilde r+\gamma} \notag \\
&&(b') \quad\quad \om=\frac{\tilde r^2-\tilde r^3}{\tilde r+\gamma/2}\,.
\ee 
Hence we have the diagrams in Fig.\ref{fig20}. 
%
%
For any given $\om$ in $0<\om<\om_c$ the intersections of the horizontal black dashed line with the red $b'$-line represent the radial coordinates of the two horizons $\tilde r_-$, $\tilde r_+$, while the intersection with the blue $a'$-line gives the radial boundary $\tilde r_2$ of the region where SEC is violated.  
Since SEC is violated for any $0<\tilde R<\tilde r_2$, we see that SEC is violated in the whole green region. As we see, $\tilde r_-<\tilde r_2<\tilde r_+$. As far as horizons exist (namely for $0<\om<\om_c$), the region where SEC is violated is always protected by the outer event horizon $\tilde r_+$\footnote{Notice that the above proposition is true because the $a'$-line crosses the $b'$-line always in the $b'$-line maximum point, for any $\gamma\geq 0$. In fact the radial coordinate of the $b'$-line maximum coincides with the radial coordinate of the intersection of the two lines $a'$ and $b'$. They both obey the equation 
$4x^2+(3\gamma-2)x-2\gamma=0$.}. Once $\om>\om_c$, namely for very small masses, horizons disappear and a violation of SEC might be in principle observed directly. As we said in the previous Section, since for $\om>0$ we deal with ``matter'' with a negative pressure, we reasonably expect some violation of an energy condition. And in fact we verified the violation of SEC. Notice however that DEC, WEC, and NEC are always fulfilled for $\om>0$.

\subsection{\textbf{Case $\om \ < \ 0$.}}

\begin{figure}[h]
\centering
\includegraphics[scale=0.46]{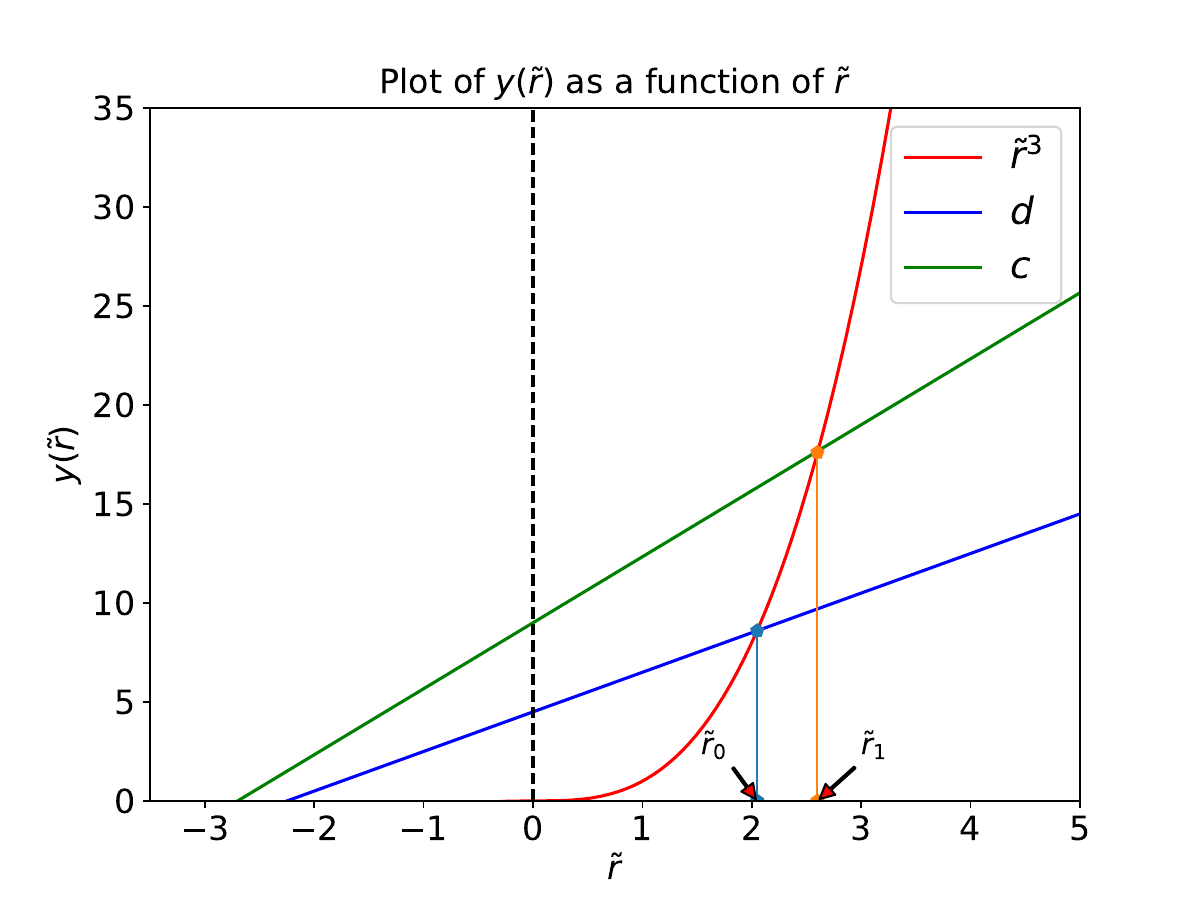}
\caption{Graphical analysis of the singularity equation $\tilde{r}_0^3-|\om|(\tilde{r}_0+\gamma/2)=0$ (line $d$), and of the DEC validity equation $\tilde r_1^3 - (5|\om|/3)(\tilde r_1+3\gamma/5)=0$ (line $c$). For any $\om<0$ DEC is violated when $\tilde r_0(\om) \leqslant \tilde R < \tilde r_1(\om)$.}
\label{fig21}
\end{figure}

\noi$\bullet$ DEC requires $\tilde\rho \geq |\tilde p|$, which now in terms of Eqs.\eqref{ax,cpk},\eqref{edeidneeo} reads
\be
\frac{3}{8\pi \tilde{R}^3} \ \geq \ 
\frac{1}{8\pi \tilde{R}^3}\cdot\frac{|\om|(2\tilde{R}+3\gamma/2)}{\tilde{R}^3-|\om|(\tilde{R}+\gamma/2)}
\ee
where we put $\om=-|\om|$ since $\om<0$. Of course we assume $\tilde R \geq \tilde r_0$, where $\tilde r_0(\om)$ is the single positive solution of the singularity equation $\tilde{r}_0^3-|\om|(\tilde{r}_0+\gamma/2)=0$. The above inequality is equivalent to 
\be
\tilde{R}^3 \ \geq \ \frac{5}{3}|\om|\left(\tilde{R}+\frac{3}{5}\gamma\right)
\label{deceq}
\ee
If we name $\tilde r_1(\om)$ the single positive root of Eq.\eqref{deceq}, again a straightforward graphical analysis (see Fig.\ref{fig21}) shows that for any $\om<0$ we always have
\be
0< \tilde r_0(\om) < \tilde r_1(\om)\,,
\ee
since line ($c$) lies always above line ($d$) for any $\om<0$. Or, in terms of equations,
\be
\frac{5}{3}|\om|\left(\tilde{R}+\frac{3}{5}\gamma\right) > |\om|(\tilde{R}+\gamma/2) \quad \forall \ \om<0, \ \ \forall \ \tilde R>0\,. \notag
\ee
%
%

Thus we conclude that 
\begin{description}
	\item[-] For $\tilde R \geqslant \tilde r_1(\om)$ DEC is fulfilled;
	\item[-] For $\tilde r_0(\om) \leqslant \tilde R < \tilde r_1(\om)$ DEC is violated.
\end{description}
From the physical point of view, the violation of DEC is not a surprise, since we learned in the previous Section that, during the collapse, when the surface $\tilde R$ is approaching the singularity $\tilde r_0$, then the density approaches a finite value, $\tilde\rho \to \tilde\rho_c$, while the pressure diverges to $+\infty$, $\tilde p \to +\infty$. So, at a certaint point it should be $\tilde p \gg \tilde\rho$.

\begin{figure}[h]
\centering
\includegraphics[scale=0.46]{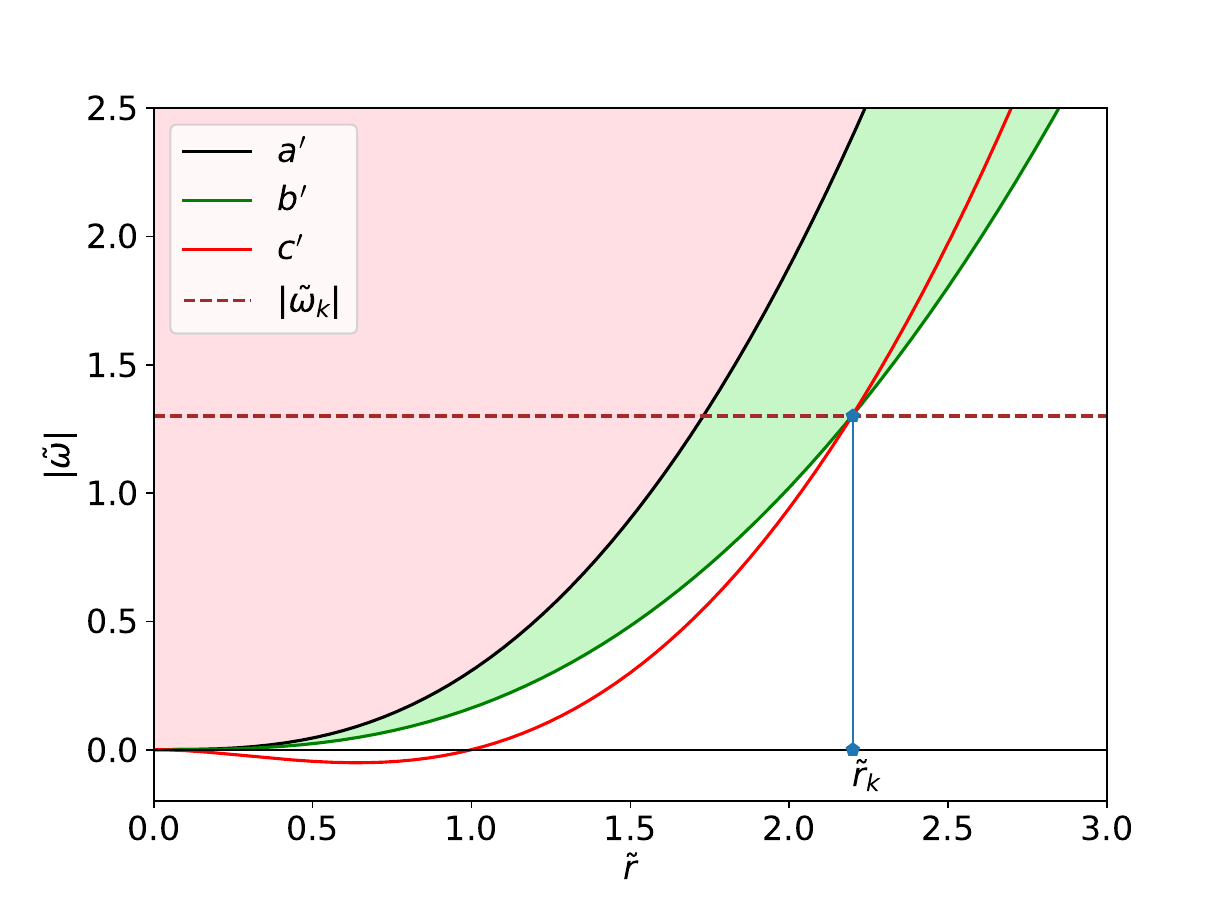}
\caption{Black line $a'$: location of the singularity $\tilde r_0$. 
Green line $b'$: border of the DEC violation green region $\tilde r_1$. 
Red line $c'$: location of the event horizon $\tilde r_+$. 
When $0<|\om|<|\om_k|$ (i.e. when $\om_k<\om<0$) the green region, where DEC is violated, is actually protected by the outer horizon $\tilde r_+$. 
But for $|\om|>|\om_k|$ (i.e. for $\om<\om_k<0$) the DEC violation can happen also outside the horizon $\tilde r_+$, in observable regions.} 
\label{fig22}
\end{figure}

As before, it is of great importance to learn where $\tilde r_1(\om)$ falls in respect to the horizon $\tilde r_+(\om)$ defined by Eq.\eqref{HH}. Arguing as in the previous subsection, we can observe that, for a given $\om<0$, the equations
\be
&&(a) \quad\quad \tilde r^3-|\om|(\tilde r + \gamma/2)=0 \notag \\
&&(b) \quad\quad \tilde r^3 - (5|\om|/3)(\tilde r+3\gamma/5)=0 \notag \\
&&(c) \quad\quad \tilde r^3-\tilde r^2-|\om|(\tilde r+\gamma/2)=0
\ee 
are solved by, respectively, the radial coordinate $\tilde r_0$ of the singularity, by the radial coordinate $\tilde r_1$ of the region where DEC is fulfilled, and by the radial coordinate $\tilde r_+$ of the (outer) horizon. By inverting  the above equations for $|\om|$, we get the functions
\be
&&(a') \quad\quad |\om|=\frac{\tilde r^3}{\tilde r+\gamma/2} \notag \\
&&(b') \quad\quad |\om|=\frac{\tilde r^3}{5\tilde r/3 + \gamma} \notag\\
&&(c') \quad\quad |\om|=\frac{\tilde r^3-\tilde r^2}{\tilde r+\gamma/2}\,.
\label{DECeqs}
\ee 
By plotting the above functions, it becomes easy to see where the points $\tilde r_0(\om)$, $\tilde r_1(\om)$, $\tilde r_+(\om)$ fall for any given $\om<0$. In fact, we have the diagrams in Fig.\ref{fig22}, where for simplicity we depict $|\om|$ directly.
%
%

It appears clear that there exists a critical value $\om_k<0$ such that
\be
&&{\rm for} \ \ \om_k<\om<0 \ \ {\rm then} \ \ \tilde r_0(\om) < \tilde r_1(\om) < \tilde r_+(\om) \notag\\
&&{\rm while}\notag \\
&&{\rm for} \ \ \om<\om_k<0 \ \ {\rm then} \ \ \tilde r_0(\om) < \tilde r_+(\om) < \tilde r_1(\om) \notag
\ee
The critical pair $(\tilde r_k, \om_k)$ can be computed for example by equating the DEC-line equation \eqref{DECeqs}($b'$) with the horizon equation \eqref{DECeqs}($c'$). Then, $\tilde r_k$ is the positive solution of the $2^{nd}$ degree equation (for any $\gamma \geq 0$)
\be
4\tilde r_k^2 + (3\gamma-10)\tilde r_k - 6\gamma = 0\,,
\ee
and from there $\om_k$ can be easily computed using, for example, Eq.\eqref{DECeqs}($b'$).\\ 
Some physical considerations are now in order. We know that sound waves travel in a relativistic fluid with a speed $v_s$ given by
\be
v_s^2 \ = \ \left.\frac{\partial p}{\partial \rho}\ \right|_{\sigma \, const}
\ee
where the propagation happens at a constant entropy per particle $\sigma$ (see e.g. Ref.\cite{Weinberg:1972kfs}). Since no signal can propagate (locally) faster than light, it should be $v_s \leq 1$, which implies 
\be
p \leq \rho\,.
\ee 
Thus, a violation of DEC in a fluid signals that the speed of sound can exceed the speed of light in that fluid 
(see Ref.\cite{Hawking:1973uf}). Now, according to Fig.\ref{fig22}, DEC is violated in the green region. For $\om$ in the range $\om_k<\om<0$ this violation happens inside the event horizon $\tilde r_+$, so it shouldn't in principle affect the ``external'' world. But when $\om<\om_k$ the violation of DEC could happen also in regions of the fluid \textit{external} to the event horizon $r_+$. So the speed of sound could exceed the speed of light in observable regions. The pathological behavior of such a kind of ``matter'' could therefore, in principle, be detected by, or affect, the distant observer. This critical situation could be a good argument in favor of imposing an inferior bound to $\om$: to avoid the observability of such a weird fluid we request $\om_k\leq\om<0$.\\

\noi$\bullet$ WEC requires $\tilde\rho \geq 0$ (which is always fulfilled for any $\om<0$, see Eq.\eqref{ax,cpk}) and 
$\tilde\rho + \tilde p \geq 0$. This second inequality in terms of Eqs.\eqref{ax,cpk},\eqref{edeidneeo} reads
\be
\frac{3}{8\pi \tilde{R}^3} \ + \ 
\frac{1}{8\pi \tilde{R}^3}\cdot\frac{|\om|(2\tilde{R}+3\gamma/2)}{\tilde{R}^3-|\om|(\tilde{R}+\gamma/2)} \ \geq \ 0
\ee
which is always true for any $\om<0$ and any $\tilde R \geqslant \tilde r_0$. So, WEC and therefore NEC are always fulfilled for any $\om<0$.\\

\noi$\bullet$ SEC requires $\tilde\rho + \tilde p \geq 0$, which we just saw is obeyed, and 
$\tilde\rho + 3\tilde p \geq 0$, which translates into
\be
\frac{3}{8\pi \tilde{R}^3} \cdot \frac{\tilde{R}^3-\om(\tilde{R}+\gamma)}{\tilde{R}^3+\om(\tilde{R}+\gamma/2)} \ \geq \ 0\,.
\ee
The above is trivially true for any $\om<0$ and any $\tilde R \geqslant \tilde r_0$. So for negative values of $\om$, the strong energy condition is satisfied outside the singularity. Inside the singularity, for $0<\tilde r<\tilde r_0$, as we have seen in previous Sections, the radius $\tilde R$ of the star's surface takes unphysical complex values, so we called this area a ``forbidden region'' (FR).

\section{Possible future investigations}\label{pfi}

Below we formulate some educate guesses on possible future investigations, in particular on the so-called 
\textit{forbidden region} (FR). We learned that the FR is surrounded by a curvature singularity, which divides spacetime into two separate parts, where the outer is an asymptotically flat spacetime. 

In Sec.\ref{EC} we proved that outside the singularity the classical energy conditions, in particular the DEC and the SEC, are to some specific extent violated. However, it remains to be investigated whether some different modified energy conditions, such as the averaged WEC \cite{Roman:1988vv, Morris:1988tu}, the averaged NEC \cite{Roman:1986tp}, and others \cite{Borde:1987qr}  are violated or not outside the singularity region. 

Of course it is tempting to try to extend such investigations also inside the FR. 
To this aim, a technical tool that can perhaps be useful is the introduction of an imaginary proper time coordinate which naturally emerges from Eqs. \eqref{pt} or \eqref{dmdkd}. In fact it is easy to see that in the FR, where $\tilde{R}<\tilde{r}_{0}$, Eq. \eqref{pt} becomes 
	\be
	\label{ddwbdwe}
	\tilde\tau = \frac{2}{3}\frac{B}{\tilde r_0}\left[(\tilde R_0 - \tilde r_0)^{3/2} - i(\tilde r_0-\tilde R )^{3/2} \right]\,.
	\ee
This suggests that in the FR a useful proper time coordinate could be an imaginary time. Somehow, this would resemble the model proposed by Hartle and Hawking in Ref.\cite{Hartle:1983ai}, where they defined the \textit{no-boundary proposal} in order to have a regular beginning for the Universe. To avoid the initial singularity, they introduced an analytic continuation from a Lorentzian manifold to a smooth Riemmannian manifold by considering an imaginary time rotation $\tau\to i\tau$. The history in time of the Universe would then start when a (quantum) fluctuation induces a transition from the imaginary time to the real time, and flips the metric signature from $(+1,1,1,1)$ to $(-1,1,1,1)$. By proceeding in analogy with the above, the region inside the black hole we proposed (with $\om<0$, see also Ref.\cite{Scardigli:2022jtt}), when described with an imaginary time, could be compared with the early universe models, such as the no-boundary proposal, and could in principle yield some insights about important features of quantum gravity.

\section{FINAL OUTLOOKS}\label{nddj}

In this paper, we have studied the gravitational collapse of a massive object in the general framework of an OS-like model. The inner geometry of the collapsing body is described by a spatially flat FLRW metric, which is then joined smoothly to the outer asymptotically-safe/scale-dependent black hole metric. To describe the stellar matter, we assume that the cosmological dynamics is governed by the standard FLRW equations. The star's density and pressure are determined by the smooth matching of the geometries at the surface of the star resulting from the usual GR junction conditions. 

With reference to Eq.\eqref{CC}, it has been shown (in Sec.\ref{omneg}) that, when $\om<0$, after a finite time the star radius goes to a finite value $\tilde r_0$, the energy density \eqref{ax,cpk} takes the large but finite value $\tilde{\rho}_{c}$ \eqref{rhocr}, and the Kretschmann scalar diverges. Indeed, in this limit the BH metric \eqref{dwdlmskp} has a singularity at a positive radial coordinate $\tilde r_0>0$. According to the modified OS gravitational collapse model presented here, the collapsing matter reaches the singularity with an infinite final velocity (see again Eq.\eqref{CC}). Therefore, the speed of the collapse in the SDG framework (case $\om<0$) is analogous to the standard OS collapse \cite{Oppenheimer:1939ue} (case $\om=0$), and results infinitely faster than the collapse leading to a regular BH formation \cite{Shojai_2022} (case $\om>0$).

It is worth mentioning that for a regular BH (in our case, $\om>0$), the curvature scalar and metric remain finite in the limit of zero radius, and the gravitational collapse of the star takes an infinite lapse of proper time 
\cite{Shojai_2022}. 

We have obtained the dynamics of the interior apparent and event horizons and also of the stellar surface as functions of the proper time of star. In all the cases studied ($\om$ positive, zero, or negative), when the star surface intersects the event horizon, then the apparent horizon starts to decrease from the outer horizon $\tilde r_+$ towards the inner regions, with a dynamics depending on the specific value of $\om$.

The equation of state that describes the perfect fluid of our ASG/SDG star is given in Eq.\eqref{smkpeoo}. We studied its behavior, and we showed that when $\om>0$ the pressure is always negative for any $\tilde\rho$, while when $\om<0$, there exists a maximum value $\tilde\rho_c$ for the density. For values of the density smaller than $\tilde\rho_c$ the pressure is positive. When $\tilde\rho \to \tilde\rho_c$, a singularity in the equation of state pushes the value of pressure to infinity. Coherently with our other findings, this infinity for pressure happens at the curvature singularity, 
when $\tilde R \to \tilde r_0$.

Finally, we extensively discussed the validity of the energy conditions in both cases $\om>0$ and $\om<0$. In the case $\om>0$,
we found that a possible violation of SEC is anyway protected by the outer horizon $\tilde r_+$, at least until horizons exist.  
In the case $\om<0$, we found that a violation of DEC strongly suggests to impose a lower bound for the negative values of $\om$.

\begin{acknowledgments}
\noi F.Shojai is grateful to the University of Tehran for supporting this work under a grant provided by the university research council.\\
F.Scardigli is grateful to the University of Salerno, Department of Physics ``E.R. Caianiello'', and to INFN-Salerno, for the support and the nice hospitality received during the development of this work.
\end{acknowledgments}

\appendix

\section{Running gravitational constant}
\label{Appendix A}

The average Einstein-Hilbert action is given by \cite{PhysRevD.62.043008}

\begin{align}
\label{anAOS[sk}
\Gamma_{k}[g,\bar{g}]=\frac{1}{16\pi G(k)}\int d^{4}x\sqrt{g}(-R+2\Lambda(k))+S_{gf}[g,\bar{g}]
\end{align}

where $G(k)$ and $\Lambda(k)$ are respectively the running Newton constant and running cosmological constant, and $S_{gf}$ is the classical background gauge fixing term. The flow equation describes the evolution of the scale-dependent couplings \cite{PhysRevD.62.043008,PhysRevD.57.971,Wetterich:1992yh} and it is written as follows
\begin{align}
\label{ddcdnk}
\partial_{t}\Gamma_{k}[g,\bar{g}] \ = \ &
\frac{1}{2}{\rm Tr}\Bigg(\frac{\partial_{t}\mathcal{R}^{grav}_{k}[\bar{g}]}{\kappa^{-2}\Gamma^{(2)}_{k}[g,\bar{g}]+\mathcal{R}^{grav}_{k}[\bar{g}]}\Bigg)\nonumber\\
&-{\rm Tr}\Bigg(\frac{\partial_{t}\mathcal{R}^{gh}_{k}[\bar{g}]}{-\mathcal{M}[g,\bar{g}]+\mathcal{R}^{gh}_{k}[\bar{g}]}\Bigg)
\end{align}
where $t=\ln k$, $\Gamma^{(2)}_{k}$ is the Hessian of $\Gamma_{k}$ with respect to $g_{\mu\nu}$, and $\mathcal{M}$ is the Faddeev-Popov ghost operator. The operators $\mathcal{R}^{grav}_{k}$ and $\mathcal{R}^{gh}_{k}$ are filtering functions, indeed they implement the infrared cutoff in the graviton and the ghost sector. They are defined as $\mathcal{R}_{k}(p^{2})\propto k^2R^{0}(z)$ with $z= p^2/k^2$ and $R^0(z)=z/(e^{z}-1)$ \cite{PhysRevD.62.043008,PhysRevD.57.971}. By putting the equation \eqref{anAOS[sk} in \eqref{ddcdnk} and projecting the flow onto the subspace spanned by the Einstein-Hilbert truncation, one can obtain a coupled system of differential equations for the dimensionless Newton's constant $g(k)\equiv k^{2}G(k)$ and the dimensionless cosmological constant $\lambda(k)\equiv \Lambda(k)/k^2$. In this investigation the cosmological constant plays no r\^ole, therefore in our calculations we considered $\Lambda(k)\simeq 0$. So, the differential equation related to dimensionless Newton's constant \cite{PhysRevD.62.043008,PhysRevD.57.971} is written as
\begin{align}
\label{mdkkd}
\frac{dg(t)}{dt}=\Big(2+\frac{B_{1}g(t)}{1-B_{2}g(t)}\Big)g(t)\,.
\end{align}
The constants $B_{1}$ and $B_{2}$ are given by
\begin{align}
\label{mxakxOS}
&B_{1}=-\frac{1}{3\pi}\big(24\Phi^{2}_{2}(0)-\Phi^{1}_{1}(0)\big)\notag\\
&B_{2}=\frac{1}{6\pi}\big(18\tilde\Phi^{2}_{2}(0)-5\tilde\Phi^{1}_{1}(0)\big)\,,
\end{align}
where $\Phi^{p}_{n}(w)$ and $\tilde\Phi^{p}_{n}(w)$ are two auxiliary functions given by
\begin{align}
\label{x[mspdpw}
&\Phi^{p}_{n}(w)\equiv\frac{1}{\Gamma(n)}\int^{\infty}_{0}dzz^{n-1}\frac{R^{(0)}(z)-zR^{(0)'}(z)}{(z+R^{(0)}(z)+w)^p}\notag\\
&\tilde\Phi^{p}_{n}(w)\equiv\frac{1}{\Gamma(n)}\int^{\infty}_{0}dzz^{n-1}\frac{R^{(0)}(z)}{(z+R^{(0)}(z)+w)^p}\,.
\end{align}
With the exponential cutoff $R^0(z)=z/(e^{z}-1)$ we have explicitly
\begin{align}
\label{sclscmnls}
&\Phi^{1}_{1}(0)=\frac{\pi^2}{6},\hspace{0.5cm}\Phi^{2}_{2}(0)=1\notag\\
&\tilde\Phi^{1}_{1}(0)=1,\hspace{0.7cm}\tilde\Phi^{2}_{2}(0)=\frac{1}{2}\,.
\end{align}
If we define $\omega\equiv-\frac{1}{2}B_{1}$ and $\omega'=\omega+B_{2}$, then the analytical solution of the differential equation \eqref{mdkkd} can be written as \cite{PhysRevD.62.043008,PhysRevD.57.971}
\begin{align}
\label{snddnoem}
\frac{g}{(1-\omega' g)^{\frac{\omega}{\omega'}}}=\frac{g(k_{0})}{[1-\omega' g(k_{0})]^{\frac{\omega}{\omega'}}}\left(\frac{k}{k_{0}}\right)^2\,.
\end{align}
We observe that the ratio $\omega'/\omega$ is actually very close to unity. Numerically one finds $\omega'/\omega \approx 1.18$.
Considering then $\omega\simeq\omega'$, one can write the solution \eqref{snddnoem} in terms of the dimensionful Newton's constant $G(k)\equiv g(k)/k^2$ as follows \cite{PhysRevD.62.043008,PhysRevD.57.971}
\begin{align}
\label{dwodnwdweo}
G(k)=\frac{G_{0}}{1+\omega G_{0}k^2/\hbar}
\end{align}
where we restored $\hbar$ for sake of clarity, and where $G_{0}=G(k=0)$ is the experimentally observed value of the Newton's constant, as measured at the scale of the solar system.

\section{$k(r)$ as a position-dependent quantity}
\label{Appendix B}

As shown in details in Ref.\cite{PhysRevD.62.043008}, we summarize here how the momentum $k$ can be converted to a position-dependent quantity in a black hole spacetime. We can write this position-dependent infrared cutoff in the form 
\begin{align}
\label{dcndpwdc}
k(P)=\frac{\xi}{d(P)}
\end{align}
where $\xi$ is a numerical constant and $d(P)$ is the distance scale which provides the relevant cutoff for the Newton's constant when a given test particle is located at the point $P$ of the black hole space-time. 
The relevant asymptotic limits that $k(r)$ should respect are: $k(r)\approx 1/r$ for $r\to\infty$, and $k(r)\approx1/r^{3/2}$ for $r\to 0$ (see Ref.\cite{PhysRevD.62.043008}). It turned out that the qualitative features of these ``quantum'' black hole spacetimes are rather insensitive to the specific form of the function $k(r)$ interpolating between $1/r^{3/2}$ and $1/r$. Therefore, considering the mentioned asymptotic limits we write, as in Ref.\cite{PhysRevD.62.043008},
\begin{align}
\label{dwdwod}
k(r)\equiv\hbar\left(\frac{r+\gamma G_{0}M}{r^3}\right)^{1/2}
\end{align}
Indeed, this function interpolates smoothly the behavior of the proper distance close to $r\to0$ and to $r\to\infty$. Substituting the above equation into \eqref{dwodnwdweo} gives Eq.\eqref{G}.

\section{Legendre Elliptic Functions}
\label{Appendix S}

In this section we present a short summary about elliptic integrals. For further explanations see Ref.\cite{hancock1917elliptic}.\\ Consider an integration of the form
	\begin{equation}
		\label{swwke}
		A=\int \frac{Q(t)}{\sqrt{R(t)}}dt
	\end{equation}
	where $Q(t)$ is an arbitrary rational function of $t$, and $R(t)$ is a fourth degree polynomial in $t$. It is given as
	\be
	\label{dcwdj}
	R(t)=a_{0}t^4+a_{1}t^3+a_{2}t^2+a_{3}t+a_{4}\,,
	\ee
	and all the $a_{i}$ are real numbers. By decomposing $R(t)$ into its factors we get
	\be
	\label{asdsd}
	R(t)=\pm a_{0}(t-\alpha)(t-\beta)(t-\gamma)(t-\delta)\,.
	\ee
	Following Ref.\cite{hancock1917elliptic}, if the roots are all real numbers we suppose $\alpha>\beta>\gamma>\delta$; if two are complex, take $\alpha$ and $\beta$ real, and write the other two as complex numbers $\gamma=\rho+i\sigma$, and $\delta=\rho-i\sigma$; and if all four of the roots are complex, denote them by $\alpha=\mu+i\nu$ , $\beta=\mu-i\nu$, $\gamma=\rho+i\sigma$, and $\delta=\rho-i\sigma$. \\\\
Further, it has been shown that, by considering the real substitutions
	\begin{equation}
		\label{dwdwd}
		t=\frac{p+q\tau}{1+\tau},~~ \tau=\frac{a + b x^2}{c+d x^2}
	\end{equation}
	the integral in Eq.\eqref{swwke} can be simplified to
	\begin{equation}
		\label{dwdwd}
		A=\int \frac{f(x)dx}{\sqrt{(1-x^2)(1-k^2x^2)}}
	\end{equation}
	where $f(x)$ is a rational function of $x$. The evaluation of this integral may be expressed through the evaluation of the three types of integrals defined as
	\begin{align}
		&	\label{sbjqwooe}
		F(k,x)=\int \frac{dx}{\sqrt{(1-x^2)(1-k^2x^2)}}
		\\
		&\label{sbjqwooe}
		E(k,x)=\int \frac{\sqrt{1-k^2x^2}}{\sqrt{1-x^2}}dx
		\\
		&
		\label{sbjqwooe}
		\Pi(n,k,x)=\int \frac{dx}{(1+nx^2)\sqrt{(1-x^2)(1-k^2x^2)}}
	\end{align}
	Considering $x=\sin (\theta)$, we can rewrite the above integrals in the Legendre notation as 
	\textit{elliptic integrals} of the first kind, 
	\begin{equation}
		\label{snswszq}
		F(\phi,k)=\int^{\phi}_{0} \frac{d\theta}{\sqrt{1-k^2\sin^2(\theta)}} \,,
	\end{equation}
	of the second kind,
	\begin{equation}
		\label{wsjws}
		E(\phi,k)=\int^{\phi}_{0} \sqrt{1-k^2\sin^2(\theta)}d\theta \,,
	\end{equation}
	and of the third kind,
	\begin{equation}
		\label{sxss}
		\Pi(n,\phi,k)=\int^{\phi}_{0} \frac{d\theta}{\big(1+n\sin^2(\theta)\big)\sqrt{1-k^2\sin^2(\theta)}} .
	\end{equation}
	Elliptic integrals can also be written in different notations
	\begin{align}
		&\label{ddhbd}
		F(\phi,k)=F(\phi| k^2)=F(\sin(\phi);k)	
		\\
		&	E(\phi,k)=E(\phi| k^2)=E(\sin(\phi);k)\\
		&\Pi(n,\phi,k)=\Pi(n;\phi|k^2)=\Pi(n,\sin\phi;k^2)
\end{align}

\section{Exact integration of equation \eqref{CC} for the star's surface collapse}
\label{Appendix C}

We note that we can write, in general,
\be
\label{R3}
\tilde R^3+\om(\tilde R+\frac{\gamma}{2})=(\tilde{R}-a)(\tilde{R}-b)(\tilde{R}-c)
\ee 
where $a,b,c$ are the roots (real or complex) of the so-called ``depressed'' cubic equation 
$\tilde R^3+\om(\tilde R+\frac{\gamma}{2})=0$. It is easy to show that $a+b+c=0$ should hold.
The differential equation \eqref{CC} describing the surface of the collapsing star can then be written as
	\begin{equation}
		\label{CCvv}
		\dot{\tilde R}(\tilde \tau) =-\frac{\tilde R}{\sqrt{(\tilde{R}-a)(\tilde{R}-b)(\tilde{R}-c)}}\,.
	\end{equation}
	This is a separable equation which originates integrals similar to \eqref{swwke}, so we expect its solutions to contain some kind of elliptic integrals. In fact, the differential equation Eq. \eqref{CCvv} with the usual initial condition $\tilde{R}(\tilde{\tau}=0)=\tilde{R}_0$ can be integrated exactly, and we get
	\begin{align}
		\label{dmdkd}
		&\frac{3}{2}\tilde\tau=-\sqrt{(\tilde R-b) (\tilde R-c)(\tilde R-a)}+\Phi(\tilde{R},a,b,c)\notag\\
		&+\sqrt{(\tilde R_{0}-b) (\tilde R_{0}-c)(\tilde R_{0}-a)}-\Phi(\tilde{R}_{0},a,b,c)
	\end{align}
	where
	\begin{align}
		\label{sxaskss}
		&\Phi(\tilde{R},a,b,c)=-i\sqrt{a-b}(b+2 c)\notag\\
		&F\left(i \sinh ^{-1}\left(\frac{\sqrt{a-b}}{\sqrt{\tilde R-a}}\right)|\frac{a-c}{a-b}\right)\notag\\
		&-\frac{3ibc}{\sqrt{a-b}}\Pi \left(\frac{a}{a-b};i \sinh ^{-1}\left(\frac{\sqrt{a-b}}{\sqrt{\tilde R-a}}\right)|\frac{a-c}{a-b}\right)\,.
	\end{align}
	$F$ and $\Pi$ are the elliptic integrals defined in Appendix \ref{Appendix S}\,.
	Reminding the Vi\'ete formula for the three roots of the depressed cubic equation
	\be
	X^3 + pX + q = 0 \,,\notag
	\ee
	namely,
	\be
	X_k = 2\sqrt{-\frac{p}{3}}\cos\left[\frac{1}{3}\arccos\left(\frac{3q}{2p}\sqrt{\frac{-3}{p}}\right)-\frac{2k\pi}{3} \right]
	\ee
	for $k=0,1,2$, we can write explicitly for the roots $a,b,c$
	\begin{align}
	\label{cccsp;s;}
		&a = 2 \sqrt{\frac{-\om}{3}} \cos \left[\frac{1}{3}\arccos\left(\frac{3\gamma}{4}\sqrt{\frac{-3}{\om}}\right)\right]\notag\\
		&b = 2 \sqrt{\frac{-\om}{3}} \cos \left[ \frac{1}{3} \arccos \left( \frac{3 \gamma}{4} \sqrt{\frac{-3}{\om}} \right) - \frac{2 \pi}{3} \right]\notag\\
		&c = 2 \sqrt{\frac{-\om}{3}} \cos \left[ \frac{1}{3} \arccos \left( \frac{3 \gamma}{4} \sqrt{\frac{-3}{\om}} \right) - \frac{4 \pi}{3} \right]\,.
\end{align}
	Making use now of some basic formulae of complex analysis, namely
	\be
	&&\cos iy = \cosh y\,; \quad\quad \sin iy = i\sinh y\,;\notag\\ 
	&&\arccos y = i\,{\rm arccosh}\,y \quad {\rm for} \quad y\geq1\,; \notag\\
	&&\arcsin iy = i\,{\rm arcsinh}\,y -\frac{\pi}{2} \quad {\rm for\,\,\,any} \quad y\in\mathbb{R} \notag\,,
	\ee
	it is possible to compute the approximation $\om \to 0^\pm$ in Eqs.\eqref{cccsp;s;}, and therefore to get 
	\be
	\label{dnkkd}
	&&a \approx -\sqrt[3]{\frac{\om\gamma}{2}}\,, \notag\\
	&&b \approx -\sqrt[3]{\frac{\om\gamma}{2}}\left(-\frac{1}{2} + i\frac{\sqrt{3}}{2}\right)\,,\notag\\
	&&c \approx -\sqrt[3]{\frac{\om\gamma}{2}}\left(-\frac{1}{2} - i\frac{\sqrt{3}}{2}\right)\,.
	\ee
Notice that the real root $a$ can be obtained directly from Eq.\eqref{R3} in the approximation $\om\to0$.	
We can now study the solution \eqref{dmdkd} in the limit $\tilde{\omega} \to 0$. Substituting \eqref{dnkkd} into \eqref{sxaskss} and using Eqs.\eqref{snswszq},\eqref{wsjws},\eqref{sxss}, 
	the solution \eqref{dmdkd} becomes
	\begin{equation}
		\label{dnfwewe}
		\frac{3}{2}\tilde{\tau}\approx\tilde{R}_{0}^{\frac{3}{2}}-\tilde{R}^{\frac{3}{2}}
		+\frac{3\om}{2}\left(\frac{1}{\sqrt{\tilde{R}}}-\frac{1}{\sqrt{\tilde{R}_{0}}}\right)+...
	\end{equation}
	This is immediately equivalent to Eq.\eqref{approx}, and also to Eq.\eqref{pt} (if we consider that in Eq.\eqref{pt} $\tilde r_0\to 0^+$, and $B/\tilde r_0 \to 1$ when $\om\to 0^-$). Obviously for $\tilde\omega=0$, Eq.\eqref{dnfwewe} becomes $3\tilde{\tau}/2=\tilde{R_0}^{3/2}-\tilde{R}^{3/2}$, which holds for the conventional OS collapsing model in a Schwarzschild metric (see Eq.\eqref{SchColl}, and also Ref.\cite{blau2011lecture}).  
	When $\om<0$, it can be checked that the solution \eqref{dmdkd} is real-valued outside the singularity, 
	namely for $\tilde R>\tilde r_{0}$. 
	\\ 
	Finally, according to solution \eqref{dmdkd}, the star's radius decreases with time and eventually reaches the singularity (if there is one, i.e. in the case $\om \leq 0$).

\section{Horizon as an explicit solution of the $3^{rd}$ degree equation \eqref{HH}}
\label{Appendix D}

\begin{figure}[h]
		\centering
		\includegraphics[scale=0.7]{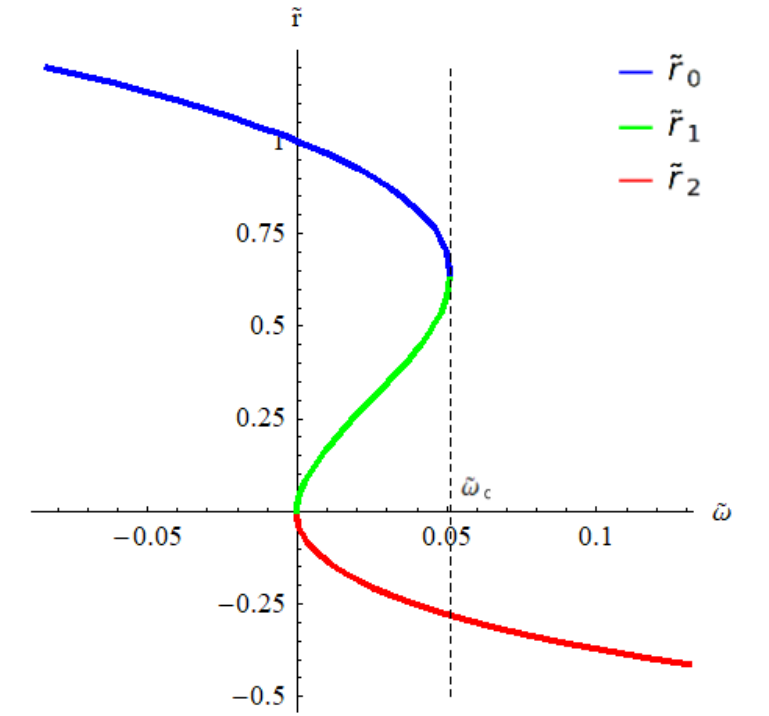}
		\caption{For $\om>0$, the behavior of the roots in Eq.\eqref{cwdwe} is plotted against $\om$. For 
		$0<\om<\om_c$ we have two positive roots, i.e. the horizons $\tilde r_+$ ($n=0$, blue line) and $\tilde r_-$ ($n=1$, green line), and one negative unphysical root $\tilde r_\times$ ($n=2$, red line). For $\om>\om_c$ there is one negative root $\tilde r_\times$ ($n=2$), and two complex conjugate roots, therefore no horizon is present.}
		\label{fig23}
	\end{figure}

\begin{figure}[h]
		\centering
		\includegraphics[scale=0.7]{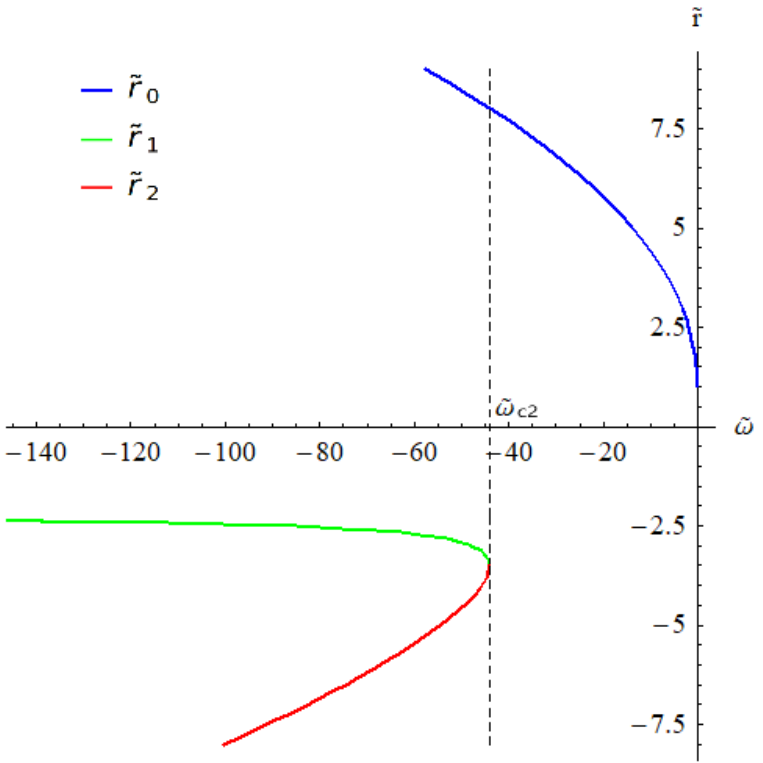}
		\caption{For $\om<0$, the behavior of the roots in Eq.\eqref{cwdwe} is plotted against $\om$. For any $\om<0$ the root $\tilde r_+$ ($n=0$, blue line) is always positive and coincides with the event (outer) horizon, $\tilde r_+=\tilde r_h$. When $\om_{c2}<\om<0$ the other two roots $\tilde r_-$ ($n=1$), $\tilde r_\times$ ($n=2$) are complex conjugate. 
		When $\om\le\om_{c2}<0$, $\tilde r_+$ is still positive, while $\tilde r_-$ (green line), $\tilde r_\times$ (red line) are real but negative, therefore unphysical.} 
		\label{fig24}
	\end{figure}

The explicit algebraic solutions of Eq.\eqref{HH},
	\be
	\tilde r^3-\tilde r^2+\om\left(\tilde r+\frac{\gamma}{2}\right)=0 \,,\notag
	\ee 
	can be obtained by changing the variable to $X$, where $\tilde r=X+\frac{1}{3}$, in order to get a new cubic equation in $X$ that has no term in $X^2$ (depressed cubic). Hence we get
	\be
	\label{II}
		X^3+(\tilde\omega-\frac{1}{3})X+q=0, \quad \ q=\om\left(\frac{1}{3}+\frac{\gamma}{2}\right)-\frac{2}{27}.
	\ee
	This yields us the solutions of \eqref{HH} as follows
	\be
	\label{cwdwe}
		\tilde r_k &=& \frac{2\sqrt{1-3\om}}{3} \cos \left(\frac{1}{3}\arccos A(\om)
		-\frac{2k\pi}{3}\right)+\frac{1}{3} \notag\\
		 && \quad k=+,-,\times \quad {\rm or} \quad k=0,1,2\,,
	\ee
	where 
	\begin{equation}
	\label{Aom}
		A(\om)=\frac{9\om(2+3\gamma)-4}{4(3\om-1)}\,\frac{1}{\sqrt{1-3\,\om}}\,.	
	\end{equation}
	We explicitly note that the above formula works for both cases of three real roots, or one real and two complex conjugate roots. We can now study the three solutions displayed in Eq.\eqref{cwdwe}, $k=0, 1, 2$,  for different values of $\tilde{\omega}$.\\
	From the analysis in Sec.\ref{ddjdjjd} (see Fig.\ref{fig5}) we know that for $0<\om\le\om_c$ (where $\om_c=0.051$ for 
	$\gamma=9/2$) only $\tilde r_+$ and 
	$\tilde r_-$ have real positive values, while $\tilde r_\times \equiv \tilde r_2$ is negative. When $\om>\om_c$ the only real root is $\tilde r_\times \equiv \tilde r_2$ which remains negative.\\
Fig.\ref{fig23} shows that $\tilde r_+ \equiv \tilde r_0$ and $\tilde r_- \equiv \tilde r_1$  have real positive values (for $0<\om\le\om_c$), and therefore can be interpreted as horizon radii, while $\tilde r_\times \equiv \tilde r_2$ is always negative for any $\om>0$, and therefore unphysical.\\
	%
	%
	For any $\om<0$, Eq.\eqref{cwdwe} and Fig.\ref{fig24} show that only $\tilde r_+$ ($k=0$) is a positive root. The other two roots, $\tilde r_-$ ($k=1$), $\tilde r_\times$ ($k=2$), are either both complex conjugate (for $\om_{c2}<\om<0$, where $\om_{c2}=-44.0979$ for $\gamma=9/2$), or both negative, $\tilde r_-<0$ and $\tilde r_\times<0$ (for $\om\le\om_{c2}<0$). So, for any $\om<0$ only $\tilde r_+$ is physically acceptable, and represents the event horizon. The other two roots, $\tilde r_-$, $\tilde r_\times$ are unphysical. See Fig.\ref{fig24} to visualize more details.
	
	%
	%
		%

	According with Eqs.\eqref{cwdwe},\eqref{Aom} for any $\om$ we can write the event horizon radius 
	$\tilde r_h \equiv \tilde r_+ \equiv \tilde r_0$ as follows
	\be
	\label{rh}
		\tilde r_h=\frac{2\sqrt{1-3\om}}{3}\cos\left(\frac{1}{3}\arccos A(\om)\right) + \frac{1}{3}
		\ee
	Making use of the asymptotic behavior
	\be
	\arccos y \sim \sqrt{2(1-y)} \quad {\rm for} \quad y\sim1,  
	\ee
	we can see that in the limit $\om\to 0$, then $A(\om)\to1$ and Eq.\eqref{rh} becomes 
	\be
	\label{rhapprox}
	\tilde r_h\approx 1-\left(1+\frac{\gamma}{2}\right)\,\om+...
	\ee
	so we recover the already known result (Sec.\ref{ddjdjjd}) that $\tilde r_h \to 1^{\mp}$ for $\om\to 0^{\pm}$.
	The same result can be proved by plugging $\tilde r_h = 1+\varepsilon$ back into Eq.\eqref{HH} and then neglecting 
	$2^{nd}$ or higher order infinitesimals.

\section{Implicit analytic form of $\tilde r_{ah}(\tilde \tau)$.}
\label{Appendix E}

To find the evolution of the apparent horizon, using Eqs.\eqref{Hubble0},\eqref{CC},\eqref{AH},\eqref{dmdkd}, we write 
	$\tilde\tau$ as a function of $\tilde{r}_{ah}$,
	\begin{align}
		\label{edwempd}
		\frac{3}{2}\tilde\tau=-(\tilde{r}_{ah}+\frac{1}{\tilde{H}_{0}})+\Psi(\tilde{r}_{ah},a,b,c)-\Psi(-\frac{1}{\tilde{H}_{0}},a,b,c)
	\end{align} 
	where
	\be
	\label{f2}
	-\frac{1}{\tilde H_0} = (\tilde{r}_{ah})_0 = -\frac{\tilde R_0}{\dot{\tilde R}_0} = 
	\sqrt{\tilde R_0^3+\om(\tilde R_0+\gamma/2)}
	\ee
	and
	\begin{align}
	\label{dwlwmwpd}
		&\Psi(\tilde{r}_{ah}, a,b,c)=\frac{i}{\sqrt{a-b}}\Bigg((b-a) (b+2 c)\notag\\
		&F\left(i \sinh ^{-1}\left(\frac{\sqrt{a-b}}{\sqrt{\Xi_k-a}}\right)|\frac{a-c}{a-b}\right)\notag\\
		&-3 b c \Pi \left(\frac{a}{a-b};i \sinh ^{-1}\left(\frac{\sqrt{a-b}}{\sqrt{\Xi_k-a}}\right)|\frac{a-c}{a-b}\right)\Bigg)
	\end{align}
	and
	\begin{align}
		&\Xi_k=2\sqrt{\frac{-\om}{3}}\cos\left[\frac{1}{3}\arccos\left(\left(\frac{3\gamma}{4}-\frac{3\tilde{r}_{ah}^2}{2\tilde\omega }\right)\sqrt{\frac{-3}{\tilde\omega}}\right)-\frac{2\pi k}{3}\right]\notag\\
		&k=0,1,2 \,.
	\end{align}
	Note that we choose the initial conditions at $\tilde{\tau}=0$ as $\tilde R(\tilde\tau=0)=\tilde R_0$,  
	$\dot{\tilde R}(\tilde\tau=0)=\dot{\tilde R}_0$, and therefore 
	$(\tilde r_{ah})_0 = -\tilde R_0/\dot{\tilde R}_0 = -1/\tilde H_0$. \\
	We study the solution \eqref{edwempd} in the limit $\om\to 0$. By considering Eqs.\eqref{dnkkd} and Eqs.\eqref{snswszq}, \eqref{wsjws}, \eqref{sxss} it can be shown that, in such a limit, 
	Eq. \eqref{edwempd} can be written as 
	\begin{equation}
		\label{dcdkwld}
		\tilde{r}_{ah}\approx-\frac{1}{\tilde{H}_{0}}-\frac{3}{2}\tilde\tau\approx\tilde{R_0}^{\frac{3}{2}}-\frac{3}{2}\tilde\tau\,,
	\end{equation}
	where we used Eq.\eqref{f2} to evaluate $-1/\tilde H_0$.
	So, we note that for $\om=0$, Eq.\eqref{edwempd} coincides with Eq.\eqref{ahtau}, which describes the evolution of $r_{ah}$ for the Schwarzschild collapse.
	On the other hand, in the limit $\om\to 0$ from the definition of apparent horizon we get
	\be
	\label{f7}
	\tilde{r}_{ah}(\tilde\tau) \ = \ -\frac{\tilde R(\tilde\tau)}{\dot{\tilde R}(\tilde\tau)} \ \approx \
	\tilde R(\tilde\tau)^{3/2} + \frac{\om}{2}\frac{1}{\sqrt{\tilde R(\tilde\tau)}}+\dots
	\ee
	For $\om\to 0$, equation \eqref{f7} becomes $\tilde{r}_{ah}=\tilde{R}(\tilde\tau)^{3/2}$ which coincides with equation \eqref{dnkebdedf}.
	 Finally, it is easy to see that $\tilde{r}_{ah}(\tilde{\tau}_{f})=\tilde r_{+}$, as expected.

\section{Explicit integration of Eq.\eqref{MM} for the evolving event horizon inside the star.}
\label{Appendix F}

Imposing the initial condition $\tilde{r}_{eh}(\tilde r_{+})=\tilde r_{+}$, we get
	\begin{align}\label{NN}
		\tilde{r}_{eh}=\tilde R+\Upsilon(\tilde R,a,b,c)-\frac{\tilde R}{\tilde r_{+}}\Upsilon(r_{+},a,b,c)
	\end{align}
	where
	\vspace{5mm}
	\begin{align}\label{dkdcmw}
		&\Upsilon(\tilde R,a,b,c)=-(a+2 \tilde R) \sqrt{\frac{(\tilde R-b) (\tilde R-c)}{\tilde R-a}}\notag\\
		&-\frac{i \tilde R}{a \sqrt{a-b}}\Bigg((a-b)\Bigg(3 a E\Big(i \sinh ^{-1}(\frac{\sqrt{a-b}}{\sqrt{\tilde R-a}})|\frac{a-c}{a-b}\Big)-\notag\\
		&(2 a+c) F\Big(i \sinh ^{-1}(\frac{\sqrt{a-b}}{\sqrt{\tilde R-a}})|\frac{a-c}{a-b}\Big)\Bigg)-(ab+ac+bc)\notag\\
		&\Pi \Big(\frac{a}{a-b};i \sinh ^{-1}\big(\frac{\sqrt{a-b}}{\sqrt{\tilde R-a}}\big)|\frac{a-c}{a-b}\Big)\Bigg)
	\end{align}
	Notice that, for $\om<0$, the solutions \eqref{dwlwmwpd} and \eqref{dkdcmw} always get real values whenever 
	$\tilde R(\tilde\tau)\ge \tilde r_0$, namely when the star surface is outside the singularity.\\
	We can study the limit $\om\to 0$ of Eq.\eqref{NN} with the help of Eqs.\eqref{dnkkd} 
and Eqs.\eqref{snswszq},\eqref{wsjws},\eqref{sxss}.
	Remembering also Eq.\eqref{rhapprox}, namely that $\tilde r_+\equiv \tilde r_h \approx 1-(1+\gamma/2)\om$ when $\om\to0$, then Eq.\eqref{NN} becomes 
	\begin{align}
	\label{fverfg}
		\tilde r_{eh}\approx -2\tilde{R}^{3/2}+3\tilde{R}+\frac{\om}{3}\left(\frac{1}{\sqrt{\tilde R}}-\left(4+\frac{3\gamma}{2}\right)\tilde{R}\right) + \dots
	\end{align}
	Notice that Eq.\eqref{MM}, and therefore its solution Eq.\eqref{NN}, has a physical meaning as long as the surface of the star is outside the event horizon, namely $\tilde R > \tilde r_+$. Hence there is no danger that the term 
	$1/\sqrt{\tilde R}$ may explode when $\tilde R\to 0$, simply because $\tilde R > \tilde r_+$ must hold. 
	Finally, we see that for $\om=0$ the above expression coincides with Eq.\eqref{dmdkjdj}, which describes the evolution of the event horizon, inside the star, in the Schwarzschild collapse.


\nocite{*}

\bibliography{RaminFabioFinal_PRD_R1}

\end{document}